\documentclass[reprint, pra,twocolumn,showkeys,showpacs,superscriptaddress,notitlepage]{revtex4-1}
\usepackage{pifont,amsmath,amssymb,amsbsy,times,color,graphicx,multirow,amsfonts,color}
\usepackage[colorlinks=true,citecolor=blue,linkcolor=blue]{hyperref}
\usepackage{physics,bm,here}
\usepackage{silence}
\usepackage[normalem]{ulem}
\usepackage{notes2bib}
\newcommand{\ve}{{\bm e}} 
\newcommand{\vk}{{\bm k}} 
\newcommand{\vp}{{\bm p}} 
 
\newcommand{\xx}{x}
\newcommand{\yy}{y}
\newcommand{\zz}{z}
 
\newcommand{\vR}{{\mathbf{R}}}
\newcommand{\ZZ}{\Bbb{Z}}
\newcommand{\boundary}{\textrm{b}}
\newcommand{\cmark}{\checkmark}%
\newcommand{\xmark}{\times}%
\newcommand{\vex}[1]{\bm{\mathrm{#1}}}
\newcommand{\sigh}{{\sigma}} 
\newcommand{\tauh}{{\tau}}
\renewcommand{\ket}[1]{|#1\rangle}
\renewcommand{\bra}[1]{\langle #1|}

\begin{document}
\title{Fragility of surface states in non-Wigner Dyson topological insulators}
\author{Alexander Altland}
\affiliation{Institute for Theoretical Physics, University of Cologne, 50937 Cologne, Germany}

\author{Piet W. Brouwer}
\affiliation{Dahlem Center for Complex Quantum Systems and Physics Department, Freie Universit\"at Berlin, Arnimallee 14, 14195 Berlin, Germany}

\author{Johannes Dieplinger}

\affiliation{Institute for Theoretical Physics, Universität Regensburg, 93040 Regensburg, Germany
}

\author{Matthew S. Foster}
\affiliation{Department of Physics and Astronomy, Rice University, Houston, Texas 77005, USA}
\affiliation{Rice Center for Quantum Materials, Rice University, Houston, Texas 77005, USA}

\author{Mateo Moreno-Gonzalez}

\affiliation{Institute for Theoretical Physics, University of Cologne, 50937 Cologne, Germany}

\author{Luka Trifunovic}
\affiliation{Laboratoire de Physique Théorique, CNRS, Université Paul Sabatier, 31400 Toulouse, France}
 
\begin{abstract}
Topological insulators and superconductors support extended surface states
protected against the otherwise localizing effects of static disorder.
Specifically, in the Wigner-Dyson insulators belonging to the symmetry classes
A, AI, and AII, a band of extended surface states is continuously connected to a
likewise extended set of bulk states forming a ``bridge'' between different
surfaces via the mechanism of spectral flow. In this work we show that this
mechanism is absent
in the majority of non-Wigner-Dyson topological
superconductors and chiral topological insulators. In these systems, there is
precisely one point with granted extended states, the center of the band, $E=0$.
Away from it, states are spatially localized, or can be made so by the addition of
spatially local potentials. Considering the three-dimensional insulator in class
AIII and winding number $\nu=1$ as a paradigmatic case study, we discuss the
physical principles behind this phenomenon, and its methodological and applied
consequences. In particular, we show that low-energy Dirac approximations in the
description of surface states can be treacherous in that they tend to conceal
the localizability phenomenon. We also identify markers defined in terms of
Berry curvature as measures for the degree of state localization in  lattice
models, and back our analytical predictions by extensive numerical simulations.
A main conclusion of this work is that the surface phenomenology of
non-Wigner-Dyson topological insulators is a lot richer than that of their
Wigner-Dyson siblings, extreme limits being spectrum-wide quantum critical
delocalization of all states vs. full localization except at the $E=0$ critical
point. As part of our study we identify possible experimental signatures
distinguishing between these different alternatives in transport or tunnel
spectroscopy.        
\end{abstract}
\maketitle
\tableofcontents

\section{Introduction}

Topological insulators are subject to a powerful bulk-boundary principle
according to which their insulating (yet topologically nontrivial) bulk implies
conducting boundaries \cite{bernevig2013,ando2015,hasan2010,qi2011}. Examples
include the chiral edge states of the quantum Hall effect, the helical edge
states of the quantum spin-Hall effect, or the single Dirac cones in the surface
spectrum of a three-dimensional topological insulator. In these three cases, the
boundary states extended along surfaces are continuously connected to a band of
likewise delocalized bulk states at high energies. The presence of such ``bulk
bridges'' between surface states is behind numerous physical phenomenana that
make topological insulators stand out against conventional ones. Examples
include anomalous transport, i.e.\ forces applied along one surface driving
currents directed towards another, as in the integer quantum Hall effect, or
topological protection against Anderson localization, safeguarding intra-surface
conduction at arbitrary values of the chemical potential. These phenomena relate
to the principle of \textit{spectral flow}, whereby adiabatic transport between
disconnected surfaces is enabled by a bulk bridge.

\begin{figure}[t]
    \centering
    \includegraphics[width=.9\linewidth]{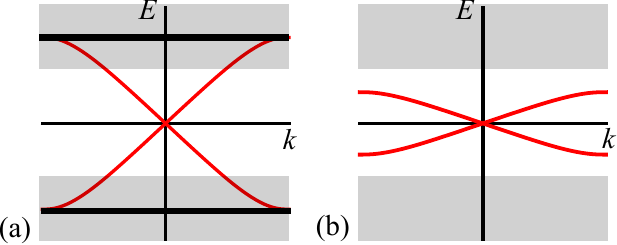}
    \caption{
    (a) A topological free-fermion phase with spectral flow possesses a
    robust continuous attachment of anomalous boundary states (red) to 
    delocalized bulk states (black) residing in a Wannier non-localizable bulk band (grey).    
    (b) Without a spectral-flow principle, anomalous surface bands (red) may in principle be detached from the bulk spectrum by a spectral gap.
    In this case there is no obstruction to Wannier localization of all bulk states (grey). The figure shows a schematic one-dimensional boundary spectrum, assuming translation symmetry parallel to the boundary.}
    \label{fig:spectralflow}
\end{figure}

In this work we show that spectral flow is not as closely tied to the physics of
topological  insulators as one might think. On the contrary, we will demonstrate
that for the majority of three-dimensional topological insulators outside the
three Wigner-Dyson classes A, AI, AII, the two prerequisites for the spectral
flow principle --- a robustly delocalized bulk state and an uninterruptible band
of boundary states continuously connected to it --- are absent. (In the
tenfold-way nomenclature, the quantum Hall effect is class A, whereas  
the three-dimensional topological insulator resides in class AII. In practice,
the non-Wigner-Dyson classes refer to topological superconductors
\cite{schnyder2008,kitaev2009,ryu2010}.) The two opposing scenarios --- with and without spectral flow --- are
illustrated schematically in Fig.~\ref{fig:spectralflow}. The fragility of the
surface bands in the absence of a spectral flow principle leaves room for a
wider phenomenology of surface physics than in the Wigner-Dyson classes. It has
profound consequences for the Anderson localization of surface states and,
hence, for the observable transport characateristics of topological matter in
the presence of disorder.

The absence of a spectral flow principle can be seen from both bulk and boundary
perspectives. From the bulk perspective, it is ruled out if the bulk is
``Wannier localizable,'' {\em i.e.}, if the bulk admits a complete basis of
exponentially localized quantum states. From the boundary perspective, its
absence means that boundary states can be localized by disorder or that they may
be gapped out by a local boundary perturbation. In the topological
non-Wigner-Dyson classes, such ``fragility'' of the boundary states exists away
from the distinguished energy $E=0$ only. The existence of delocalized
boundary states at $E=0$ is robustly protected so long as the bulk remains
topologically nontrivial \cite{essin2015,Schulz-BaldesBook}.
Conversely, a non-localizable phase possesses at least
some Anderson delocalized bulk states above and below the Fermi energy, as well as an uninterruptable \emph{band} of Anderson
delocalized boundary states that are continuously connected (in energy) to
these bulk states.
The spectral flow principle exists in non-localizable phases only.

The bulk and boundary perspectives are closely intertwined in topological matter.
Examples of Wannier localizable phases include topologically trivial band insulators, but also all topological bands in one dimension \cite{kohn1959}, including
topological superconductors \cite{kitaev2001}.
(In one dimension, a continuous band of boundary states is trivially ruled out by phase-space considerations.)
By contrast, two-dimensional topological insulators are non-localizable
\cite{bernevig2013,ando2015,hasan2010,qi2011}.
A case in point is the integer quantum Hall insulator, where the celebrated
Laughlin gauge argument \cite{laughlin1981} shows that a quantized Hall
conductance necessitates 
the existence of delocalized bulk states at energies below the Fermi energy, 
continuously connected to the edge states.
The bulk bridge states in this case are critically Anderson-delocalized in the presence of disorder, 
and associated to the topological Hall plateau transitions \cite{Huckestein95,evers2008}.
Similar arguments have been made for the quantum spin-Hall effect and the three-dimensional topological insulator~\cite{prodan2011}.

In this article, we demonstrate that most topological phases of non-Wigner-Dyson
type are Wannier localizable and, hence, do not possess a spectral flow
principle. In three dimensions, the only exception is the topological class-DIII
superconductor for odd values of its integer invariant. In all other cases ---
classes AIII, CII, CI, and DIII for even values ---
surface states in these phases are fragile in the
sense defined above.

Boundary fragility implies that the surface band {\em can} be detached from the
bulk band and that surface states at $E \neq 0$ {\em can} Anderson localize in
the presence of weak disorder, but not that this {\em must} happen. In fact,
numerical studies of surface states of three-dimensional topological
superconductor classes AIII, CI, and DIII revealed surface states that remain
delocalized in the presence of disorder at all energies
\cite{SWQC-CI,sbierski2020,SWQC-DIII,SWQC-Rev}. Delocalization was observed as a
robust feature of the effective two-dimensional Dirac surface theories of class
CI, AIII, and DIII
superconductors~\cite{SWQC-CI,sbierski2020,SWQC-DIII,SWQC-Rev} and it was backed
up by numerical studies of a three-dimensional lattice model for class
AIII~\cite{sbierski2020}. On the other hand,  topological
arguments~\cite{essin2015}, or even  rigorous mathematical
proof~\cite{Schulz-BaldesBook} for surface state delocalization exist for states
at zero energy only --- consistent with the notion that boundary states in these
classes are fragile for $E \neq 0$.

To answer the question of when fragile boundary states localize in the presence
of weak disorder and when they do not, we have considered the case of a
``minimal'' topological insulator in class AIII in detail, which has a single
surface Dirac cone at zero energy. Using a field-theoretic analysis, we show
that whether or not surface states localize in the presence of disorder is
intimately tied to the presence of {\em surface Berry curvature}
\cite{moreno2023topological}: The surface states localize for $E \neq 0$ if the
surface Berry curvature is nonzero, but not if it is zero. The surface Berry
curvature must necessarily be nonzero if the surface band is detached (in
energy) from the bulk, in which case the fragility of the surface states is
manifest, but it can vanish if there is a continuous connection between surface
and bulk bands.

Without exception, models of class-AIII insulators studied in the literature ---
not only Refs.\ \cite{SWQC-CI,sbierski2020,SWQC-DIII,SWQC-Rev}, but also earlier
work on class AIII \cite{ryu2010} --- have vanishing surface Berry curvature and
a continuous connection between surface and bulk bands. For these models, it
takes the addition of a particular ``surface potential'' to the Hamiltonian to
impose surface Berry curvature and, hence, expose the fragility of the surface
bands. We present numerical evidence that, once such a potential is included,
all surface states away from zero energy are localized in the presence of
disorder, albeit with a localization length diverging at zero energy. 

Even if the surface band structure without disorder does not possess Berry
curvature, a sufficiently random disorder potential itself contains terms that
would induce it. A spatially inhomogeneous potential will then parcel the
surface into domains of positive and negative Berry curvature. Whereas states
inside each domain may localize, domain boundaries host one-dimensional chiral
edge states, similar to the chiral edge states between domains of different
filling fraction in the quantum Hall effect. These edge states form a
percolating network if the spatial average of the Berry curvature is zero,
leading to critical delocalization of surface states at all energies. This is
precisely the ``spectrum-wide quantum criticality'' of quantum-Hall
plateau-transition type previously observed numerically in Refs.\
\cite{SWQC-CI,sbierski2020,SWQC-DIII,SWQC-Rev}. The spectrum-wide delocalization
of Refs.\ \cite{SWQC-CI,sbierski2020,SWQC-DIII,SWQC-Rev} therefore reflects a
{\em statistical symmetry} of a model with zero average surface Berry curvature,
not a topological obstruction to Anderson localization.

Physical properties of  boundary states --- such as their spatial structure at a
given energy and the resulting conduction properties --- are commonly addressed
in terms of effective low-energy approaches that zoom in on linear crossing
points in the boundary spectra. Employing such Dirac, or ``$\vk\cdot \vp$''
approximations, physically relevant parts of the boundary spectrum are
thus described by minimal models of manageable complexity. Our findings show
that such Dirac descriptions can be a dangerously oversimplification
if they are used to capture localization properties of topological surface
states in the non-Wigner-Dyson classes.
Case in point is the minimal Dirac
theory with chiral symmetry, which is the effective description of
topological surface states of an insulator in class AIII used in the
numerical studies of Refs.\ \cite{SWQC-CI,sbierski2020,SWQC-DIII,SWQC-Rev}. This effective two-dimensional description
is strictly {\em without} Berry curvature and, therefore, unable to capture the
geometric effects responsible for surface-state localization of the full
three-dimensional insulator.

\subsection{Outline}

The remainder of this article is organized as follows: 
In Sec.~\ref{sec:2} we present a general argument
showing that Wannier localizability of the bulk
implies the possibility to spectrally
detach surface states from the bulk --- thus underlining the
equivalence the bulk and boundary perspectives on the prerequisites for
the spectral flow principle. We also state the Wannier
localizability status of all
tenfold-way symmetry classes in dimensions up to three. In Sec.~\ref{sec:3}
we derive the same conclusion for the Wannier-localizable class AIII
from the boundary perspective.
In Sec.\ \ref{sec:4} we
consider a canonical four-orbital topological-insulator lattice model for the
class-AIII insulator in three dimensions~\cite{ryu2010}, analogous to the model
that was analyzed in Ref.~\cite{sbierski2020} to demonstrate spectrum-wide
delocalization. For this model, we show that surface Berry curvature
can be induced and the surface bands can be detached from the bulk
bands at a high energy by the addition of a suitably chosen potential. 

Sections \ref{sec:2}, \ref{sec:3}, and \ref{sec:4} consider topological
insulators without disorder. Yet, the implications of the findings of
these Sections have profound consequences for the localization
properties of surface states in the presence of disorder. 
Disorder is
considered 
in Sec.\ \ref{sec:num}, where we present
numerical evidence for Anderson localization of the surface
states of a 3D class-AIII insulator, if (and only if)
the average surface Berry curvature is nonzero.
We also discuss ramifications of our results for the three-dimensional
topological superconductors.
The field-theoretic analysis, which relates surface-state localization
in the presence of disorder to the presence of surface Berry curvature,
follows in Sec.~\ref{sec:FT}. We conclude in Sec.~\ref{sec:6}.

\section{Spectral flow: Bulk perspective} \label{sec:2}

In the following, we show that Wannier localizability of a topological phase
necessarily implies absence of the spectral flow principle. Furthermore,
for class AIII in three dimensions we demonstrate its Wannier localizabilty,
whereas for the remaining symmetry classes we only state the final results. In
the next section we will then discuss gapability of surface states.

\subsection{Wannier localizability implies absence of spectral flow} \label{sec:2a}

A free-fermion insulator is described in terms of a set of Bloch bands. It
is  defined to be \emph{Wannier localizable} if the subspaces defined by
conduction and valence bands admit bases of states, 
$\ket{\Psi_{\mathbf{R}\alpha}}$,
exponentially localized around centers $\vR$, where $\alpha$ 
is an additional index.
Wannier localizability implies the absence of spectral flow, as the ground state
defined by the occupied bands can be represented in terms of individually
flux-insensitive states.  Wannier localizability also is in contradiction to the
presence of protected delocalized states. Finally, Wannier localizability is a
weaker condition than retractability to an ``atomic
limit''~\cite{thouless1984,kuchment2009,ludewig2022,thonhauser2006,soluyanov2011,winkler2016,cornean2017,cornean2017b}.
The possibly small but finite exponential overlap between neighboring unit cells
remains essential to the definition of ground state topology~\cite{read2017},
and to the stabilization of anomalously delocalized surface states.

While the Wannier states $\ket{\Psi_{\mathbf{R}\alpha}}$ are not, in general,
eigenstates of the parent Hamiltonian, it is straightforward to
show \footnote{The corresponding protocol proceeds in three steps: first deform
the Hamiltonian in such a way that its bands are individually flat. Second, in
the projections of the Hilbert space to the band subspaces, apply unitary
transformations (commuting with the flattened Hamiltonian) to the Wannier basis.
Third, if desired, fan out the spectrum to that the Hamiltonian assumes the form
of Eq.~\eqref{eq:Hbulkprime}. }  that a symmetry and gap preserving deformation
can be employed to bring the latter into the form  
\begin{equation}
    H^\prime=\sum_{\vR,\alpha}\varepsilon_{\vR \alpha}\ket{\Psi_{\vR \alpha}} \bra{\Psi_{\vR \alpha}},
        \label{eq:Hbulkprime}
\end{equation}
with parameters $\epsilon_{\textbf{R}\alpha}$ assuming the role of state energies. 

\begin{figure}[t]
    \centering
    \includegraphics[width=\linewidth]{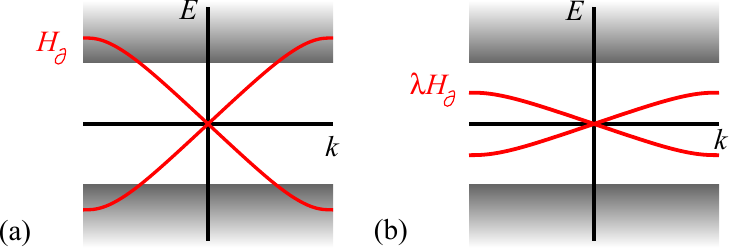}
\caption{
Schematic picture of surface (red) and bulk (black) spectra of the topological insulator Hamiltonian $H' = H_{\partial} \oplus H_{\rm bulk}$, see Eq.\ (\ref{eq:HH}), before (a) and after (b) a rescaling $H_{\partial} \to \lambda H_{\partial}$ to shrink the surface band to a width narrower than the bulk gap.
\label{fig:spectra}
}
\end{figure}

Now consider a system  with surfaces, Wannier localizable in directions
transverse to them. (The existence of states extended \emph{along} the surfaces
in topological insulators excludes unconditional  localizabilty.) Referring to
the above representation, we consider the decompostion~\cite{trifunovic2020b}
\begin{equation}
    H^\prime = H_{\partial} \oplus H_{\rm bulk},
    \label{eq:HH}
  \end{equation}
where the surface Hamiltonian, $H_\partial$, is the  contribution of states   to
\eqref{eq:Hbulkprime} with  centers within a Wannier localization radius of the
surface, and $H_\textrm{bulk}$ its complement. We now have the option to rescale
$H_\partial \to \lambda H_\partial$ to shrink the surface band defined by
$H_\partial$ to a width narrower than the band gap implied by $H_\textrm{bulk}$,
see
Fig.~\ref{fig:spectra}(b). This construction demonstrates the topological
equivalence of our system to one with surface bands detached from the bulk bands
by gaps of adjustable width. Referring to section ~\ref{sec:4} for a microscopic
realization of this construction, we anticipate that 
detachable
surface bands 
in the above sense are best suited to address the surface
phenomenology of insulators without protected spectral flow.

\subsection{Case study: AIII insulator in three dimensions}\label{sec:AIII}

The arguments of Sec.\ \ref{sec:2a} suggest that a continuous attachment of
boundary states to the bulk spectrum and the existence of bulk delocalized
states are flipsides of the same coin. The two-dimensional class A insulator is
case in point for a situation where both exist. We now turn to the opposite
situation, as realized in the three-dimensional AIII insulator, where spectral
flow and delocalized bulk states are generically \emph{absent}. 

A Hamiltonian in class AIII can be written as
\begin{align}
    H&=\begin{pmatrix}
        0 & A\\
        A^\dagger & 0
    \end{pmatrix},
    \label{eq:HAIII}
\end{align}
where $A$ is a complex square matrix acting on the subspaces defined by the
condition $\Gamma=\pm1$, and $\Gamma=\tau_\zz$ defines the chiral symmetry ${\cal S}$. For
example, in a lattice system the subspaces corresponding to $\Gamma=\pm1$ may
define two bipartite sublattices. We assume that $H$ and, hence, $A$ are local
matrices: The matrix elements $\langle \vR \alpha | A |\vR' \alpha' \rangle$
between atomic orbitals $\alpha$ and $\alpha'$ at lattice sites $\vR$ and $\vR'$
decay exponentially with the distance $|\vR-\vR'|$.

For an  insulator subject to periodic boundary conditions the spectrum of $H$
has a finite gap around zero energy. Hence, following Ref.~\cite{ryu2010}, we
may deform the Hamiltonian~(\ref{eq:HAIII}) by sending the positive (negative)
eigenvalues of $H$ to $1$ ($-1$). Such a flattening deformation does not change
the bulk topology and preserves the locality of the Hamiltonian matrix. It
defines the  Hamiltonian
\begin{align}
    H_\textrm{f}&=\begin{pmatrix}
        0 & U\\
        U^\dagger & 0
    \end{pmatrix}.
    \label{eq:HAIIIf}
\end{align}
Locality of $H_{\rm f}$ implies that $U$ is a local unitary operator. ($U$ is
unitary because $H_{\rm f}^2 = \openone$.) We may then easily construct a basis
of localized eigenstates at energy $\pm 1$ \cite{read2017},
\begin{equation}
	\ket{\Psi_{\mathbf{R}\alpha}^\pm} =\frac{1}{\sqrt{2}}
	\begin{pmatrix}
		\ket{\mathbf{R}\alpha}\\
		\pm U^\dagger \ket{\mathbf{R}\alpha}
	\end{pmatrix},
	\label{eq:ev}
\end{equation}
where $\alpha$ is an additional index. (Eigenstates $\ket{\Psi^\pm_{\vR
\alpha}}$ are localized near lattice site $\vR$ because the unitary operator $U$
is local.) This simple construction proves the existence of a basis of localized
eigenstates of $H_{\rm f}$, regardless of the underlying topology~\footnote{This
result is in conflict with the conclusions of Ref.~\cite{song2014}, while it
agrees with the conclusions of Ref.~\cite{hastings2011}.}. Since the
(many-body) ground states of $H$ and $H_{\rm f}$ are the same, Eq.~(\ref{eq:ev})
proves Wannier localizability of this topological phase. The discussion of the
previous subsection then implies the corollary that surface and bulk spectra can be separated from each other.

For later reference we mention that the Wannier functions $\{\ket{\Psi^\eta_{\mathbf{R}\alpha}}\}$ 
[$\eta \in \pm$, Eq.~(\ref{eq:ev})]
and, hence, the
decomposition of $H_{\rm f}$ into surface and bulk contributions as in Eq.~(\ref{eq:HH}) is not unique: Each local unitary matrix $V$
generates another set of Wannier functions $\{\ket{\Psi^{\eta,V}_{\mathbf{R}\alpha}}\}$ by
multiplying both elements of the two-component spinor in Eq.~(\ref{eq:ev}) by
$V$. In Sec.~\ref{Sec:UFO_Surface}, we demonstrate that the surface states
described by  $H_{\partial}$ in Eq.~(\ref{eq:HH}) may
have a nonzero Chern number $\text{Ch}$. This number is constrained to have
the same \emph{parity} as the winding number of the bulk Hamiltonians $H$ and $H_{\rm
f}$. However, for a given parity its (integer) value depends on the choice of the localized basis for
the bulk states: basis changes $\ket{\Psi^\pm_{\mathbf{R}\alpha}} \to \ket{\Psi^{\pm,V}_{\mathbf{R}\alpha}}$
lead to a change~\cite{lapierre2021}
\begin{align}
    \delta\text{Ch}&=2\,\nu[V],
    \label{eq:deltaCh}
\end{align}
where, for a periodically extended definition of the local transformation, 
$\nu[V]
\in \mathbb{Z}$ is the third winding number of $V$ over momentum space.
(An analogous relation was found for the surface response theory of a
three-dimensional topological insulator~\cite{qi2008}.)

\subsection{Other tenfold-way classes}

\begin{table}[t]
\begin{center}
\begin{tabular}[t]{l@{\extracolsep{\fill}} ccc} \hline\hline 
                                class  & $\,d=1$ & $d=2$ & $d=3$\\ %& $d=4$ & $d=5$ & $d=6$ & $d=7$\\ \hline
                                A      & $0$ & $\ZZ^\cmark$ & $0$\\ %& $\ZZ^\cmark$ & $0$ & $\cmark$ & $0$\\
                                AIII   & $\ZZ^\xmark$ & $0$  & $\ZZ^\xmark$\\% & $0$ & $\ZZ^\xmark$ & $0$ & $\xmark$\\   
                                \hline
                                AI     & $0$ & $0$ & $0$\\ %&$\cmark$ & $0$ & $\cmark$ & $\cmark$\\  
                                BDI    & $\ZZ^\xmark$  & $0$ & $0$ \\%& $0$ & $\xmark$ & $0$ & $\xmark$\\
                                D      & $\ZZ_2^\xmark$ &  $\ZZ^\cmark$ & $0$ \\%& $0$ & $0$ & $\cmark$ & $0$\\
                                DIII   & $\ZZ_2^\xmark$ & $\ZZ_2^\cmark$ & $\ZZ^{\cmark/\xmark}$\\
                                %$\begin{cases}
                                AII    & $0$ & $\ZZ_2^\cmark$ & $\ZZ_2^\cmark$\\ %& $\cmark$ & $0$ & $0$ & $0$\\
                                CII    & $2\ZZ^\xmark$ & $0$ & $\ZZ_2^\xmark$\\ %& $\xmark$ & $\xmark$ & $0$ & $0$\\
                                C   & $0$ & $2\ZZ^\cmark$ & $0$\\ %& $\xmark$ & $\xmark$ & $\cmark$ & $0$\\
                                CI   & $0$ & $0$ & $2\ZZ^\xmark$\\ %& $0$ & $\xmark$ &
                                \hline\hline
                        \end{tabular}
                        \caption{Wannier localizability of topological insulators and superconductors. The absence of topological phases is
                        denoted by $0$, entries where topological phases
                        exist are labeled by ``$\xmark$'' for Wannier localizable and
                        ``$\cmark$'' for non-localizable phases. For class DIII in three dimensions, superconductors with
                        even bulk invariant are Wannier localizable, whereas superconductors with odd bulk invariant are non-localizable. The spectral flow principle only applies to non-localizable classes (denoted with $\cmark$). \label{tab:TF}}
        \end{center}    
\end{table}

The discussion above shows that there are two different classes of topological
insulators: those with and without a spectral flow principle. Given that this
dichotomy presides over the spectrum-wide robustness of boundary states, it
seems necessary to tag each entry in the   periodic table of topological
insulators and superconductors~\cite{kitaev2009,schnyder2008,ryu2010} according
to its Wannier localizability status. Referring for the full classification
program to the upcoming publication~\footnote{P.~W.~Brouwer, B.~Lapierre,
T.~Neupert, L.~Trifunovic, in preparation}, Table~\ref{tab:TF} summarizes the
result for dimensions up to three. It adds to the  topological status of a given
symmetry class and dimensionality ($\mathbb{Z}, \mathbb{Z}_2$, or $0$)
information on the localizability of its  states: The absence of topology is
denoted by the entry ``$0$,'' topological classes with symmetry-compatible
exponentially Wannier localizable bulk states are labeled by ``$\xmark$,'' and
those with non-localizable bulk by ``$\cmark$.''

The information provided by Table \ref{tab:TF} confirms that the three
Wigner-Dyson classes A, AI, and AII are always non-localizable, provided they
are topological in the first place. Indeed, since for the Wigner-Dyson classes
one is free to choose a reference energy inside the bulk gap, the existence of a
topologically protected boundary state at one energy implies that such states
must exist for all energies inside the gap. On the other hand, in the
non-Wigner-Dyson classes, the presence of charge-conjugation symmetry
$\mathcal{C}$ and/or chiral symmetry $\mathcal{S}$ forces the spectrum to be
mirror-symmetric around  the distinguished energy $E=0$. The equivalence of all
energies inside the gap therefore no longer holds, and the presence of a
protected uninterruptable band of boundary states must be reconsidered.

For the non-Wigner-Dyson classes, what determines Wannier localizability is
whether  $\mathcal{C}$ or $\mathcal{S}$  are essential for the topology, or
whether they are ``spectator symmetries''  and the topology of the bulk
Hamiltonian remains nontrivial if all constraints imposed by $\mathcal{C}$
and/or $\mathcal{S}$ are lifted. In three dimensions, class AIII is an example
of the former category, which we refer to as ``genuine'' non-Wigner-Dyson
classes. Other examples of genuine non-Wigner-Dyson classes are classes BDI and
D in one dimension and classes CI and CII in three dimensions. Examples of
non-genuine non-Wigner-Dyson classes are classes C and D in two dimensions,
which have chiral edge states \cite{senthil2000,read2000}, and which remain
topological if the particle-hole symmetry is lifted. Class DIII in three
dimensions, which describes time-reversal-invariant superconductors with broken
spin-rotation symmetry is a special case, as it is a localizable genuine
non-Wigner-Dyson class only if the topological invariant is even.

The Wannier localizability of one-dimensional topological insulators (second
column in Table~\ref{tab:TF}) is well known in the literature \cite{kohn1959}.
Examples are the Su-Schrieffer-Heeger model (class AIII)
\cite{su1979,kivelson1982,bradlyn2017} and the topological superconductor 
Kitaev chain
(class D) 
\cite{kitaev2001}, 
which in their nontrivial phases have
localized bases stretching across adjacent unit cells. The possibility of
Wannier localizable topological superconductors in dimensions larger
than one was mentioned by Ono, Po, and Watanabe~\cite{ono2020} in the context of
topological superconductors with additional crystalline symmetries (see also
Ref.~\cite{zhang2020}). Furthermore, the existence of exponentially localized
Wannier basis for topological phases with (possibly higher-order) winding number
invariant was discussed in Ref.~\cite{hastings2011}.

In Sec.\ \ref{sec:2a}, we have shown that Wannier localizability of the
insulating bulk implies that a gap in the surface spectrum may be opened by a
perturbation at the surface. Similarly, away from the energy 
$E = 0$, 
surface states of a Wannier localizable insulator may undergo Anderson
localization in the presence of disorder. We refer to both properties as the {\em
fragility} of surface states. It is important to point out that such fragility only
signals a possibility of a gap opening or of Anderson localization; it does not
imply that a gap in the surface state spectrum {\em must} open or that surface
states {\em must} localize in the presence of disorder. The question, whether or
not surface states of a specific band structure will Anderson localize in the
presence of disorder remains to be answered.  Either way, fragility leaves room for a wider spectrum of surface phenomenologies than in insulators with protected surfaces. We will consider such phenomena in concrete  detail for
the case of a ``minimal'' topological insulator in class AIII in Secs.\
\ref{sec:num} and \ref{sec:FT}.

\section{Spectral-flow: Boundary perspective}
\label{sec:3}

In the previous section, we considered the status of surface states on the basis of a connection to bulk states via the spectral flow correspondence. We here approach the question from a complementary perspective, which focuses entirely on the surfaces themselves.

\subsection{General considerations}

The vicinity of the Fermi crossing points  at topological insulator surfaces is
often described in terms of effective Dirac Hamiltonians, which in the
two-dimensional case assume the form 
\begin{align}
    H_{\boundary} = k_1\Gamma_1+k_2\Gamma_2,
    \label{eq:Dirac}
\end{align}
where $k_i$ are two momenta along the surface measured relative to the crossing
point $\vk = 0$, and the two Gamma matrices satisfy
$\{\Gamma_1,\Gamma_2\}=\delta_{ij}$. Additional Gamma matrices may appear at
higher order in $k$ or as prefactors of a random surface potential. At large
momenta, the Hamiltonian (\ref{eq:Dirac}) is ultraviolet divergent. 
These divergences cannot be cured by embedding $H_\boundary$ into the periodic
structure of a two-dimensional Brillouin zone, reflecting that the  surface theory does not define the low-energy limit of a
two-dimensional stand-alone lattice model.

In the context defined by Eq.~\eqref{eq:Dirac}, the spectral flow principle
translates to the statement of an anomaly: Coupled to an external vector
potential, $H_{\boundary}$, supplemented with an ultraviolet regularization,
lacks gauge invariance. The absence of gauge invariance 
indicates (quasi-)particle number non-conservation. Specifically, under adiabatic
insertion of a flux quantum through the bulk, high-lying boundary states get
pushed up in energy, leading to a drain out of the window of momentum states
below a fixed cutoff. If the spectral flow principle applies, overall particle
number conservation is restored by conversion of boundary states into bulk
states and eventually to states at the opposite surface. The observable
consequence is adiabatic transport from one boundary to the other, {\it i.e.},
the quantized transverse conductance characteristic of topological insulators.

In the previous Section, we have argued that the spectral flow principle does
not apply to all symmetry classes. 
How can this
be reconciled with the intrinsic absence of an ultraviolet closure of the
 Dirac surface theory? To
resolve the  anomaly of the latter there must exist
a ``sink'' of high-lying states absorbing spectral weight being pushed up by an
anomalous gauge operation. If the spectral flow principle applies, these are
extended bulk states. If it does not apply, implying that the
boundary states can be detached from the bulk, these states must be
supported by the boundaries themselves. 

In the following we illustrate these  complementary scenarios on two case studies, class A
in two, and class AIII in three dimensions. In either case, the focus will be
entirely on the boundaries, no explicit reference to bulk states is made. 

\subsection{Case study: Class-A Chern insulator in two dimensions}
\label{sec:Chernboundary}

Two-dimensional class A is the paradigmatic example of an insulator with spectral flow, as realized, e.g., in the physics of the integer quantum Hall effect. In this case, the boundary theory (linearized around any Fermi energy) is governed by the effective Hamiltonian 
\begin{equation}
    H_{\boundary} = k, \label{eq:HC1}
\end{equation}
describing a single branch of chiral fermions. 

\begin{figure}[t]
   \centering
    \includegraphics[width=0.9\linewidth]{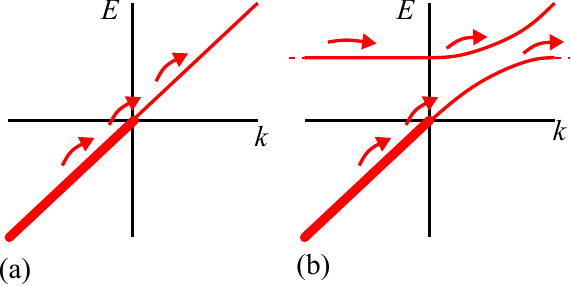}
    \caption{Schematic dispersions of the minimal edge Hamiltonian (\ref{eq:HC1})
    (a) and of the non-minimal model (\ref{eq:HC2}), which has an additional
    band of localized states at the edge (b). Since it derives from a lattice model, the asymptotically flat band from
    the localized states must be continuously connected for $k \to \infty$ and
    $k \to -\infty$, to reflect the periodicity of the edge Brillouin zone.
    Unbounded bands are continuously connected to the bulk spectrum for $k \to
    \infty$ and $k \to -\infty$. The arrows indicate how the occupation of
    states is changed after insertion of a flux quantum through the bulk.
    \label{fig:Chernboundary}}
\end{figure}

Assuming zero temperature, and the Fermi energy at $E=0$ for convenience, states with $k<0$ are occupied. Coupling the system to a gauge potential representing adiabatic
magnetic flux insertion through the bulk, $k \to k+ A$, causes an upward shift
of all levels. After the insertion of one flux quantum, the full quantized
single particle spectrum is restored, but the occupations of the states have
changed, as shown schematically in Fig.~\ref{fig:Chernboundary}(a).
Specifically, one occupied state previously at $k<0$ now occupies the lowest
state at $k>0$. By repeated insertion of flux quanta, the range of occupied
states will extend up to arbitrarily high energies and, eventually, a state
previously sitting at the upper cutoff of the low-energy theory gets
pushed beyond it [see the arrows in Fig.~\ref{fig:Chernboundary}(a)]. 

Topological features must be stable with respect to arbitrary perturbations at the
boundary. To see  how this  robustness manifests  itself in the anomaly of the
boundary Hamiltonian (\ref{eq:HC1}), consider adding a band of trivial localized
boundary states at energy $\varepsilon_\mathrm{c} > 0$. Assuming weak coupling $\gamma$ to the chiral band, the boundary Hamiltonian generalizes to   
\begin{equation}
  H_{\boundary}' = \begin{pmatrix} k & \gamma \\ \gamma^* & \varepsilon_{\rm c} \end{pmatrix}.
  \label{eq:HC2}
\end{equation}
The coupling matrix element $\gamma$ now generates an avoided crossing between
the chiral and the flat band,  see Fig.~\ref{fig:Chernboundary}(b), and a
\emph{local} gap close to the momentum $k\sim \varepsilon_\mathrm{c}$ . However, the
\emph{global} spectrum of the boundary Hamiltonian remains gapless, the reason
being that the band structure of the localized states must be continuous
throughout the boundary Brillouin zone. For the same reason, the addition of the
localized band does not resolve the ultraviolet anomaly of the boundary theory.
Indeed, if flux quanta are inserted repeatedly through the bulk, the occupied
states will eventually completely fill the flat band of localized boundary
states 
and continue to reach the upper cutoff
[see arrows in Fig.~\ref{fig:Chernboundary}(b)].

We now turn to  three-dimensional class AIII and discuss how a construction similar to the one above leads to very different conclusions.

\subsection{Case study: Class AIII insulator in three dimensions}\label{sec:AIII2}

The minimal surface theory of a class AIII insulator with winding number one is described by a two-dimensional generalization of Eq.~\eqref{eq:HC1},
\begin{equation}
  H_{\boundary} 
  = k_x \tau_\xx + k_y \tau_\yy,
  \label{eq:HA1}
\end{equation}
where the chiral symmetry is realized as $H_{\boundary} = -\Gamma H_{\boundary}
\Gamma$ with $\Gamma = \tau_\zz$. Again, this Hamiltonian has no ultraviolet closure. (To see why, note that the off-diagonal element $k_x -i k_y$ defines a winding number around the origin in two-dimensional $k$-space. This is  incompatible with the
$k$-space periodicity required by  a genuine two-dimensional lattice Hamiltonian.) 

A continuous deformation of this \emph{two-band} Hamiltonian 
cannot open a gap at zero energy. The required perturbation would have to be proportional to $\tau_\zz$, in violation of the chiral symmetry.  The  conduction and valence band then  connect the single touching point $\vk=\mathbf{0}$ to the ultraviolet divergences at large momentum; Within this two-band representation, the dispersion is continuous without gap openings at finite momenta.     
It is for this Hamiltonian, augmented with a chiral symmetry respecting random
vector potential, that Refs.~\cite{SWQC-CI,sbierski2020,SWQC-Rev} established a
spectrum-wide resilience to Anderson localization.

\begin{figure}[t]
   \centering
    \includegraphics[width=0.9 \linewidth]{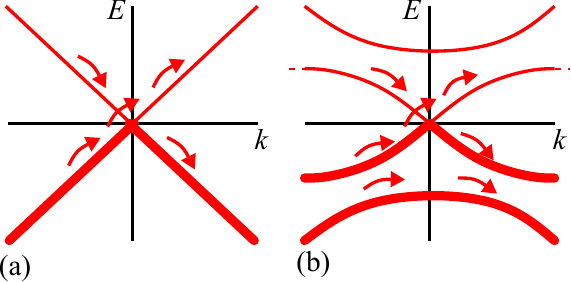}
    \caption{Schematic dispersions of the minimal edge Hamiltonian (\ref{eq:HA1}) (a) and the non-minimal model (\ref{eq:HA2}), which has two additional bands of localized states at the edge (b). The asymptotically flats band from the localized states must be continuously connected for $k \to \infty$ and $k \to -\infty$, to reflect the periodicity of the edge Brillouin zone. The ultraviolet divergent bands from the chiral edge states are continuously attached to the bulk spectrum for $k \to \infty$ and $k \to -\infty$. In the non-minimal model, the high-energy band is detached from the low-energy band containing the linear crossing at zero energy. The arrows indicate how the occupation of states is changed after insertion of a flux quantum through the bulk. \label{fig:Aboundary}}
\end{figure}

As in our previous discussion of class A, we now introduce a band of localized surface states at
energies $\pm \varepsilon_{\rm c}$. As a two-dimensional surface Hamiltonian analogous to Eq.~(\ref{eq:HC2}) we consider
\begin{equation}
  H_{\boundary}' = \begin{pmatrix} k_x \tau_\xx + k_y \tau_\yy & \gamma \tau_- \\
  \gamma^* \tau_+ & \varepsilon_{\rm c}\tau_\xx \end{pmatrix},
  \label{eq:HA2}
\end{equation}
where $\tau_{\pm} = \tau_\xx \pm i \tau_\yy$. As in class A, the band-coupling
$\gamma \tau_-$ defines an  avoided crossing between the localized and the
linearly dispersive bands, see Fig.~\ref{fig:Aboundary}(b). However, unlike in
class A, the continuous interpolation of the bands at the boundaries of the
Brillouin zone no longer presents an obstruction to the opening of a
\emph{global} gap, see Fig.~\ref{fig:Aboundary}(b). This option to disrupt the
surface spectrum was to be expected from the discussion of bulk Wannier
localizability in Sec.~\ref{sec:2}, but here follows from inspection of the surface alone.

We note that  the Hamiltonian~(\ref{eq:HA2}) still has, and needs to have, an
ultraviolet divergence. However, unlike in class A the repeated insertion of
bulk flux quanta no longer leads to a spectral flow anomaly: the occupied states
get shifted as indicated by the arrows in Fig.~\ref{fig:Aboundary}(b), and
eventually fill the large-$k$ part of band of localized boundary states. However
no particles reach the upper energy cut-off. This observation  indicates that
the ultraviolet divergence of effective surface Hamiltonians --- which reflects
their non-existence without a supporting bulk --- does \emph{not} necessarily
imply spectral flow from the surface into the bulk.

In the literature, topological properties that are robust to the addition of
trivial bands are called ``stable''. The sensitivity of the minimal model to the
addition of bands, \emph{i.e.}, the opening of gaps and the disruption of
spectral flow, should therefore be considered a manifestation of ``unstable'' or
``fragile'' topology: the conclusion is that minimal models are insufficient to fathom the full
spectrum of phenomenologies displayed by the surfaces of three-dimensional AIII
insulators.  

Although the concrete worked-out example here is for an AIII
insulator with winding number $\nu=1$, the general conclusion about the fragility
of the spectrum-wide protection of the surface
bands in a minimal effective $2 \times 2$ theory equally applies to higher winding
numbers. A $2 \times 2$ surface theory with a winding number larger than one
may be realized as a nodal point involving higher powers of the momentum $\vk$ 
\cite{SWQC-CI}
or by having multiple nodal points at different locations in reciprocal space. 
In either case, the
combination of chiral symmetry and the
restriction to two-component spinors precludes the interruption of the surface
states by a spectral or mobility gap.

Below Eq.~\eqref{eq:HA1}, we linked the absence of intrinsic UV regularization
of effective surface theories to the presence of a winding number. It is
interesting to observe that the isolation of 
a
detached surface band introduces
another invariant, namely a two-dimensional surface Chern number~\cite{alexandradinata2021,lapierre2021}. Describing the
momentum space topology of the band through the map $\vk\to |\alpha_\vk\rangle$,
where $\vk$ runs through the two-dimensional Brillouin zone, and
$|\alpha_\vk\rangle$ are the positive energy states of the finite band indicted
in Fig.~\ref{fig:Aboundary}b, we define the  Berry curvature 
\begin{equation}
  \label{eq:BerryCurvatureDef}
  \Omega_\vk=i\langle d \alpha_\vk\vert\wedge d \alpha_\vk\rangle,
\end{equation}
and from it the Chern number,
\begin{align}
    \label{eq:ChernNumberDef}
    \textrm{Ch} \equiv \frac{2}{2\pi}\int_\mathrm{BZ}\Omega_\vk.
\end{align}
This is the surface Chern number mentioned previously in section~\ref{sec:AIII};
the prefactor of $2$ accounts for the identical contributions of the positive- and negative-energy surface bands, which are related by chiral symmetry.
While 
the numerical value of $\textrm{Ch}$
may vary depending on the realization of the coupling between the trivial and the chiral bands, its parity is determined by that of the bulk winding number. For further discussion of this point, we refer to section~\ref{Sec:UFO_Surface}.

\section{AIII insulator in three dimensions} \label{sec:4}

In the previous two sections we discussed generic features of topological insulator bulk states and of their asymptotically linearizable surface spectra. We will now turn to a more concrete 
analysis 
and discuss the band structure of a microscopic  model in class AIII.  In the next section we then generalize to the presence of static disorder and discuss localization properties of the model's surface states.

We start our discussion with the definition of the model in subsection~\ref{sec:protlattmodel}. Following standard
protocol, we then project down to its low-energy Dirac approximation in
subsection~\ref{sec:SurfaceDirac}. [Impatient readers may just take notice of the two
principal definitions~\eqref{eq:ham_realspace} of the lattice model
and~\eqref{eq:LudwigDirac} of its Dirac approximation, and directly proceed to
Sec.~\ref{Sec:UFO_Surface}.] On the basis of these model definitions, we then discuss 
the consequences of the absence of a spectral flow principle in
Sections~\ref{Sec:UFO_Surface}--\ref{sec:domain}.

\subsection{Lattice model}
\label{sec:protlattmodel}
We consider a cubic lattice model with four orbitals per site 
defined by the Bloch Hamiltonian~\cite{ryu2010}
\begin{align}\label{HLudwig}
  H(\vk) =&\, \left( M - \sum_{a=\xx,\yy,\zz} \cos k_a \right) \tau_{\yy} \sigma_0
  \nonumber \\ &\, \mbox{}
  + \sum_{a=\xx,\yy,\zz} \tau_{\yy} \sigma_a \sin k_a,
\end{align}
where we set the hopping strength, which is parametrically of the same order as the total band width, to unity for convenience. The Pauli matrices $\sigh_a$ and $\tauh_a$
act on two independent degrees of freedom of the $4=2\times 2$ orbitals in the unit cell. An application of the
standard mapping~\cite{ryu2010} between chiral lattice Hamiltonians and
three-dimensional winding number invariants, 
$\nu$,
shows that
\begin{align*}
    \left. \begin{array}{l}
        \nu = 1 
        \cr 
        \nu = - 2 
        \cr 
        \nu = 0
    \end{array}\right\}\qquad \textrm{for}\qquad \left. \begin{array}{l}
       1< |M| < 3 \cr |M|<1\cr \textrm{else}
    \end{array}\right\}.
\end{align*}
The Hamiltonian~(\ref{HLudwig}) is invariant under the symmetry operations $\mathcal{S}$  and $\mathcal{C}$, 
with $\mathcal{C}^2=-1$,
\begin{align}\label{eq:syms}
    H(\vk) =&\, - M_{\mathcal{S}} \, H(\vk) \, M_{\mathcal{S}} \nonumber \\ =&\,
  - M_{\mathcal{C}} \, H^{\mathsf{T}}(-\vk) \, M_{\mathcal{C}}.
\end{align}
Here $\mathsf{T}$ denotes the matrix transpose, $M_{\mathcal{S}} = \tau_\zz$, and
$M_{\mathcal{C}} = \tau_\yy \sigma_\yy$. We note that the combination
$\mathcal{C}\mathcal{S}=\mathcal{T}$ of  charge-conjugation and chiral symmetry
satisfies $\mathcal{T}^2=-1$, putting our system into class DIII~\footnote{
The model in Eq.~(\ref{HLudwig}) can be viewed as the mean-field description 
of a bulk topological superconductor, see also Sec.~\ref{sec:exp}.
In particular, the Hamiltonian~(\ref{HLudwig}) defines a lattice
representation of the DIII topological superfluid $^3$He-$B$, treated at the
level of static mean-field theory~\cite{volovikbook}. In this interpretation,
the Pauli matrices $\tauh_a$ and $\sigh_b$ act on particle-hole and spin-1/2
space, respectively, and the $p$-wave pairing is encoded by the sine terms. The
pairing amplitude has been set equal to unity, and the topology is controlled by
the ``chemical potential'' $M$. The winding numbers are consistent with those of
the continuum representation, where $\nu = 0$ corresponds to a topologically trivial
BEC phase. In the $^3$He-$B$ interpretation,  the $\mathcal{C}$-symmetry
follows from the reality of the Balian-Werthammer spinor used to define the
Bogoliubov-de Gennes Hamiltonian~\cite{ColemanBook,Foster2014}.}.

To define a  class AIII model, we break $\mathcal{C}$, while
preserving $\mathcal{S}$. For our purposes, it will be convenient to realize
this symmetry breaking by adding $\mathcal{C}$-breaking disorder.  
To this end, we turn to a real space representation of the model~\eqref{HLudwig}, which reads 
\begin{align}
\label{eq:ham_realspace}
  H = H_0 + \sum_{a=\xx,\yy,\zz} H_a,
\end{align}
with
\begin{align}
  H_0 =&\, 
  M
  \sum_{\vR} \ket{\vR} \tau_{\yy} \sigma_0 \bra{\vR}, \\ \nonumber 
  H_a =&\, 
  \frac{1}{2} 
  \sum_{\vR} 
  \Big[ 
        t_{\vR}^a \ket{\vR + \ve_a} (\tau_{\yy} \sigma_0 -i \tau_{\xx} \sigma_a) \bra{\vR} + \mbox{h.c.}
    \Big],  
\end{align}
where the $\vR$ are lattice vectors on the cubic lattice, 
and the $\mathbf{e}_a$, $a=\xx,\yy,\zz$ unit vectors in the lattice directions.
The amplitudes $t_{\vR}^a$, constant in the  clean
case, are now chosen as,
\begin{align}
    t_{\vR}^a \rightarrow t \, e^{- i \phi^a_{\vex{R}}},
    \label{eq:Peierls}
\end{align}
where $a\in\{\xx,\yy,\zz\}$ specifies the direction of the nearest-neighbor bond, and
the $\{\phi^a_{\vex{x}}\}$ are static random phase variables with
variance~\footnote{In the numerical simulations disorder is introduced both in
the bulk and surface layers. On the surface layers, however, it is stronger by a
factor of $\times 5$. This is to enhance surface multifractality of the $E=0$
state.}
 \begin{equation}
     \langle\phi_{\mathbf{R}}^a\phi_{\mathbf{R^{\prime}}}^{a^{\prime}}\rangle=W^2\delta_{\mathbf{R}\mathbf{R}^{\prime}}\delta_{aa^\prime}.
\end{equation}
The effects of this disorder on the (de)localization properties of the surface
states will be discussed in Sec.~\ref{sec:num}.

\subsection{Surface Dirac approximation}

\label{sec:SurfaceDirac}
A bulk winding number 
$\nu$
generically implies the appearance of 
$|\nu|$
species
of gapless Dirac fermions at the surface~\cite{ryu2010}. Specifically, we consider a 
$\nu = 1$
realization of  the model~(\ref{HLudwig}) with $1<M<3$,  a vacuum interface at $x=0$ in the $\xx$-direction, and infinite extension in  $\yy$- and
$\zz$-directions. In this case, a Dirac surface state appears in the
surface Brillouin zone at $(k_\yy,k_\zz) = (0,0)$.

Considering  $\mu \equiv 3 - M$, with $0<\mu\ll 1$, a continuum approximation near the bottom of the band leads to the effective Hamiltonian
\begin{align}
    {H}
    \simeq&\,
    {H}_0
    +
    {H}_\xx,
\end{align}
with
\begin{align}
    {H}_0
    =&\,
        \left(- \frac{1}{2} \frac{d^2}{d x^2} - \mu\right)
        \tauh_\yy  \sigh_0
        +
        \tauh_\xx  \sigh_\xx
        \left(
        -i \frac{d }{d x}
        \right),
\\
    {H}_\xx
    =&\,
        \left(\frac{k_\yy^2 + k_\zz^2}{2}\right)
        \tauh_\yy  \sigh_0
        +
        k_\yy \, \tauh_\xx  \sigh_\yy
        +
        k_\zz \, \tauh_\xx  \sigh_\zz.
\end{align}
The zero modes of ${H}_0$ are 
\begin{align}\label{ZMProj}
    \ket{0,m}
    \equiv  
    \ket{m_z}_\tau
    \ket{m_x}_\sigma
    \ket{\psi},
\end{align}
where $m \in \{\uparrow,\downarrow\}$ denotes the polarization of the surface
state, 
$\ket{m_z}_{\tau}$ and $\ket{m_x}_{\sigma}$ are eigenspinors of $\tau_\zz$ and $\sigma_\xx$, respectively, and $\ket{\psi}$ is an envelope function  decaying exponentially into
the bulk in the $\xx$-direction. The projection of the transverse part of the
Hamiltonian ${H}_\xx$ into the space of zero modes gives
\begin{align}
    {H}_\xx
    \rightarrow 
    \begin{pmatrix}
    0 & k_\zz - i k_\yy \\
    k_\zz + i k_\yy & 0    
    \end{pmatrix}
    =
    \vex{k} \cdot \vex{\Gamma},
    \label{eq:LudwigDirac}
\end{align}
where $\vex{k} \equiv \{k_\zz,k_\yy\}$ and 
$
    {\vex{\Gamma}} 
    =
    \{{\Gamma}_\xx,{\Gamma}_\yy\}
$
are the standard Pauli matrices acting in the
space of zero modes. 

Equation~(\ref{eq:LudwigDirac}) defines the minimal two-component massless Dirac approximation
to the surface states of the bulk model in Eq.~(\ref{HLudwig}) with winding number $\nu = 1$. 
As discussed in Sec.~\ref{sec:3}, the minimal Dirac surface theory has a fragile obstruction 
to localization, which is lifted if additional degrees of freedom are added to the surface
theory. In Sec.~\ref{sec:3} the additional degrees of freedom were added in the form of a
trivial band of localized states at the surface. In the next subsections, we will show that 
the inclusion of the quantum geometric structure of the high-lying states in the full 
three-dimensional theory has the same effect.

\subsection{Detaching and characterizing surface bands \label{Sec:UFO_Surface}}

\begin{figure}
	\centering
	\includegraphics[width=\columnwidth]{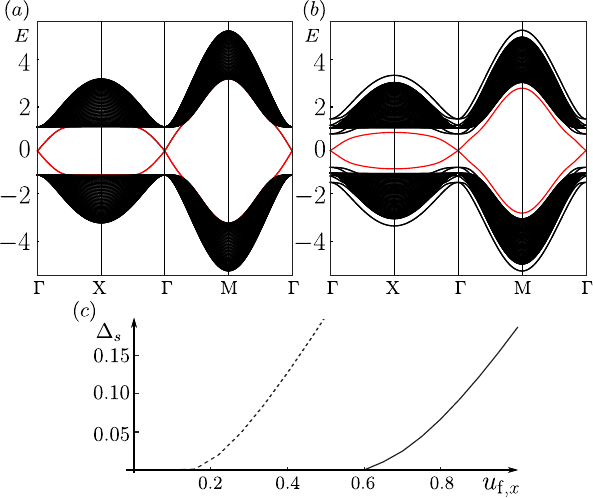}
	\caption{(a) Spectrum of the topological insulator (\ref{HLudwig})
		with $M=2$ for open boundary conditions the $\xx$-direction. The
		bulk (surface) spectrum is shown in black (red). There is a single
		low-energy Dirac cone at $(k_{\yy},k_{\zz}) = \Gamma = (0,0)$. Surface and bulk 
        bands merge at high energy. (b) The same as panel (a) but with the
        additional perturbation $U_{\rm f}$ of Eq.~(\ref{eq:UFO}) with 
        $u_{{\rm f},\xx} = 0.5$ and $n=3$. (c) A  minimal value of the strength $u_{{\rm f},\xx}$ of the fragmenting potential~(\ref{eq:UFO}) required
    to detach surface and bulk bands, for a perturbation with support on the outermost surface layer only ($n=1$, solid) and for a perturbation with support on the three outermost surface layers ($n=3$, dashed). The vertical axis shows the minimal value of the indirect gap $\Delta_{\rm s}$ between surface and bulk bands.}
	\label{fig:bandstructure}
\end{figure}

The  model~(\ref{HLudwig}) has a surface band
that is continuously connected to the bulk, see Fig.~\ref{fig:bandstructure}(a).
We add a term
\begin{align}
    U_{\rm f} &= \sum_{\vR\in \textrm{surface}} \sum_{a=\xx,\yy,\zz} u_{{\rm f},a}
    \ket{\vR}
    \tauh_y \, \sigh_a
    \bra{\vR}
    ,
    \label{eq:UFO}
\end{align}
where the 
$\vR$-summation runs
over the outermost $n$ surface layers.
(We set $n=1$ or $n=3$ in our numerical
calculations.) The perturbation $U_{\rm f}$ breaks $\mathcal{C}$ and
$\mathcal{T}=\mathcal{C}\mathcal{S}$ symmetries, but preserves the chiral symmetry $\mathcal{S}$. We
refer to this perturbation as 
the
``fragmenting surface potential.'' 
For values
 $\vert u_{{\rm f},\xx}\vert>u_{\rm f}^{\rm c}$, we observe the opening of an indirect global gap between the $x$-surface band and bulk bands, see Fig.~\ref{fig:bandstructure}(b). (Detaching surface and bulk bands for surfaces perpendicular to the $\yy$- and $\zz$-directions 
 would require
 nonzero $u_{{\rm f},\yy}$ and $u_{{\rm f},\zz}$, respectively.)
The threshold parameter  $u_{\rm f}^{\rm c}$  
depends on how many surface layers are perturbed by the fragmenting surface potential, see
Fig.~\ref{fig:bandstructure}(c). However, regardless of its 
particular
value, the present construction demonstrates the 
absence of spectral flow.

The band geometry of the surface band is characterized by its Berry curvature $\Omega_{\vk}$, see 
Eq.\ (\ref{eq:BerryCurvatureDef}).  
If the surface band is fully detached from the bulk, the integral of $\Omega_{\vk}$ over the Brillouin zone is well-defined and gives the integer Chern number $\mbox{Ch}$ 
in 
Eq.\ (\ref{eq:ChernNumberDef}).
The parity of the Chern number must be the same as that of the winding number
$\nu$ 
\footnote{To see this, consider two gapped class-AIII Hamiltonians $H$ and $H'$ with
the the same winding numbers $\nu$ and with detached surface bands,  having
Chern numbers $\mbox{Ch}$ and $\mbox{Ch}'$, respectively. The difference
$\mbox{Ch} - \mbox{Ch}'$ is the Chern number of the surface band of the direct sum
$\delta H = H \oplus \overline{H'}$, where $\overline{H'}$ is the
topological ``inverse'' of $H'$, {\em i.e.}, a gapped Hamiltonian with detached 
surface band 
possessing the bulk
winding number $-\nu$ and surface Chern number $-\mbox{Ch}'$.
Since, by construction, 
$\delta H$ has vanishing bulk winding number, its surface band
can be continuously deformed to separate bands at positive and
negative energy without closing the gap between surface band and bulk states
during the deformation process. As the surface spectrum of $\delta H$ is
symmetric around $E=0$, the Chern number $\delta \mbox{Ch}$ of its (full)
surface band is twice that of its positive-energy band and, hence, even. [Note
that the same conclusion was reached in Eq.\ (\ref{eq:deltaCh}) using an
explicit construction of the surface states.]},
but its (integer) numerical value depends on the sign of the potential (\ref{eq:UFO}) 
used to separate the surface band from the bulk,
\begin{equation}
\mbox{Ch} = - \mbox{sign}\,(u_{{\rm f},\xx}).
\label{eq:Ch}
\end{equation}
In the language of Sec.\ \ref{sec:AIII}, the sign of $u_{{\rm f},\xx}$ thus
corresponds to two different bulk gauge choices for the detachment of surface
and bulk, see Eq.~(\ref{eq:deltaCh}). Alternatively, in the language of the
phenomenological surface theory of Sec.\ \ref{sec:AIII2}, different signs of
$u_{{\rm f},\xx}$ represent different perturbations coupling the
dispersing surface band and the degrees of freedom of the external band, see
Eq.~(\ref{eq:HA2}).
The parity constraint on the Chern
number of the surface band implies that for the minimal class-AIII insulator,
detaching the surface band is not possible without inducing surface Berry
curvature.

In Sec.\ \ref{sec:FT} we show that the question whether or not states in a surface band with winding number one (as is the case for the model we investigate here) are localized at energy $E$ in the presence of weak disorder can be answered by considering the integrated Berry curvature carried by the states with energy $\varepsilon_k$ between $0$ and $E$. Hereto we define
\begin{equation}
  \theta(E) = \pi + \int_{0 \le \varepsilon_{\vk} \le E} \Omega_{\vk}.
      \label{eq:ThetaDef}
\end{equation} 
The field theoretical analysis of such a system in the presence of disorder, see
Sec.~\ref{sec:FT}, then shows that states are delocalized at energy $E$ if 
\begin{equation}
  \theta(E) = \pi \mod 2 \pi.
  \label{eq:BerryCriterion}
\end{equation}
Since $\theta(0) = \pi$, this condition is consistent with the topological surface states at $E=0$ being delocalized.

\begin{figure}
	\centering
	\includegraphics[width=\columnwidth]{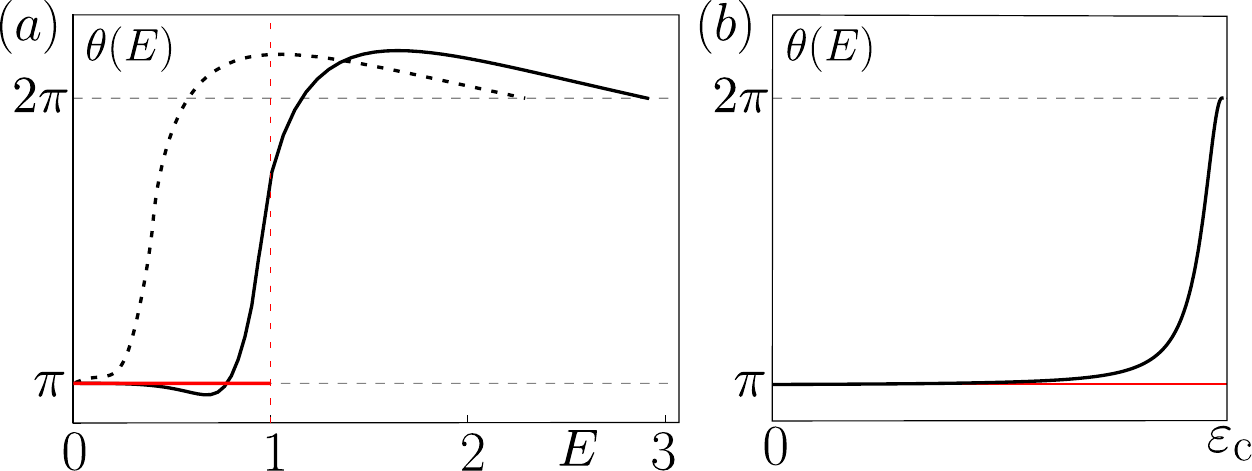}
	\caption{(a) Integrated Berry curvature $\theta(E)$ of the surface
    band of the model~(\ref{HLudwig}) vs.~energy $E$ for the case of zero
    fragmenting potential 
    $u_{{\rm f},\xx}= 0$ 
    (red), 
    $u_{{\rm f},\xx}=-0.3$
    (black, solid), and $u_{{\rm f},\xx}=-1$ (black, dotted). The
    fragmenting potential is added on the three outermost layers adjacent to the
    $x$-surface, $n=3$ and the vertical red line marks the position of the bulk
    gap. For $\vert u_{{\rm f},\xx}\vert>u_{\rm f}^{\rm c}$, surface and bulk
    bands are detached, so that $\theta(E)$ can be determined for the full
    surface band; for 
    $u_{{\rm f},\xx}= 0$, 
    % $\bm u_{\rm f}=0$, 
    $\theta(E)$ can be calculated only for
    energy $E$ inside the bulk gap. (b) The same as panel (a) but for the
    minimal surface theory~(\ref{eq:HA1}) (red) and the non-minimal
    model~(\ref{eq:HA2}) (black) with $|\gamma|/\varepsilon_{\rm c} = 0.05$.
    According to the criterion of Ref.~\cite{moreno2023topological}, surface
    states at energy $E$ are delocalized if $\theta(E) =\pi \mod 2 \pi$.}
	\label{fig:sigma}
\end{figure}

Fig.~\ref{fig:sigma}(a)  shows the angle $\theta(E)$ as a function of $E$ for
$u_{{\rm f},\xx}= - 0.3$ and $u_{{\rm f},\xx} = - 1.0 $
and with
the fragmenting perturbation supported on the three outermost layers ($n=3$). We
note that  the presence of $U_{\rm f}$ has little effect on the surface
Berry curvature for energies close to $E=0$ \footnote{Close inspection
of Fig.~\ref{fig:sigma} (solid curve) shows that $\theta(E) = \pi$ for a finite
value $E \approx 0.8$ for $u_{{\rm f},x} = -0.3 $, corresponding to
delocalization of surface states at that energy according to the
criterion~(\ref{eq:BerryCriterion}). This delocalization does not contradict our
general conclusion that states are localizable in principle at $E \neq 0$, as
one finds, {\em e.g.}, that all states with $E \neq 0$ can be localized upon
increasing the value of $ u_{{\rm f},x}$ [dashed curve in
Fig.~\ref{fig:sigma}(a)].}. This is consistent with the absence of Berry
curvature in the two-component Dirac approximation, which in turn is a
consequence of the chiral symmetry (i.e., the absence of terms on the diagonal of
the $2\times 2$ matrix operator.) However, we observe significant deviations
from $\theta(E)=\pi$ for  energies approaching the edges of the bulk bands. In
the light of our discussion of section \ref{sec:AIII2}, these reflect the onset
of effective hybdridization with extraneous bands, whose role in the present
context is assumed by the bulk.

For comparison, in Fig.~\ref{fig:sigma}(b) we also show $\theta(E)$ for the
two-dimensional band that was obtained by coupling a surface Dirac cone to a
localized surface band, see Eq.~(\ref{eq:HA2}). Analogous to the detached
surface band of the full 3d model shown in Fig.~\ref{fig:sigma}(a), the Berry
curvature is concentrated mainly near the band edges.

\subsection{Chiral-symmetric chiral modes: surface Hall conductance}
\label{sec:domain}

We have seen that addition of the 
perturbation
(\ref{eq:UFO}) detaches the surface bands
from the bulk and that the now isolated surface band has a nonzero Chern number
$\mbox{Ch}$, which for the present $\nu=1$ model is given by Eq.~\eqref{eq:Ch}.
For a surface perpendicular to the $x$ direction, it is interesting to ask what happens if $u_{{\rm f},\xx}$ changes sign, for example along an intra-surface domain wall. As we will see, the ensuing phenomenology is key to the
understanding of the disordered system discussed
in Secs.\ \ref{sec:num} and \ref{sec:FT}.

To explore this situation in the simplest possible setting, we consider a
flattened version of the model~(\ref{HLudwig}). The latter is obtained from the
Hamiltonian~(\ref{HLudwig}) by keeping its Bloch states unchanged, while sending
the energy eigenvalues to $\pm 1$.  To describe an insulator with a surface, we
then switch to the position representation and impose open (or vacuum) boundary
conditions at two coordinates in the $x$-direction. More specifically, we  consider
an annular cylinder geometry with two surfaces in the radial $x$-direction, periodic
boundary conditions in circumferential $\yy$-direction, and the cylinder axis in
$\zz$-direction.  

\begin{figure}
	\centering
	\includegraphics[width=\columnwidth]{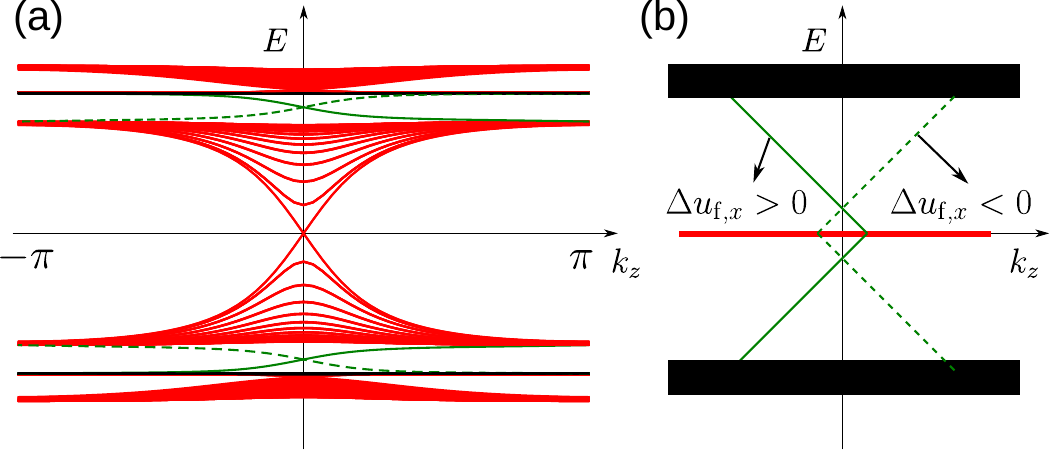}
	\caption{(a) Band structure for a slab geometry of the flattened version of
	the model~(\ref{HLudwig}) with open boundary conditions along $x$ and two
	symmetrically placed domain walls and periodic boundary boundary conditions
	along $\yy$. The fragmenting potential is of the form $u_{{\rm f},\xx}(x,y)$, where $u_{{\rm f},\xx}(x,y)$ switches from $0.2$ to $-0.2$ and back at the two domain walls in the $n=1$ outermost surface layers, and is zero otherwise. 
    The bulk, surface, and domain-wall part of the
	spectrum are indicated in black, red, and green, respectively. Solid and
	dashed green curves are for the domain wall for which $u_{{\rm f},\xx}$ goes from negative
	to positive and from positive to negative upon increasing $\yy$, respectively.
	Note that the addition of the fragmenting potential to the surface raises part of the
	surface spectrum above the flattened bulk band. (b) In the presence of
	disorder, only surface states at $E=0$ are delocalized. In this case, the
	chiral domain-wall modes  are expected to extend from the bulk spectrum all
	the way down to zero energy.
    }
	\label{fig:DW}
\end{figure}

Figure~\ref{fig:DW} shows that for a constant fragmenting potential $u_{{\rm f},\xx} $, see Eq.~(\ref{eq:UFO}), the flattened
model shows a \emph{global} gap between surface (red) and flat bulk (black)
bands. However, the spectrum becomes more interesting, once we introduce two
surface domain walls parallel to the $\zz$-direction where $u_{{\rm f},\xx}$ switches sign.
(Periodicity in $\yy$-direction requires the presence of two of these.) In this
case, we observe the formation of two counterpropagating chiral modes bound to
the respective domain walls. The spectrum of these modes, indicated in green in
Figure~\ref{fig:DW}(a),
connects the surface and the bulk spectrum.

In the clean model, these chiral modes are supported only by states inside the
high-lying band gap. Below it, they hybridize with the extended surface states.
In the presence of disorder, the surface states at $E \neq 0$ will localize, but 
the chiral edge modes do not, so that, for sufficiently strong disorder, the
chiral modes will eventually extend their support over the entire surface spectrum, 
as shown in
Fig.~\ref{fig:DW}(b).

At energies $E \neq 0$ the chiral symmetry is effectively broken, so that we may
think of each surface as a two-dimensional system in class A. In this reading,
the counterpropagating chiral modes at the domain walls surrounding a surface region
with a different sign of the fragmenting potential acquire a status equivalent to the edge
modes of a quantum Hall insulator, and Laughlin's gauge argument applies. It
requires that the branches of chiral modes \emph{must} eventually hybridize with
extended states, at $E=0$ as well as at high energies. The extended states at $E=0$ 
are the topologically protected delocalized surface states of the class-AIII 
insulator. At high energies, 
the chiral modes must hybridize with delocalized bulk states or with energetically
high-lying delocalized surface states. Either way, the presence of the domain wall 
modes prevents a full localizability of all states at large energies.

In anticipation of our later in-depth discussion of disorder, it will be  rewarding to link the presence of chiral edge modes to transport coefficients. To this end, consider a surface geometry where a puddle of given value of $u_{{\rm f},\xx}$  is surrounded by an outer region with $u_{{\rm f},\xx}$ of opposite sign. The presence of a chiral edge mode at the puddle boundary implies that a fictitious four-terminal measurement of its Hall conductance would yield the result $\sigma(E)=\mathrm{sgn}(E) \,\mathrm{sgn}(u_{{\rm f},\xx})/ 2$  for all energies inside the mobility gap where the chiral mode exists. (The factor $1 /2$ reflects the fact that the surface is governed by a single Dirac fermion species, with its characteristic half-integral transverse conductance. Chiral symmetry requires $\sigma(E)$ to be an odd function of $E$~\cite{trifunovic2019b}. At the band center, $\sigma(0)=0$, again by chiral symmetry.).

Now imagine a profile $u_{{\rm f},\xx}(y,z)$ (in the $n$ outermost layers) smoothly varying in such a way that
the spatial average 
of $u_{{\rm f},\xx}$ vanishes and puddles with fragmenting potentials of
opposite sign form with equal probability.  Since each puddle is surrounded by
its own chiral mode, we expect the formation of a network in which co- and
counterpropagating loops occur with equal likelihood. This system is topologically equivalent to the Chalker-Coddington
network~\cite{chalker1988} of the integer quantum Hall effect \emph{at
criticality}. At this point, the network model predicts the percolation of
quantum states evading Anderson localization in the presence of even strong
disorder. This simple picture --- which is the mechanism behind the ``statistical topological insulator'' \cite{fulga2014} --- is compatible with the observation of model
realizations with a spectrum-wide existence of delocalized surface
states~\cite{sbierski2020}. (The same statistical mechanism underlies delocalization of
a topological-insulator surface in a random magnetic field \cite{nomura2008} and of the surface states of a weak topological insulator \cite{ringel2012,mong2012,Fu2012}.) On the other hand, we expect localization of 
finite-energy surface states if a non-vanishing average 
of the fragmenting potential causes  an 
imbalance between puddles with opposite signs of  $u_{{\rm f},\xx}$. Note that these
predictions are consistent with the expectation that it is the presence or
absence of Berry curvature, corresponding to the presence or absence of an
average surface 
fragmenting potential,
that decides over localization. In the next section, we
will back these hypotheses by a quantitative analysis of disorder.

\section{Surface localization properties \label{sec:num}} Previous sections
demonstrated that the surface states of a class AIII topological insulator can
be detached from the bulk. Concomitant with the opening of the spectral gap, the
2D surface acquires a nonzero Chern number due to induced surface Berry
curvature, as explicated above in Secs.~\ref{sec:AIII2} and
\ref{Sec:UFO_Surface}.

Here and in the next section, we consider the implications of surface
Berry curvature for the Anderson localization properties of the surface states
in the presence of symmetry-preserving disorder. Since the minimal 2-component
surface Dirac theory is void of curvature, we work with a slab of the 3D lattice
model defined in Eq.~(\ref{eq:ham_realspace}). Surface Berry curvature is
induced along the slab boundary via the fragmenting potential introduced in
Eq.~(\ref{eq:UFO}), above.

We demonstrate below (see Fig.~\ref{fig:mfa}) that in the presence of a
nonzero uniform fragmenting potential all surface states are localized by weak
disorder, except the zero-energy one, which remains topologically protected
from Anderson localization. By contrast, spectrum-wide criticality ---
{\em i.e.}, critical delocalization linked to the plateau transition of the
quantum Hall effect \cite{sbierski2020} --- survives when either (a) the surface
fragmenting potential is set to zero, or (b)
the surface fragmenting potential is
randomly distributed with zero mean. Scenario (b) explains the origin of
spectrum-wide criticality as the percolation of chiral edge modes discussed in
the end of the previous section.

We discuss detectable ramifications of our results for experiment in Sec.~\ref{sec:exp}.

\subsection{Surface Berry curvature and disorder: Numerics}

We perform a numerical study of  localization in a disordered version of the
model~(\ref{HLudwig}), with and without a uniform fragmenting potential $U_{\rm f}$, see Eq.\ (\ref{eq:UFO}). Disorder is
implemented by random Peierls phases as introduced in Eq.~(\ref{eq:Peierls}). We
apply multifractal analysis to decide whether the surface wave functions are
localized or delocalized~\cite{evers2008,rodriguez2011}. 

It turns out that the  spectral weight  of surface wave functions is dominantly ($>75 \%$) 
concentrated on the outermost  surface layer. 
We define the surface {\em inverse participation ratios} (IPR) of these wave functions via the moments, 
\begin{equation}
    P^E_q=\frac{\sum_{y,z}\left(\sum_{\sigma} \mid \psi^E_\sigma (y,z) \mid^{2}\right)^q}{[\sum_{\sigma,y,z} \mid \psi^E_\sigma (y,z) \mid^{2}]^q},
    \label{eq_IPR}
\end{equation}
where $\psi_\sigma^E(y,z)\equiv \psi_\sigma^E(x=0,y,z)$ are 3d wave functions of
energy $E$ evaluated  at  $x=0$. The IPRs are normalized such that $P^E_1=1$. 
To improve statistics, we consider $P^E_q$  averaged over disorder and a small
window of energy. (For system sizes from $N_y = N_z = L=16$ to $L=128$, 
the number of wave functions over which the moments are averaged ranges from $3\times10^5$ to $10^3$.) Details concerning the
convergence of our data with the slab thickness and the distribution functions of the IPR are provided in Appendices~\ref{app:distros} and~\ref{app:conv}.

\begin{figure*}[t]
    \centering
    \includegraphics[width=\linewidth]{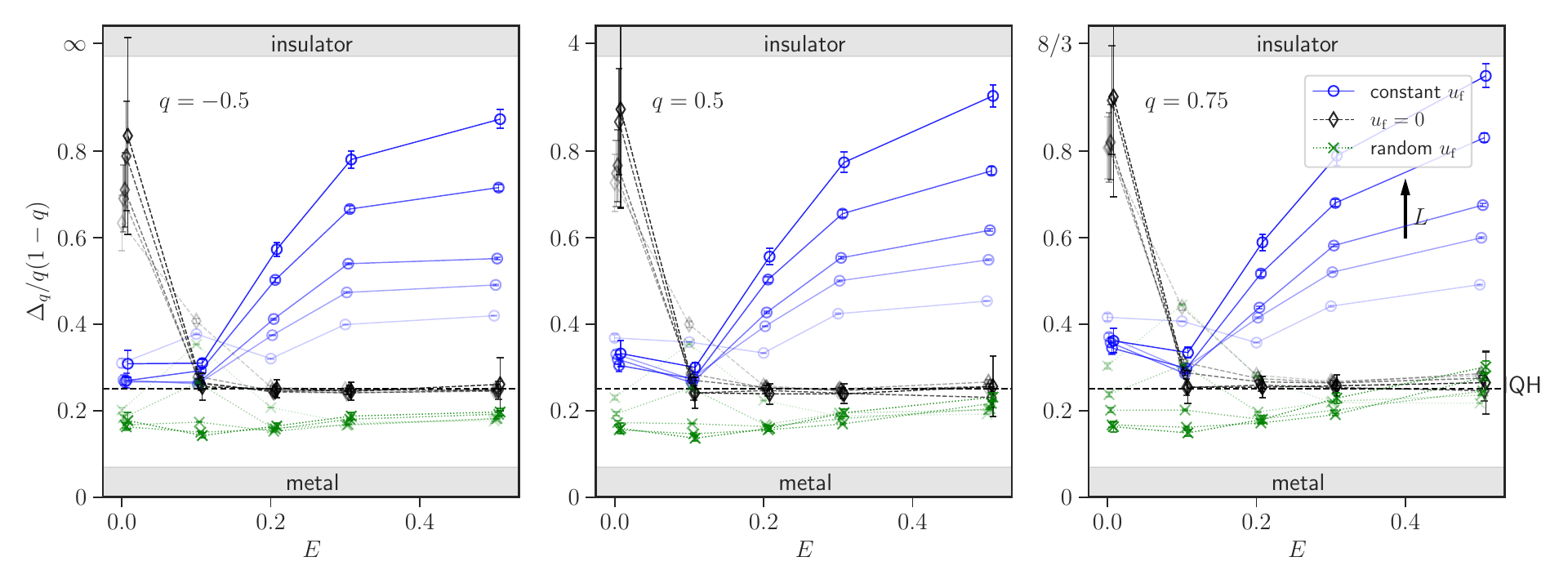}
    \caption{The scaled exponent, $\Delta_q/ q(1-q)$  for $q=-0.5,0.5,0.75$
    (left to right) and system sizes $N_x=8$ and $N_y=N_z=L=24$ to $L=128$ of
    the model~(\ref{HLudwig}). Black data: no fragmenting potential $u_{\rm
    f}=0$ [Eq.~(\ref{eq:UFO})] and disorder strength $W=0.15$; Blue data: a
    constant potential with $u_{\rm f}=0.3$ applied to the outermost surface
    layers, and disorder strength $W=0.2$; Green data: random fragmenting
    potential with zero average and standard deviation $u_{\rm f}=0.3$
    [Eq.~(\ref{eq:variance})], and disorder strength $W=0.2$. The horizontal
    dashed lines marks the value $0.25$ of quantum Hall criticality, and the
    localization limit, $\tau_q=0$ corresponds to $\Delta_q/q(1-q)\to \infty, 4,
    8/3$ for $q=-0.5,0.5,0.75$ . (The value $\infty$ reflects the formal
    divergence of the IPR in the localized limit for negative $q$.)}
    \label{fig:mfa}
\end{figure*}

For surfaces of large linear extension $L$ \footnote{The computational cost of
the sparse matrix diagonalization, carried out using the ARPACK library scales
with the Hilbert space dimension of the 3d system, {\em i.e.}, with $4\times
N_x\times L^2\leq 5\times 10^5$.} the IPRs are expected to asymptotically scale as 
\begin{align}
    P^E_q\propto L^{-\tau^E_q},
    \label{eq:powerlawIPR}
\end{align}
with an effective dimension $\tau^E_q$ \cite{evers2008}. Multifractality  manifests itself in the appearance of a non-trivial
anomalous dimension
\begin{equation}
    \label{eq:DeltaqDef}
    \Delta_q^E=\tau_q^E-d(q-1),
\end{equation}
measuring deviations from the naive dimension $d(q-1)$ expected for uniformly
distributed states. The opposite extreme is that of localized states, for which
$\tau_q=0$, reflecting the absence of scaling in system size. 
Presently, we are discussing a system with two distinct realizations of critical points. The first is the mirror symmetric point, $E=0$, marking the position of a topologically protected critical state. In the vicinity of this point, the system is expected to show multifractality with the anomalous dimensions~\cite{ludwig1994,evers2008}, 
\begin{align}
    \label{eq:AIIIAnomalousDimension}
 \Delta_q^\text{AIII}=\Theta^\mathrm{AIII}q(1-q),
\end{align}
where  $\Theta^\mathrm{AIII}$ is a non-universal coefficient depending on the disorder strength. 
The second is the quantum criticality otherwise realized by states sitting at the center of Landau levels in quantum Hall 
systems. For these states, the spectrum of scaling dimensions is approximately parabolic
\cite{Huckestein95}
\begin{equation}
    \Delta_q^\text{QH}
    \simeq
    \Theta^\text{QH} q (1-q),
    \label{eq:qh}
\end{equation}
with $\Theta^\text{QH} \simeq 0.25$
\cite{evers2008,Obuse2008,evers2008a,zirnbauer2019}. In either case, the spectrum  is expected to be approximately parabolic up to a threshold  $|q|\simeq q_c=\sqrt{2/\Theta}$~\cite{evers2008,ludwig1994,Foster2014}.

Figure \ref{fig:mfa}  shows the anomalous multifractal exponent
$\Delta_q/q(1-q)$ for $q=-0.5,0.5,0.75$ (left to right) and vanishing (dashed),
constant  (blue) and random (green) fragmenting potential. The different
curves show data obtained for increasing system sizes $N_y=N_z=L=24$ to $128$ as
a function of energy, $E$ . Numerically, we calculate an effective $L$-dependent
multifractal exponent, with a finite logarithmic difference between IPRs of
increasing system size $L$. The details of this procedure are delegated to
Appendix \ref{app:Numerics}. At $E=0$ we obtain $\Theta^\text{AIII}\approx
0.85\pm 0.2$ ($u_{\rm f}=0$), $0.32\pm0.05$ (constant $u_{\rm f}$), and $0.18\pm0.03$ (random
$u_{\rm f}$). 
Here $u_{\rm f}\equiv u_{{\rm f},x}$ determines the fragmenting potential $U_{\rm{f}}$~(\ref{eq:UFO}).
The approximate independence of these values on the value of $q$ and $L$
indicates that we are observing the anomalous dimension
\eqref{eq:AIIIAnomalousDimension} of the $E=0$ quantum critical point. Away from
$E=0$, our results sensitively depend on the chosen model for $u_{\rm f}$:

\paragraph*{Zero fragmenting surface potential:} For $u_{\rm f}=0$ the data quickly drops
to the value Eq.~\eqref{eq:qh} expected at quantum Hall criticality. This value
is maintained, including for large energies inside the bulk gap. In this way we
confirm the  observation of spectrum-wide criticality of
Ref.~\cite{sbierski2020}.  

\paragraph*{Constant fragmenting surface potential:} Upon application of a
constant $u_{\rm f}=0.3$, we observe a clear tendency away from criticality and towards
localized behavior upon increasing the system size. We note that the
perturbation of strength $u_{\rm f}=0.3$, presently applied to only one surface layer,
is by a factor of two below the threshold $u_{\rm f}^{\rm c} \simeq 0.6$ required to induce an
indirect gap below surface and bulk band,  see Fig.~\ref{fig:bandstructure}(c).
(We restrict our attention  to $|u_{\rm f}| < u_{\rm f}^{\rm c}$, because larger values require
stronger disorder to observe effects in finite size.) This finding is consistent
with the expectation that  fragility of the  surface-bulk connection --- and 
consequently
eigenstate 
localization at large length scales --- will be induced by
any constant non-vanishing $u_{\rm f}$. 

\paragraph*{Random fragmenting surface potential:}
Motivated by the scenario laid out at the end of Sec.~\ref{sec:domain}, we
consider a spatially varying surface deformation $u_{\rm f}(x,y,z)$ of unit layer depth, vanishing average, and variance  
\begin{equation}
  \label{eq:variance}
    \langle u_{\rm f}(x,y,z) \, u_{\rm f}(x,y^\prime,z^\prime)\rangle_{y,z}
    =
    u_{\rm f}^2
    \,
    \delta_{y,y^\prime}
    \,
    \delta_{z,z^\prime}
\end{equation}
for $x=0$ and $x=N_x$. The green data shows that this perturbation leads to delocalized and quantum Hall critical behavior at finite energies, much as that of the unperturbed model. However, the convergence towards the quantum-Hall exponent is
 slower than in the absence of a fragmenting potential. For further discussion of this point we refer to Appendix~\ref{app:distros}.

We conclude from the three sets of data (black, blue, green) presented in
Fig.~\ref{fig:mfa} that an arbitrary perturbation of the surfaces is not
necessarily sufficient to localize the finite-energy states. Instead, a
deformation that induces a nonzero average surface Berry curvature [e.g.,
$\langle u_{\rm f}(x,y,z) \rangle_{y,z}\neq 0$] is needed. This conclusion is consistent
with the considerations of Sec.~\ref{Sec:UFO_Surface}, where it was argued that
a uniform perturbation of the form in Eq.~(\ref{eq:UFO}) induces surface Berry curvature. 

As discussed in Sec.~\ref{sec:domain}, spatial fluctuations of the fragmenting potential with zero average, on the other hand, lead to a percolating network of
chiral domain-wall modes. This percolating network appears at all nonzero
surface-state energies as quantum-Hall criticality. 

Localization was not observed in previous continuum studies
\cite{SWQC-CI,sbierski2020,SWQC-Rev} that employed a 2D minimal Dirac
description, as this carries exactly zero Berry curvature as long as chiral
symmetry is preserved. The fragmenting potential \emph{projects to zero} in the
minimal Dirac approximation. Although we consider a phase with only a single
surface Dirac cone (2 surface bands), the Berry curvature necessary to localize
the surface states appears in the full 4-component description of the
surface-state wave functions, when the fragmenting potential is applied to the
lattice Hamiltonian in Eq.~(\ref{eq:ham_realspace}). Alternatively, localization
should occur in the continuum Dirac description when the latter is wedded to a
trivial band in such a way so as to induce surface Berry curvature, see
Sec.~\ref{sec:AIII2}. In both cases, it is essential to retain additional
degrees of freedom beyond the minimal Dirac description in order to decide the
fate of the surface states in the presence of disorder.

\subsection{Implications for experiment \label{sec:exp}}

Interpreted as a topological insulator with sublattice symmetry, the model in Eqs.~(\ref{HLudwig}) and (\ref{eq:ham_realspace}) is rather artificial. Although clean systems with approximate sublattice chiral symmetry appear in nature (e.g., graphene), simple onsite impurity potentials destroy the symmetry and revert the system to a Wigner-Dyson class. For this reason, topological phases in classes CI, AIII, and DIII have received far less attention than the quantum-Hall and quantum-spin Hall insulators. 

However, the non-Wigner-Dyson classes admit natural interpretations as 3D topological superconductors \cite{schnyder2008}. Then a lattice model as in Eq.~(\ref{eq:ham_realspace}) can be viewed as the Bogoliubov-de Gennes quasiparticle Hamiltonian in static mean field theory \cite{ryu2010,Foster2014}; indeed Eq.~(\ref{eq:ham_realspace}) can be interpreted as a lattice regularization of the topological superfluid ${}^3$He-$B$~\cite{Note4}. Although quantum fluctuations are inevitable in a non-$s$-wave topological superconductor, the notion of topology is expected to carry through to fully interacting phases of matter~\cite{chiu2016}.

Three-dimensional topological superconductors are protected by physical time-reversal symmetry (which transmutes into the chiral $\mathcal{S}$ condition in Eq.~(\ref{eq:syms}) within the Bogoliubov-de Gennes language \cite{schnyder2008}), and varying degrees of spin SU(2) symmetry. 
Class CI, AIII, and DI superconductors respectively possess SU(2), U(1), and no spin rotational symmetry. Beyond time-reversal and spin symmetries, no other restrictions are placed upon lattice structure or disorder realizations;
see Appendix~\ref{sec:AIIISC} for the explicit mapping in class AIII.
This means that generic non-magnetic impurities do not alter the symmetry class for these superconductors.

The surface fluid of a bulk topological superconductor consists of unpaired fermion quasiparticles (``Dirac fermions'' for classes CI and AIII, ``Majorana fermions'' for class DIII). This fluid can dominate certain observables at low temperature $T$. In particular, a clean surface-Dirac cone gives a power-law-in-$T$ contribution to the Meissner effect, due to the paramagnetic current of the surface \cite{Wu2020}. The surface quasiparticles also contribute to the longitudinal thermal and (for classes CI and AIII) spin conductivities.
By contrast, the contribution of the fully gapped bulk is exponentially suppressed for these quantities in the $T \ll \Delta_0$ limit, where $\Delta_0$ is the bulk superconducting gap. 

Disorder is inevitable in real materials, and particularly at crystal boundaries. Then, the alternative scenarios of spectrum-wide criticality versus surface Anderson localization produce very different phenomenologies. Localization suppresses surface conduction, which can then be mediated at finite temperature only through inelastic processes. This should suppress the surface contribution to the Meissner effect. Without inelastic scattering, the finite-$T$ surface thermal conductivity vanishes with surface localization in the thermodynamic limit; this is because the single delocalized state at zero energy is a set of measure zero in the surface spectrum. In reality, dephasing  stabilizes a surface contribution at finite $T$, analogous to the longitudinal conductivity measured at the plateau transition of the quantum Hall effect \cite{Wang2000}.
By contrast, spectrum-wide quantum criticality should yield a universal surface thermal conductivity determined (via the Wiedemann-Franz relation) by the average \emph{zero-temperature} electrical conductivity of the quantum Hall plateau transition \cite{sbierski2020}.

Which scenario is expected to be realized experimentally? 
A main message of this paper is that microscopics are necessary to determine the presence or absence of average surface Berry curvature; the latter is responsible for surface localization with disorder. We note that for the cubic lattice model in Eq.~(\ref{HLudwig}), the fragmenting surface potential in Eq.~(\ref{eq:UFO}) used to induce the localization in Fig.~\ref{fig:mfa} breaks the average cubic rotational symmetry. Equivalently, different surface perturbations are needed to induce average surface Berry curvature on different surfaces. This suggests that point-group symmetry-breaking perturbations tailored to particular crystal terminations may be necessary to induce surface localization. 
Alternatively, as described in Sec.~\ref{sec:AIII2}, surface localization for $E \neq 0$ can be induced by coupling the surface of the AIII superconductor to a non-magnetic atomic-limit insulator. Hybridization between the insulator states and the superconductor, which 
is
most effective if the band edge of the insulator is close to the Fermi energy of the superconductor, then yields the required Berry curvature.
On the other hand, magnetic impurities or a weak external field should be sufficient to localize even the zero-energy surface state because they break the (physical) time-reversal symmetry protecting the class-AIII superconductor. This would exponentially suppress the longitudinal surface thermal conductivity at low temperatures.

\section{Field theory}\label{sec:FT}

 In this section, we discuss how the physics discussed above presents itself
from the  perspective of effective field theory. The presentation is self-contained, and
included to provide an analytical justification for the criterion
(\ref{eq:BerryCriterion}) of state delocalization in the presence of weak
disorder. Readers willing to accept the empirical application of the criterion as convincing enough may consider the section as
optional reading.

By ``field theory,'' we here mean theoretical frameworks in which averaging over
static disorder is performed at an early stage to  describe  $d$-dimensional
systems in a given symmetry class in terms of $(d+0)$ dimensional~\footnote{The
`(d+0)' means that we are considering non-dynamical and non-interacting
frameworks and thus can avoid the introduction of a time-like coordinate; we
will always be considering a fixed reference energy.} nonlinear $\sigma$-models.
Such theories have been in use for a long time in the physics of conventional
disordered metals (see the textbook~\cite{Efetbook} for review) and were
extended to the description of various  topological
insulators~\cite{gruzbergExactExponentsSpin1999,gruzbergRandombondIsingModel2001,Altland2001511}
even before the momentum space topologies of
clean insulators became understood. In parallel to that developement, the
approach was upgraded to a full classification of disordered topological
insulators~\cite{ryu2010}  alternative to, say, the mathematical framework of
non-commutative geometry~\cite{schulz-baldesTopologicalInsulatorsPerspective2016}. From an applied perspective, its strength is that it
can predict, e.g, the flow of transport coefficients~\cite{Pruisken1984a,altlandTopologyAndersonLocalization2015,Fu2012} as
a function of disorder strength or system size. It is this latter aspect that
will be important in our discussion below. 

We begin with a short review of the physics of the two-dimensional class A, and
the three-dimensional AIII insulator, extending the discussion of Sec.~\ref{sec:AIII} to the presence of disorder. While these are
known structures, included here to provide perspective, our discussion of the
AIII surface in section~\ref{sec:FieldTheorySurfaceAIII}, and specifically that
of a connection between field theoretical topological $\theta$-angles and
momentum space Berry curvature  in Appendix~\ref{sec:FieldTheoryAppendix} is new
material. 

\subsection{Two-dimensional Chern insulator}
The starting point of  field theoretical representations of topological
insulators is an intermediate action (see Appendix~\ref{sec:FieldTheoryAppendix} for a brief review of its derivation) of the form 
\begin{equation}
    \label{eq:FieldTheoryIntermediate}
    S[X] \equiv -\tr \ln{(E- \hat{H}(\textbf{k})+ i \kappa \hat{X}(\textbf{x}))},
\end{equation}
where $\textbf{k}$ and $\textbf{x}$ are momentum and position, respectively, $\hat{H}(\textbf{k})$ is the clean Hamiltonian (throughout, we will omit carets
on operators in their eigenbasis representation), $\kappa$ a parameter measuring
the effective disorder scattering rate, and $\hat{X}=\{X^{r r'}_{s s'}\}$ a matrix
valued slowly fluctuating field carrying a replica index $r=1,\dots,R$
(sometimes traded for the mathematically more rigorous internal supersymmetry
structure~\cite{Efetbook}), and a second index $s=\pm=\pm1$  distinguishing
between propagators of retarded and advanced causality. Further details of the
internal structure of $\hat{X}$ depend on the symmetry class under consideration. For
example, in class A, $\hat{X}\equiv Q=T \hat{\tau}_\zz T^{-1}$, where  $\hat{\tau}_\zz$ will be a Pauli matrix in
$s$-space throughout this section (not to be confused with the earlier sublattice/chiral ${\tau}_\zz$) or in AIII just $Q=T$. 

The further processing of the action reflects  a notion of real and momentum
space duality, according to which the momentum space symmetries and topology
encoded in $H(\textbf{k})$ determine the real space symmetries and topology of
$Q(\textbf{x})$. To demonstrate the principle, consider the first step towards a
gradient expansion in class A and use the unitary invariance
of the trace to transform the action to
\begin{equation}
    \label{eq:TrLnMoyal}
    S[Q] \equiv -\tr \ln\left(\hat{G}^{-1}(E,\textbf{k}) - [T^{-1}(\textbf{x}),\hat{H}(\textbf{k})]T(\textbf{x})\right).
\end{equation}
Here, we encounter the impurity broadened Green function of the system
$\hat{G}^{-1}(\textbf{k})\equiv E- \hat{H}(\textbf{k})+ i \kappa \hat{\tau}_3$ in conjunction with
a term in which the real- and momentum-space dependent terms of the action talk
to each other. Assuming smooth variation of its  two constituents, a first order
Wigner-Moyal expansion leads to $[T^{-1}(\textbf{x}),\hat{H}(\textbf{k})]T(\textbf{x})\approx
F_i(\textbf{k}) \Phi_i(\textbf{x})$, with $F_i=i\partial_i \hat{H}$ and $\Phi_i \equiv
(\partial_i T^{-1})T$, where the derivatives are with respect to $k_i$ and
$x_i$, respectively. The effective action describing the system at large
distance scales then is obtained by expansion of the tr ln up to second order in
the derivative terms $\Phi_i$.  Notice that the real space $\Phi_i$ always
appear in conjunction with momentum space $F_i$. Also note that to leading order
 in a derivative expansion, 
\begin{align}
    \label{eq:TrPhaseSpace}
    \tr(\hat{A}(\mathbf{x})\hat{B}(\mathbf{k}))=\int dx 
dk \,\tr(\hat{A}(\mathbf{x}) \hat{B}(\mathbf{k}) )
\end{align}
where $dx=d^dx$ and  $ dk=d^dk / (2\pi)^d$, and the trace on
the right hand side is over the internal matrix structure of the operators in
question. In this way,  terms appearing in the action naturally assume the form
of (momentum space integrals) $\times$ (real space integrals), where in the case
of topological terms, the two partners encode ``dual'' aspects of the topology of
the system.           

Specifically, for the case of the two-dimensional Chern insulator, the result of
the expansion to second order in gradients is Pruisken's nonlinear $\sigma$-model, which first
appeared in the context of the integer quantum Hall effect~\cite{Pruisken1984a},
 \begin{align}
    \label{eq:PruiskenAction}
    S[Q]=g\int d^2x \, \tr(\partial_i Q \partial_i Q) + 
    \frac{\theta \epsilon_{ij}}{16\pi} \int d^2 x \,\tr(Q\partial_i Q \partial_j Q).
 \end{align}  
Here,  the first term describes the diffusion and eventually Anderson localization of a
two-dimensional electron gas in the presence of disorder, where the bare value
of the coupling constant $g=\sigma_{x x}/8$ is set by the system's longitudinal
conductance. The second term is of topological nature and counts the number of
times the $Q$-matrix field winds around its target manifold. In the classical
reference its weighing topological angle $\theta=2\pi \sigma_{xy}$ was
identified with the Hall  conductance. Of more relevance to
our present discussion is its interpretation as a momentum space dual of the
real space term, namely $\theta=\theta(E)$ as given in
Eq.~\eqref{eq:ThetaDef}. (This expression is derived for general two-dimensional
systems in class A in Appendix~\ref{sec:FieldTheoryAppendix}.)    
 
As derived, the action Eq.~\eqref{eq:PruiskenAction} describes the system at
 ``bare length scales,'' with a minimal distance cutoff set by the scattering mean free path.
 Upon integrating out short distance fluctuations,  and for generic values of
 $E$, the coupling constant $g$ renormalizes to zero (Anderson localization),
 while the effective angle $\theta$ renormalizes to a multiple of $2\pi$ (Hall
 quantization). For these fixed point values, the topological $\theta$-action
 reduces to a boundary action $\frac{1}{8}\epsilon_{ij} \int d^2x
 \,\tr(Q\partial_i Q \partial_j Q)=S_\textrm{1d}[T]$, where
 \begin{align}
    \label{eq:ClassABOundary}
    S_\mathrm{1d}[T]\equiv \frac{1}{2}\oint dx\, \tr(T^{-1} \hat{\tau}_\zz \partial_x T),
 \end{align}
  and $x$ now is a one-dimensional boundary coordinate. This single
 derivative action describes the dissipationless chiral circulation of boundary
 currents against the protecting background of a localized bulk. As with the
 chiral Hamiltonian of the clean system it lacks gauge
 invariance, signalling spectral flow through the delocalized states at the
 energies $E_+$ or $E_-$.       

We next compare this physics to that in our reference system  without protected spectral flow.  

\subsection{Three-dimensional AIII insulator} 

As with its lower-dimensional cousin, the gradient expansion of the prototypical
action \eqref{eq:FieldTheoryIntermediate} leads to a nonlinear
$\sigma$-model~\cite{gadeReplicaLimitModels1991,altlandFieldTheoryRandom1999a} enriched by a topological term~\cite{ryu2010}, 
\begin{align}
    \label{eq:BulkAIIIAction}
    S_\textrm{3d}[T]&=\int d^3 x \,\bigg[g\,\tr(\partial_i T \partial_i T^{-1})+ E \nu \tr(T+T^{-1})\bigg]+\cr 
    &+  \frac{\vartheta \epsilon^{ijk}}{24 \pi^2} \int d^3x \,\tr(T^{-1}\partial_i TT^{-1}\partial_j TT^{-1}\partial_k T).
 \end{align} 
Compared to Eq.~\eqref{eq:PruiskenAction}, the field manifold has changed to
group-valued matrix fields, $T\in \mathrm{U}(2R)$. Otherwise, we again have a
job division between a gradient term describing bulk conduction properties, and
a topological term now measuring three-dimensional windings over the unitary
group. The second term describes the symmetry breaking induced by departures
away from $E=0$, where $\nu$ is proportional to the three-dimensional
density of states. 

There are  different physical limits that may be investigated on the basis of
this representation: at the particle-hole symmetric point $E=0$ we are sitting
inside the bulk spectral gap. The bare conduction parameter $g$ may nevertheless
be finite, due to impurity states smearing the  band gap of the clean system. At
large length scales, we expect renormalization to an Anderson insulator, $g=0$,
where a value $\vartheta = 2\pi n$ with $n$ a non-vanishing integer will signal topological
non-triviality. In this limit, and in analogy to the Pruisken action, the
topological term becomes a boundary term, $\Gamma[T]/12 \pi$, with the physical
interpretation of a Wess-Zumino term~\cite{Altland2002} of an emerging surface action. In the
immediate vicinity of the surface, the gradient term  remains finite and now
describes intra-surface conduction. The net effect is the stabilization of  a
surface Wess-Zumino action 
\begin{align}
    S_\textrm{2d}[T]&=g\int d^2 x \,\tr(\partial_i T \partial_i T^{-1})+ \frac{1}{12\pi}\Gamma[T],
\end{align} 
through the localization of the bulk. This action is the AIII analog of
Eq.~\eqref{eq:ClassABOundary} for the A system. At large length scales, this
theory renormalizes~\cite{nersesyanDisorderEffectsTwodimensional1994,Altland2002} to the conformally invariant action with $g=1 /8 \pi$ 
representing  a single two-dimensional Dirac point at zero energy; this is the field theoretical
interpretation of \emph{zero energy} surface delocalization in the AIII
insulator. 

However, we may also investigate what happens at finite deviations $E\not=0$
away from chiral symmetry. In this case, the (strongly RG relevant) ``mass term''
in Eq.~\eqref{eq:BulkAIIIAction} only admits  configurations $T\to Q=T \hat{\tau}_\zz
T^{-1}$ for which $\tr(Q+Q^{-1})$ is a constant vanishing in the
replica limit. These are the $Q$-matrices of the model of lower symmetry AIII
$\to $ A. Substitution into the bulk action annihilates the second and third
term, while the gradient term becomes the conventional action of a disordered
three-dimensional metal \emph{below} the Anderson transition point: away from
zero energy, the AIII insulator behaves like a conventional Anderson insulator.
A more interesting limit is the case of small but  finite $E \not=0$ in the
vicinity of the surface. The symmetry breaking now collapses the  Wess-Zumino
term $\Gamma[Q]$  to the Pruisken term of a two-dimensional class A action, at
topological angle $\theta=(2\Bbb{Z}+1)\pi$~\cite{Altland2002}. We conclude that
the naive extension of the zero-energy WZW action to finite energies equals  the
action Eq.~\eqref{eq:PruiskenAction} of a two-dimensional Chern insulator
\emph{fine tuned into criticality}. This is a field theoretical indication of a
tendency to extended surface quantum criticality. The question is what happens
for larger deviations $E$ away from zero. To answer it, we need to go beyond the
present level of high level reasoning and turn to a first principle approach. 

\subsection{Surface of the three-dimensional AIII insulator}
\label{sec:FieldTheorySurfaceAIII}

In order to understand the physics of the disordered surface at arbitrary
$E$, we again start from the prototypical
representation~\eqref{eq:FieldTheoryIntermediate}. For the slow field, we take
$\hat{X}(\mathbf{x})=(T \hat{\tau}_\zz T^{-1})(\mathbf{x})$, where $\mathbf{x}$ is a
two-dimensional surface coordinate. (The justification behind this surface
projection is that states of finite extension into the bulk have eigenenergies
much larger than $E$, which we assume to be way below the bulk gap.) For
the surface Hamiltonian $\hat{H}(\mathbf{k})$, we assume a  spectral decomposition
\begin{align}
    \label{eq:SurfaceDecomposition}
    \hat{H}(\mathbf{k})\equiv \sum_\alpha |\alpha_\textbf{k}\rangle
\,\epsilon_{\alpha_\mathbf{k}} \langle \alpha_\mathbf{k}|,
\end{align}
where $\{|\alpha_\textbf{k}\rangle\}$ are the system eigenstates at a given
transverse momentum. 

This formal spectral decomposition actually is less innocent than it looks:
Naively, it should include \emph{all} eigenstates at a given $\mathbf{k}$.
However, this is not the case. Going back to the tr ln
\eqref{eq:FieldTheoryIntermediate}, only eigenstates of $\hat{H}(\textbf{k})$ with a
finite spatial overlap with the surface Hubbard-Stratonovich field
$\hat{X}(\textbf{x})$ contribute to the expansion. The obvious candidates here are the
two eigenstates forming the chiral partners of the surface band. However, the
internal spinor representation space at a given $\textbf{k}$  of the lattice
model is four-dimensional, implying that two states are insufficient to span it.
We must, therefore, assume a contribution of bulk states (with finite surface
amplitude), and an associated state-dependent weight $\kappa=\kappa_\alpha$. As we do not have full access to this information, we sidestep the problem by considering Eq.~\eqref{eq:SurfaceDecomposition} as a formal complete sum. We also consider the  surface band for the
flattened model, i.e. we assume a finite spectral gap to higher-lying bands.    

In Appendix~\ref{sec:FieldTheoryAppendix} we show that under these conditions,
the surface action assumes the form of a two-dimensional class A action
Eq.~\eqref{eq:PruiskenAction}, with the topological angle given by
Eq.~\eqref{eq:ThetaDef}, or Eq.~\eqref{eq:TopologicalAction} in a more explicit
representation. As discussed above, the added curvature integrals  of the upper
and lower surface band computed in this way need not add to zero. In view of the
above discussion, this phenomenon relates back to the embedding of the surface
band into a larger Hilbert space of bulk states. Unlike with  intrinsic
two-dimensional lattice bands, whose Chern numbers would have to add to zero, we
are here considering a single two-dimensional shadow of a three-dimensional bulk
(the other lives at the opposite surface) and the cancellation principle does
not apply. 

To summarize, away from $E=0$ the surface of the AIII insulator is
described by the action otherwise describing the physics of the integer quantum
Hall effect at distance scales exceeding the mean free scattering path.
(Equivalently, we may think of it as the continuum version of the network
structure discussed in the end of Section \ref{sec:domain}.) Its two coupling
constants specify the localization properties of the system in terms of the bare
longitudinal conductance, $g$, and the topological angle $\theta$-angle,
respectively. The latter is remarkable in that it links the real-space long
range localization properties of states at a fixed energy $E$ to momentum-space
short-distance structures at \emph{all other} band energies, via the integrated
Berry curvature. Such  extreme forms of infrared-ultraviolet mixing are rare (and strictly absent in generic free fermions systems), but here enabled by topology.

\section{Discussion and conclusion} \label{sec:6}
For topological insulators in the Wigner-Dyson classes A, AI, and AII ---
the most prominent realizations being the two-dimensional class A integer
quantum Hall insulator and class AII quantum spin Hall insulator, and the
three-dimensional class AII topological insulator ---  boundary states are
continuously attached to delocalized bulk states, without interruption by a
spectral or mobility gap. In this paper, we showed that for the complementary
class of genuinely non-Wigner-Dyson class topologlogical insulators this key
principle is broken. (The attribute ``genuine'' indicates that the constraints
imposed by charge conjugation symmetry ${\cal C}$ or chiral symmetry ${\cal S}$
that define the non-Wigner-Dysnon classes and force the spectrum to be symmetric
around $E=0$ are essential for the protection of the bulk topology. By contrast,
non-genuine classes  remain topological after lifting
constraints  due to ${\cal C}$ and ${\cal S}$ 
and
behave effectively as Wigner-Dyson
insulators.) 

In the literature, nontrivial topology is often associated with an obstruction
to the construction of a localized basis of conduction and valence bands,
referred to as ``Wannierizability''. Our general results --- see
Tab.~\ref{tab:TF} --- show that, by contrast, all genuine non-Wigner-Dyson class
insulators enjoy this property, and  can be topologically non-trivial
nonetheless. We arrived at these conclusions both from a bulk perspective,
showing that Wannier localizability of the bulk  implies that the connection between
surface and bulk bands becomes fragile, and from an intrinsic boundary
perspective, showing that the effective surface theory  admits a gap-opening
perturbation. A key conclusion following from  this observation is that the
surface states of genuine non-Wigner-Dyson topological insulators themselves are
localizable, except at the center $E=0$ where state delocalization is
topologically protected.  

The existence of gapless or conducting surfaces is the key signature
distinguishing topological from conventional insulators. Our analysis shows
that, in this regard, the physics of genuine non-Wigner-Dyson topological
insulators is different from that of their Wigner-Dyson siblings:  Their surface
states can, but need not be delocalized away from one isolated energy, $E=0$. As a concrete case study, we considered the three-dimensional AIII insulator 
and showcased its rich boundary phenomenology: the surface states can be
detached from the bulk and electrons residing in them can form different phases 
of matter distinguished by their Chern number. Furthermore, at the transition 
point between any of these surface phases one observes spectrum wide quantum
critical delocalization of states.

Indeed, our study was motivated by recent work, which reported a spectrum-wide
delocalization of the surface states of a class-AIII insulator and other
non-Wigner-Dyson classes \cite{SWQC-CI,sbierski2020,SWQC-DIII,SWQC-Rev}. These
observations were based both on a numerical analysis of an effective Dirac
surface theory and a numerical study of a three-dimensional lattice model of a
class-AIII insulator. In this work we identified the principles that led to this
seemingly robust prediction. We  showed that the minimal $2\times 2$  Dirac
theory considered in these references is intrinsically protected against
localization. However this protection is \emph{topologically fragile} in that
it is lifted if additional trivial surface bands are added (which can be
achieved, e.g., by coupling  an extraneous surface layer). The absence of
localization  in the full three-dimensional lattice model considered in Ref.\
\cite{sbierski2020} results from a {\em statistical} symmetry, similar to that
realized in the ``statistical topological insulator'' \cite{fulga2014}.

For the case of minimal topological winding number $\nu=1$, we identified a
powerful indicator for the localization properties of states at energy $E$,
namely the integrated Berry curvature of all states energetically below (or
equivalently above) that energy: an integer-quantized integral implies
delocalization, departures from these values localization.  While the energetic
non-locality of this criterion may be unexpected for a model of non-interacting
particles, it reflects the importance of global momentum space quantum geometry
in a topologically nontrivial context. In the two models mentioned above, that
criterion signals global delocalization, if for different reasons: the minimal
surface Dirac theory is Berry-flat; in the lattice model, the disorder model
considered in Ref.~\cite{sbierski2020} leads to a statistical cancellation of
curvature in the integral. However, both the embedding of the minimal $2\times
2$  theory into  a four-component spinor theory (the minimal framework to
describe topological non-triviality in three-dimensional AIII), or the lifting
of the statistical symmetry in the lattice model by addition of a ``fragmenting
surface potential'' of non-zero average, lead to state localization in a manner
discussed in detail in section \ref{sec:num}. 

At the same time, our analysis
indicates that for winding numbers $|\nu|>1$,  the precise meaning of the term
``minimal model'', and the identification of quantitative localization measures
must be reconsidered.
In conclusion, the principle that a localizable bulk implies a gapable surface
spectrum (and {\em vice versa}) applies to all genuine non-Wigner-Dyson
classes. Other observations, such as the precise ways in which the  minimal
Dirac description is fragile, and the relation of localizability and Berry
curvature,  need not straightforwardly generalize beyond the $\nu=1$ AIII
context  and invite future work.

Most of the previous  work on surface state localization in the AIII insulator
considered a low-energy, continuum two-dimensional Dirac description with
two-component Dirac spinors~\cite{BernardLeClair2002,Foster2014}. Our results
imply that these theories  are fundamentally incomplete, because the surface
Berry curvature responsible for Anderson localization is strictly ruled out. To
readers who trust in the predictive power of minimal models, it is a surprising
and possibly disturbing notion that such  Lorentz-covariant and renormalizable
field theories cannot encode the most basic characteristics (localized versus
extended) of surface-state wave functions. At the same time, it may be
reassuring that the origin of the problem does not lie in the notorious and
difficult-to-handle lack of ultraviolet closure of the Dirac theories, but that
the problem can be cured by the simple addition of trivial degrees of freedom.
Allowing for the addition of trivial bands is common practice in topological
classifications based on stable equivalence, and our results show that it
is equally important when determining the existence of a topological obstruction
to Anderson localization.

\paragraph*{Acknowledgements:} 
We thank Chris Bourne, Jennifer Cano, Sayed Ali Akbar Ghorashi, Alex Kamenev, Bastien Lapierre, Herrmann
Schulz-Baldes, Haruki Watanabe, and Justin Wilson for helpful discussions. This work was
supported by the Deutsche Forschungsgemeinschaft (DFG) project grant 277101999
within the CRC network TR 183 (subproject A03) (A.A., P.W.B., M.M.-G.),
the German Academic Scholarship Foundation and the German Research and the Collaborative Research
Center SFB 1277 (J.D.),
the Welch Foundation Grant No. C-1809 (M.S.F.),
and by the
FNS/SNF Ambizione Grant No.~PZ00P2\_179962 (L.T.).

\bibliographystyle{apsrev4-1}
\bibliography{ref,refs}

\begin{thebibliography}{85}%
\makeatletter
\providecommand \@ifxundefined [1]{%
 \@ifx{#1\undefined}
}%
\providecommand \@ifnum [1]{%
 \ifnum #1\expandafter \@firstoftwo
 \else \expandafter \@secondoftwo
 \fi
}%
\providecommand \@ifx [1]{%
 \ifx #1\expandafter \@firstoftwo
 \else \expandafter \@secondoftwo
 \fi
}%
\providecommand \natexlab [1]{#1}%
\providecommand \enquote  [1]{``#1''}%
\providecommand \bibnamefont  [1]{#1}%
\providecommand \bibfnamefont [1]{#1}%
\providecommand \citenamefont [1]{#1}%
\providecommand \href@noop [0]{\@secondoftwo}%
\providecommand \href [0]{\begingroup \@sanitize@url \@href}%
\providecommand \@href[1]{\@@startlink{#1}\@@href}%
\providecommand \@@href[1]{\endgroup#1\@@endlink}%
\providecommand \@sanitize@url [0]{\catcode `\\12\catcode `\$12\catcode
  `\&12\catcode `\#12\catcode `\^12\catcode `\_12\catcode `\%12\relax}%
\providecommand \@@startlink[1]{}%
\providecommand \@@endlink[0]{}%
\providecommand \url  [0]{\begingroup\@sanitize@url \@url }%
\providecommand \@url [1]{\endgroup\@href {#1}{\urlprefix }}%
\providecommand \urlprefix  [0]{URL }%
\providecommand \Eprint [0]{\href }%
\providecommand \doibase [0]{http://dx.doi.org/}%
\providecommand \selectlanguage [0]{\@gobble}%
\providecommand \bibinfo  [0]{\@secondoftwo}%
\providecommand \bibfield  [0]{\@secondoftwo}%
\providecommand \translation [1]{[#1]}%
\providecommand \BibitemOpen [0]{}%
\providecommand \bibitemStop [0]{}%
\providecommand \bibitemNoStop [0]{.\EOS\space}%
\providecommand \EOS [0]{\spacefactor3000\relax}%
\providecommand \BibitemShut  [1]{\csname bibitem#1\endcsname}%
\let\auto@bib@innerbib\@empty
%</preamble>
\bibitem [{\citenamefont {Bernevig}\ and\ \citenamefont
  {Hughes}(2013)}]{bernevig2013}%
  \BibitemOpen
  \bibfield  {author} {\bibinfo {author} {\bibfnamefont {B.~A.}\ \bibnamefont
  {Bernevig}}\ and\ \bibinfo {author} {\bibfnamefont {T.~L.}\ \bibnamefont
  {Hughes}},\ }\href@noop {} {\emph {\bibinfo {title} {Topological Insulators
  and Topological Superconductors}}}\ (\bibinfo  {publisher} {Princeton
  University Press},\ \bibinfo {year} {2013})\BibitemShut {NoStop}%
\bibitem [{\citenamefont {Ando}\ and\ \citenamefont {Fu}(2015)}]{ando2015}%
  \BibitemOpen
  \bibfield  {author} {\bibinfo {author} {\bibfnamefont {Y.}~\bibnamefont
  {Ando}}\ and\ \bibinfo {author} {\bibfnamefont {L.}~\bibnamefont {Fu}},\
  }\href {\doibase 10.1146/annurev-conmatphys-031214-014501} {\bibfield
  {journal} {\bibinfo  {journal} {Ann. Rev. Condensed Matter Physics}\ }\textbf
  {\bibinfo {volume} {6}},\ \bibinfo {pages} {361} (\bibinfo {year}
  {2015})}\BibitemShut {NoStop}%
\bibitem [{\citenamefont {Hasan}\ and\ \citenamefont {Kane}(2010)}]{hasan2010}%
  \BibitemOpen
  \bibfield  {author} {\bibinfo {author} {\bibfnamefont {M.~Z.}\ \bibnamefont
  {Hasan}}\ and\ \bibinfo {author} {\bibfnamefont {C.~L.}\ \bibnamefont
  {Kane}},\ }\href {\doibase 10.1103/RevModPhys.82.3045} {\bibfield  {journal}
  {\bibinfo  {journal} {Rev. Mod. Phys.}\ }\textbf {\bibinfo {volume} {82}},\
  \bibinfo {pages} {3045} (\bibinfo {year} {2010})}\BibitemShut {NoStop}%
\bibitem [{\citenamefont {Qi}\ and\ \citenamefont {Zhang}(2011)}]{qi2011}%
  \BibitemOpen
  \bibfield  {author} {\bibinfo {author} {\bibfnamefont {X.-L.}\ \bibnamefont
  {Qi}}\ and\ \bibinfo {author} {\bibfnamefont {S.-C.}\ \bibnamefont {Zhang}},\
  }\href {\doibase 10.1103/RevModPhys.83.1057} {\bibfield  {journal} {\bibinfo
  {journal} {Rev. Mod. Phys.}\ }\textbf {\bibinfo {volume} {83}},\ \bibinfo
  {pages} {1057} (\bibinfo {year} {2011})}\BibitemShut {NoStop}%
\bibitem [{\citenamefont {Schnyder}\ \emph {et~al.}(2008)\citenamefont
  {Schnyder}, \citenamefont {Ryu}, \citenamefont {Furusaki},\ and\
  \citenamefont {Ludwig}}]{schnyder2008}%
  \BibitemOpen
  \bibfield  {author} {\bibinfo {author} {\bibfnamefont {A.~P.}\ \bibnamefont
  {Schnyder}}, \bibinfo {author} {\bibfnamefont {S.}~\bibnamefont {Ryu}},
  \bibinfo {author} {\bibfnamefont {A.}~\bibnamefont {Furusaki}}, \ and\
  \bibinfo {author} {\bibfnamefont {A.~W.~W.}\ \bibnamefont {Ludwig}},\ }\href
  {\doibase 10.1103/PhysRevB.78.195125} {\bibfield  {journal} {\bibinfo
  {journal} {Phys. Rev. B}\ }\textbf {\bibinfo {volume} {78}},\ \bibinfo
  {pages} {195125} (\bibinfo {year} {2008})}\BibitemShut {NoStop}%
\bibitem [{\citenamefont {Kitaev}(2009)}]{kitaev2009}%
  \BibitemOpen
  \bibfield  {author} {\bibinfo {author} {\bibfnamefont {A.}~\bibnamefont
  {Kitaev}},\ }\href {\doibase 10.1063/1.3149495} {\bibfield  {journal}
  {\bibinfo  {journal} {AIP Conference Proceedings}\ }\textbf {\bibinfo
  {volume} {1134}},\ \bibinfo {pages} {22} (\bibinfo {year}
  {2009})}\BibitemShut {NoStop}%
\bibitem [{\citenamefont {Ryu}\ \emph {et~al.}(2010)\citenamefont {Ryu},
  \citenamefont {Schnyder}, \citenamefont {Furusaki},\ and\ \citenamefont
  {Ludwig}}]{ryu2010}%
  \BibitemOpen
  \bibfield  {author} {\bibinfo {author} {\bibfnamefont {S.}~\bibnamefont
  {Ryu}}, \bibinfo {author} {\bibfnamefont {A.~P.}\ \bibnamefont {Schnyder}},
  \bibinfo {author} {\bibfnamefont {A.}~\bibnamefont {Furusaki}}, \ and\
  \bibinfo {author} {\bibfnamefont {A.~W.~W.}\ \bibnamefont {Ludwig}},\ }\href
  {\doibase 10.1088/1367-2630/12/6/065010} {\bibfield  {journal} {\bibinfo
  {journal} {New Journal of Physics}\ }\textbf {\bibinfo {volume} {12}},\
  \bibinfo {pages} {065010} (\bibinfo {year} {2010})}\BibitemShut {NoStop}%
\bibitem [{\citenamefont {Essin}\ and\ \citenamefont
  {Gurarie}(2015)}]{essin2015}%
  \BibitemOpen
  \bibfield  {author} {\bibinfo {author} {\bibfnamefont {A.~M.}\ \bibnamefont
  {Essin}}\ and\ \bibinfo {author} {\bibfnamefont {V.}~\bibnamefont
  {Gurarie}},\ }\href {\doibase 10.1088/1751-8113/48/11/11FT01} {\bibfield
  {journal} {\bibinfo  {journal} {J. Phys. A}\ }\textbf {\bibinfo {volume}
  {48}},\ \bibinfo {pages} {11FT01} (\bibinfo {year} {2015})}\BibitemShut
  {NoStop}%
\bibitem [{\citenamefont {Schulz-Baldes}\ and\ \citenamefont
  {Stoiber}(2022)}]{Schulz-BaldesBook}%
  \BibitemOpen
  \bibfield  {author} {\bibinfo {author} {\bibfnamefont {H.}~\bibnamefont
  {Schulz-Baldes}}\ and\ \bibinfo {author} {\bibfnamefont {T.}~\bibnamefont
  {Stoiber}},\ }\href@noop {} {\emph {\bibinfo {title} {Harmonic Analysis in
  Operator Algebras and its Applications to Index Theory and Topological Solid
  State Systems}}}\ (\bibinfo  {publisher} {Springer, Cham, Switzerland},\
  \bibinfo {year} {2022})\BibitemShut {NoStop}%
\bibitem [{\citenamefont {Kohn}(1959)}]{kohn1959}%
  \BibitemOpen
  \bibfield  {author} {\bibinfo {author} {\bibfnamefont {W.}~\bibnamefont
  {Kohn}},\ }\href {\doibase 10.1103/PhysRev.115.809} {\bibfield  {journal}
  {\bibinfo  {journal} {Phys. Rev.}\ }\textbf {\bibinfo {volume} {115}},\
  \bibinfo {pages} {809} (\bibinfo {year} {1959})}\BibitemShut {NoStop}%
\bibitem [{\citenamefont {Kitaev}(2001)}]{kitaev2001}%
  \BibitemOpen
  \bibfield  {author} {\bibinfo {author} {\bibfnamefont {A.~Y.}\ \bibnamefont
  {Kitaev}},\ }\href {\doibase 10.1070/1063-7869/44/10S/S29} {\bibfield
  {journal} {\bibinfo  {journal} {Physics-Uspekhi}\ }\textbf {\bibinfo {volume}
  {44}},\ \bibinfo {pages} {131} (\bibinfo {year} {2001})}\BibitemShut
  {NoStop}%
\bibitem [{\citenamefont {Laughlin}(1981)}]{laughlin1981}%
  \BibitemOpen
  \bibfield  {author} {\bibinfo {author} {\bibfnamefont {R.~B.}\ \bibnamefont
  {Laughlin}},\ }\href {\doibase 10.1103/PhysRevB.23.5632} {\bibfield
  {journal} {\bibinfo  {journal} {Phys. Rev. B}\ }\textbf {\bibinfo {volume}
  {23}},\ \bibinfo {pages} {5632} (\bibinfo {year} {1981})}\BibitemShut
  {NoStop}%
\bibitem [{\citenamefont {Huckestein}(1995)}]{Huckestein95}%
  \BibitemOpen
  \bibfield  {author} {\bibinfo {author} {\bibfnamefont {B.}~\bibnamefont
  {Huckestein}},\ }\href {\doibase 10.1103/RevModPhys.67.357} {\bibfield
  {journal} {\bibinfo  {journal} {Rev. Mod. Phys.}\ }\textbf {\bibinfo {volume}
  {67}},\ \bibinfo {pages} {357} (\bibinfo {year} {1995})}\BibitemShut
  {NoStop}%
\bibitem [{\citenamefont {Evers}\ and\ \citenamefont
  {Mirlin}(2008)}]{evers2008}%
  \BibitemOpen
  \bibfield  {author} {\bibinfo {author} {\bibfnamefont {F.}~\bibnamefont
  {Evers}}\ and\ \bibinfo {author} {\bibfnamefont {A.~D.}\ \bibnamefont
  {Mirlin}},\ }\href {\doibase 10.1103/RevModPhys.80.1355} {\bibfield
  {journal} {\bibinfo  {journal} {Rev. Mod. Phys.}\ }\textbf {\bibinfo {volume}
  {80}},\ \bibinfo {pages} {1355} (\bibinfo {year} {2008})}\BibitemShut
  {NoStop}%
\bibitem [{\citenamefont {Prodan}(2011)}]{prodan2011}%
  \BibitemOpen
  \bibfield  {author} {\bibinfo {author} {\bibfnamefont {E.}~\bibnamefont
  {Prodan}},\ }\href {\doibase 10.1088/1751-8113/44/11/113001} {\bibfield
  {journal} {\bibinfo  {journal} {Journal of Physics A: Mathematical and
  Theoretical}\ }\textbf {\bibinfo {volume} {44}},\ \bibinfo {pages} {113001}
  (\bibinfo {year} {2011})}\BibitemShut {NoStop}%
\bibitem [{\citenamefont {Ghorashi}\ \emph {et~al.}(2018)\citenamefont
  {Ghorashi}, \citenamefont {Liao},\ and\ \citenamefont {Foster}}]{SWQC-CI}%
  \BibitemOpen
  \bibfield  {author} {\bibinfo {author} {\bibfnamefont {S.~A.~A.}\
  \bibnamefont {Ghorashi}}, \bibinfo {author} {\bibfnamefont {Y.}~\bibnamefont
  {Liao}}, \ and\ \bibinfo {author} {\bibfnamefont {M.~S.}\ \bibnamefont
  {Foster}},\ }\href {\doibase 10.1103/PhysRevLett.121.016802} {\bibfield
  {journal} {\bibinfo  {journal} {Phys. Rev. Lett.}\ }\textbf {\bibinfo
  {volume} {121}},\ \bibinfo {pages} {016802} (\bibinfo {year}
  {2018})}\BibitemShut {NoStop}%
\bibitem [{\citenamefont {Sbierski}\ \emph {et~al.}(2020)\citenamefont
  {Sbierski}, \citenamefont {Karcher},\ and\ \citenamefont
  {Foster}}]{sbierski2020}%
  \BibitemOpen
  \bibfield  {author} {\bibinfo {author} {\bibfnamefont {B.}~\bibnamefont
  {Sbierski}}, \bibinfo {author} {\bibfnamefont {J.~F.}\ \bibnamefont
  {Karcher}}, \ and\ \bibinfo {author} {\bibfnamefont {M.~S.}\ \bibnamefont
  {Foster}},\ }\href {\doibase 10.1103/PhysRevX.10.021025} {\bibfield
  {journal} {\bibinfo  {journal} {Phys. Rev. X}\ }\textbf {\bibinfo {volume}
  {10}},\ \bibinfo {pages} {021025} (\bibinfo {year} {2020})}\BibitemShut
  {NoStop}%
\bibitem [{\citenamefont {Ghorashi}\ \emph {et~al.}(2020)\citenamefont
  {Ghorashi}, \citenamefont {Karcher}, \citenamefont {Davis},\ and\
  \citenamefont {Foster}}]{SWQC-DIII}%
  \BibitemOpen
  \bibfield  {author} {\bibinfo {author} {\bibfnamefont {S.~A.~A.}\
  \bibnamefont {Ghorashi}}, \bibinfo {author} {\bibfnamefont {J.~F.}\
  \bibnamefont {Karcher}}, \bibinfo {author} {\bibfnamefont {S.~M.}\
  \bibnamefont {Davis}}, \ and\ \bibinfo {author} {\bibfnamefont {M.~S.}\
  \bibnamefont {Foster}},\ }\href {\doibase 10.1103/PhysRevB.101.214521}
  {\bibfield  {journal} {\bibinfo  {journal} {Phys. Rev. B}\ }\textbf {\bibinfo
  {volume} {101}},\ \bibinfo {pages} {214521} (\bibinfo {year}
  {2020})}\BibitemShut {NoStop}%
\bibitem [{\citenamefont {Karcher}\ and\ \citenamefont
  {Foster}(2021)}]{SWQC-Rev}%
  \BibitemOpen
  \bibfield  {author} {\bibinfo {author} {\bibfnamefont {J.~F.}\ \bibnamefont
  {Karcher}}\ and\ \bibinfo {author} {\bibfnamefont {M.~S.}\ \bibnamefont
  {Foster}},\ }\href {\doibase 10.1103/j.aop.2021.168439} {\bibfield  {journal}
  {\bibinfo  {journal} {Ann. Phys.}\ }\textbf {\bibinfo {volume} {435}},\
  \bibinfo {pages} {168439} (\bibinfo {year} {2021})}\BibitemShut {NoStop}%
\bibitem [{\citenamefont {Moreno-Gonzalez}\ \emph {et~al.}(2023)\citenamefont
  {Moreno-Gonzalez}, \citenamefont {Dieplinger},\ and\ \citenamefont
  {Altland}}]{moreno2023topological}%
  \BibitemOpen
  \bibfield  {author} {\bibinfo {author} {\bibfnamefont {M.}~\bibnamefont
  {Moreno-Gonzalez}}, \bibinfo {author} {\bibfnamefont {J.}~\bibnamefont
  {Dieplinger}}, \ and\ \bibinfo {author} {\bibfnamefont {A.}~\bibnamefont
  {Altland}},\ }\href {\doibase https://doi.org/10.1016/j.aop.2023.169258}
  {\bibfield  {journal} {\bibinfo  {journal} {Annals of Physics}\ }\textbf
  {\bibinfo {volume} {456}},\ \bibinfo {pages} {169258} (\bibinfo {year}
  {2023})}\BibitemShut {NoStop}%
\bibitem [{\citenamefont {Thouless}(1984)}]{thouless1984}%
  \BibitemOpen
  \bibfield  {author} {\bibinfo {author} {\bibfnamefont {D.~J.}\ \bibnamefont
  {Thouless}},\ }\href {\doibase 10.1088/0022-3719/17/12/003} {\bibfield
  {journal} {\bibinfo  {journal} {Journal of Physics C: Solid State Physics}\
  }\textbf {\bibinfo {volume} {17}},\ \bibinfo {pages} {L325} (\bibinfo {year}
  {1984})}\BibitemShut {NoStop}%
\bibitem [{\citenamefont {Kuchment}(2008)}]{kuchment2009}%
  \BibitemOpen
  \bibfield  {author} {\bibinfo {author} {\bibfnamefont {P.}~\bibnamefont
  {Kuchment}},\ }\href {\doibase 10.1088/1751-8113/42/2/025203} {\bibfield
  {journal} {\bibinfo  {journal} {Journal of Physics A: Mathematical and
  Theoretical}\ }\textbf {\bibinfo {volume} {42}},\ \bibinfo {pages} {025203}
  (\bibinfo {year} {2008})}\BibitemShut {NoStop}%
\bibitem [{\citenamefont {Ludewig}\ and\ \citenamefont
  {Thiang}(2022)}]{ludewig2022}%
  \BibitemOpen
  \bibfield  {author} {\bibinfo {author} {\bibfnamefont {M.}~\bibnamefont
  {Ludewig}}\ and\ \bibinfo {author} {\bibfnamefont {G.~C.}\ \bibnamefont
  {Thiang}},\ }\href {\doibase 10.1063/5.0098471} {\bibfield  {journal}
  {\bibinfo  {journal} {Journal of Mathematical Physics}\ }\textbf {\bibinfo
  {volume} {63}},\ \bibinfo {pages} {091902} (\bibinfo {year} {2022})},\
  \Eprint
  {http://arxiv.org/abs/https://pubs.aip.org/aip/jmp/article-pdf/doi/10.1063/5.0098471/16563239/091902\_1\_online.pdf}
  {https://pubs.aip.org/aip/jmp/article-pdf/doi/10.1063/5.0098471/16563239/091902\_1\_online.pdf}
  \BibitemShut {NoStop}%
\bibitem [{\citenamefont {Thonhauser}\ and\ \citenamefont
  {Vanderbilt}(2006)}]{thonhauser2006}%
  \BibitemOpen
  \bibfield  {author} {\bibinfo {author} {\bibfnamefont {T.}~\bibnamefont
  {Thonhauser}}\ and\ \bibinfo {author} {\bibfnamefont {D.}~\bibnamefont
  {Vanderbilt}},\ }\href {\doibase 10.1103/PhysRevB.74.235111} {\bibfield
  {journal} {\bibinfo  {journal} {Phys. Rev. B}\ }\textbf {\bibinfo {volume}
  {74}},\ \bibinfo {pages} {235111} (\bibinfo {year} {2006})}\BibitemShut
  {NoStop}%
\bibitem [{\citenamefont {Soluyanov}\ and\ \citenamefont
  {Vanderbilt}(2011)}]{soluyanov2011}%
  \BibitemOpen
  \bibfield  {author} {\bibinfo {author} {\bibfnamefont {A.~A.}\ \bibnamefont
  {Soluyanov}}\ and\ \bibinfo {author} {\bibfnamefont {D.}~\bibnamefont
  {Vanderbilt}},\ }\href {\doibase 10.1103/PhysRevB.83.035108} {\bibfield
  {journal} {\bibinfo  {journal} {Phys. Rev. B}\ }\textbf {\bibinfo {volume}
  {83}},\ \bibinfo {pages} {035108} (\bibinfo {year} {2011})}\BibitemShut
  {NoStop}%
\bibitem [{\citenamefont {Winkler}\ \emph {et~al.}(2016)\citenamefont
  {Winkler}, \citenamefont {Soluyanov},\ and\ \citenamefont
  {Troyer}}]{winkler2016}%
  \BibitemOpen
  \bibfield  {author} {\bibinfo {author} {\bibfnamefont {G.~W.}\ \bibnamefont
  {Winkler}}, \bibinfo {author} {\bibfnamefont {A.~A.}\ \bibnamefont
  {Soluyanov}}, \ and\ \bibinfo {author} {\bibfnamefont {M.}~\bibnamefont
  {Troyer}},\ }\href {\doibase 10.1103/PhysRevB.93.035453} {\bibfield
  {journal} {\bibinfo  {journal} {Phys. Rev. B}\ }\textbf {\bibinfo {volume}
  {93}},\ \bibinfo {pages} {035453} (\bibinfo {year} {2016})}\BibitemShut
  {NoStop}%
\bibitem [{\citenamefont {Cornean}\ \emph {et~al.}(2017)\citenamefont
  {Cornean}, \citenamefont {Monaco},\ and\ \citenamefont
  {Teufel}}]{cornean2017}%
  \BibitemOpen
  \bibfield  {author} {\bibinfo {author} {\bibfnamefont {H.~D.}\ \bibnamefont
  {Cornean}}, \bibinfo {author} {\bibfnamefont {D.}~\bibnamefont {Monaco}}, \
  and\ \bibinfo {author} {\bibfnamefont {S.}~\bibnamefont {Teufel}},\ }\href
  {\doibase 10.1142/S0129055X17300011} {\bibfield  {journal} {\bibinfo
  {journal} {Reviews in Mathematical Physics}\ }\textbf {\bibinfo {volume}
  {29}},\ \bibinfo {pages} {1730001} (\bibinfo {year} {2017})},\ \Eprint
  {http://arxiv.org/abs/https://doi.org/10.1142/S0129055X17300011}
  {https://doi.org/10.1142/S0129055X17300011} \BibitemShut {NoStop}%
\bibitem [{\citenamefont {Cornean}\ and\ \citenamefont
  {Monaco}(2017)}]{cornean2017b}%
  \BibitemOpen
  \bibfield  {author} {\bibinfo {author} {\bibfnamefont {H.~D.}\ \bibnamefont
  {Cornean}}\ and\ \bibinfo {author} {\bibfnamefont {D.}~\bibnamefont
  {Monaco}},\ }\href {\doibase 10.1007/s00023-017-0621-y} {\bibfield  {journal}
  {\bibinfo  {journal} {Annales Henri Poincar{\'e}}\ }\textbf {\bibinfo
  {volume} {18}},\ \bibinfo {pages} {3863} (\bibinfo {year}
  {2017})}\BibitemShut {NoStop}%
\bibitem [{\citenamefont {Read}(2017)}]{read2017}%
  \BibitemOpen
  \bibfield  {author} {\bibinfo {author} {\bibfnamefont {N.}~\bibnamefont
  {Read}},\ }\href {\doibase 10.1103/PhysRevB.95.115309} {\bibfield  {journal}
  {\bibinfo  {journal} {Phys. Rev. B}\ }\textbf {\bibinfo {volume} {95}},\
  \bibinfo {pages} {115309} (\bibinfo {year} {2017})}\BibitemShut {NoStop}%
\bibitem [{Note1()}]{Note1}%
  \BibitemOpen
  \bibinfo {note} {The corresponding protocol proceeds in three steps: first
  deform the Hamiltonian in such a way that its bands are individually flat.
  Second, in the projections of the Hilbert space to the band subspaces, apply
  unitary transformations (commuting with the flattened Hamiltonian) to the
  Wannier basis. Third, if desired, fan out the spectrum to that the
  Hamiltonian assumes the form of Eq.~\protect \textup {\hbox {\mathsurround
  \z@ \protect \normalfont (\ignorespaces \ref {eq:Hbulkprime}\unskip
  \@@italiccorr )}}.}\BibitemShut {Stop}%
\bibitem [{\citenamefont {Trifunovic}(2020)}]{trifunovic2020b}%
  \BibitemOpen
  \bibfield  {author} {\bibinfo {author} {\bibfnamefont {L.}~\bibnamefont
  {Trifunovic}},\ }\href {\doibase 10.1103/PhysRevResearch.2.043012} {\bibfield
   {journal} {\bibinfo  {journal} {Phys. Rev. Res.}\ }\textbf {\bibinfo
  {volume} {2}},\ \bibinfo {pages} {043012} (\bibinfo {year}
  {2020})}\BibitemShut {NoStop}%
\bibitem [{Note2()}]{Note2}%
  \BibitemOpen
  \bibinfo {note} {This result is in conflict with the conclusions of
  Ref.~\cite {song2014}, while it agrees with the conclusions of Ref.~\cite
  {hastings2011}.}\BibitemShut {Stop}%
\bibitem [{\citenamefont {Lapierre}\ \emph {et~al.}(2021)\citenamefont
  {Lapierre}, \citenamefont {Neupert},\ and\ \citenamefont
  {Trifunovic}}]{lapierre2021}%
  \BibitemOpen
  \bibfield  {author} {\bibinfo {author} {\bibfnamefont {B.}~\bibnamefont
  {Lapierre}}, \bibinfo {author} {\bibfnamefont {T.}~\bibnamefont {Neupert}}, \
  and\ \bibinfo {author} {\bibfnamefont {L.}~\bibnamefont {Trifunovic}},\
  }\href {https://link.aps.org/doi/10.1103/PhysRevResearch.3.033045} {\bibfield
   {journal} {\bibinfo  {journal} {Phys. Rev. Research}\ }\textbf {\bibinfo
  {volume} {3}},\ \bibinfo {pages} {033045} (\bibinfo {year}
  {2021})}\BibitemShut {NoStop}%
\bibitem [{\citenamefont {Qi}\ \emph {et~al.}(2008)\citenamefont {Qi},
  \citenamefont {Hughes},\ and\ \citenamefont {Zhang}}]{qi2008}%
  \BibitemOpen
  \bibfield  {author} {\bibinfo {author} {\bibfnamefont {X.-L.}\ \bibnamefont
  {Qi}}, \bibinfo {author} {\bibfnamefont {T.~L.}\ \bibnamefont {Hughes}}, \
  and\ \bibinfo {author} {\bibfnamefont {S.-C.}\ \bibnamefont {Zhang}},\ }\href
  {\doibase 10.1103/PhysRevB.78.195424} {\bibfield  {journal} {\bibinfo
  {journal} {Phys. Rev. B}\ }\textbf {\bibinfo {volume} {78}},\ \bibinfo
  {pages} {195424} (\bibinfo {year} {2008})}\BibitemShut {NoStop}%
\bibitem [{Note3()}]{Note3}%
  \BibitemOpen
  \bibinfo {note} {P.~W.~Brouwer, B.~Lapierre, T.~Neupert, L.~Trifunovic, in
  preparation}\BibitemShut {NoStop}%
\bibitem [{\citenamefont {Senthil}\ and\ \citenamefont
  {Fisher}(2000)}]{senthil2000}%
  \BibitemOpen
  \bibfield  {author} {\bibinfo {author} {\bibfnamefont {T.}~\bibnamefont
  {Senthil}}\ and\ \bibinfo {author} {\bibfnamefont {M.~P.~A.}\ \bibnamefont
  {Fisher}},\ }\href {\doibase 10.1103/PhysRevB.61.9690} {\bibfield  {journal}
  {\bibinfo  {journal} {Phys. Rev. B}\ }\textbf {\bibinfo {volume} {61}},\
  \bibinfo {pages} {9690} (\bibinfo {year} {2000})}\BibitemShut {NoStop}%
\bibitem [{\citenamefont {Read}\ and\ \citenamefont {Green}(2000)}]{read2000}%
  \BibitemOpen
  \bibfield  {author} {\bibinfo {author} {\bibfnamefont {N.}~\bibnamefont
  {Read}}\ and\ \bibinfo {author} {\bibfnamefont {D.}~\bibnamefont {Green}},\
  }\href {\doibase 10.1103/PhysRevB.61.10267} {\bibfield  {journal} {\bibinfo
  {journal} {Phys. Rev. B}\ }\textbf {\bibinfo {volume} {61}},\ \bibinfo
  {pages} {10267} (\bibinfo {year} {2000})}\BibitemShut {NoStop}%
\bibitem [{\citenamefont {Su}\ \emph {et~al.}(1979)\citenamefont {Su},
  \citenamefont {Schrieffer},\ and\ \citenamefont {Heeger}}]{su1979}%
  \BibitemOpen
  \bibfield  {author} {\bibinfo {author} {\bibfnamefont {W.~P.}\ \bibnamefont
  {Su}}, \bibinfo {author} {\bibfnamefont {J.~R.}\ \bibnamefont {Schrieffer}},
  \ and\ \bibinfo {author} {\bibfnamefont {A.~J.}\ \bibnamefont {Heeger}},\
  }\href {\doibase 10.1103/PhysRevLett.42.1698} {\bibfield  {journal} {\bibinfo
   {journal} {Phys. Rev. Lett.}\ }\textbf {\bibinfo {volume} {42}},\ \bibinfo
  {pages} {1698} (\bibinfo {year} {1979})}\BibitemShut {NoStop}%
\bibitem [{\citenamefont {Kivelson}(1982)}]{kivelson1982}%
  \BibitemOpen
  \bibfield  {author} {\bibinfo {author} {\bibfnamefont {S.}~\bibnamefont
  {Kivelson}},\ }\href {\doibase 10.1103/PhysRevB.26.4269} {\bibfield
  {journal} {\bibinfo  {journal} {Phys. Rev. B}\ }\textbf {\bibinfo {volume}
  {26}},\ \bibinfo {pages} {4269} (\bibinfo {year} {1982})}\BibitemShut
  {NoStop}%
\bibitem [{\citenamefont {Bradlyn}\ \emph {et~al.}(2017)\citenamefont
  {Bradlyn}, \citenamefont {Elcoro}, \citenamefont {Cano}, \citenamefont
  {Vergniory}, \citenamefont {Wang}, \citenamefont {Felser}, \citenamefont
  {Aroyo},\ and\ \citenamefont {Bernevig}}]{bradlyn2017}%
  \BibitemOpen
  \bibfield  {author} {\bibinfo {author} {\bibfnamefont {B.}~\bibnamefont
  {Bradlyn}}, \bibinfo {author} {\bibfnamefont {L.}~\bibnamefont {Elcoro}},
  \bibinfo {author} {\bibfnamefont {J.}~\bibnamefont {Cano}}, \bibinfo {author}
  {\bibfnamefont {M.~G.}\ \bibnamefont {Vergniory}}, \bibinfo {author}
  {\bibfnamefont {Z.}~\bibnamefont {Wang}}, \bibinfo {author} {\bibfnamefont
  {C.}~\bibnamefont {Felser}}, \bibinfo {author} {\bibfnamefont {M.~I.}\
  \bibnamefont {Aroyo}}, \ and\ \bibinfo {author} {\bibfnamefont {B.~A.}\
  \bibnamefont {Bernevig}},\ }\href {\doibase 10.1038/nature23268} {\bibfield
  {journal} {\bibinfo  {journal} {Nature}\ }\textbf {\bibinfo {volume} {547}},\
  \bibinfo {pages} {298} (\bibinfo {year} {2017})}\BibitemShut {NoStop}%
\bibitem [{\citenamefont {Ono}\ \emph {et~al.}(2020)\citenamefont {Ono},
  \citenamefont {Po},\ and\ \citenamefont {Watanabe}}]{ono2020}%
  \BibitemOpen
  \bibfield  {author} {\bibinfo {author} {\bibfnamefont {S.}~\bibnamefont
  {Ono}}, \bibinfo {author} {\bibfnamefont {H.~C.}\ \bibnamefont {Po}}, \ and\
  \bibinfo {author} {\bibfnamefont {H.}~\bibnamefont {Watanabe}},\ }\href
  {\doibase 10.1126/sciadv.aaz8367} {\bibfield  {journal} {\bibinfo  {journal}
  {Sci Adv}\ }\textbf {\bibinfo {volume} {6}},\ \bibinfo {pages} {eaaz8367}
  (\bibinfo {year} {2020})}\BibitemShut {NoStop}%
\bibitem [{\citenamefont {Zhang}\ \emph {et~al.}(2020)\citenamefont {Zhang},
  \citenamefont {Hsu},\ and\ \citenamefont {Das~Sarma}}]{zhang2020}%
  \BibitemOpen
  \bibfield  {author} {\bibinfo {author} {\bibfnamefont {R.-X.}\ \bibnamefont
  {Zhang}}, \bibinfo {author} {\bibfnamefont {Y.-T.}\ \bibnamefont {Hsu}}, \
  and\ \bibinfo {author} {\bibfnamefont {S.}~\bibnamefont {Das~Sarma}},\ }\href
  {\doibase 10.1103/PhysRevB.102.094503} {\bibfield  {journal} {\bibinfo
  {journal} {Phys. Rev. B}\ }\textbf {\bibinfo {volume} {102}},\ \bibinfo
  {pages} {094503} (\bibinfo {year} {2020})}\BibitemShut {NoStop}%
\bibitem [{\citenamefont {Hastings}\ and\ \citenamefont
  {Loring}(2011)}]{hastings2011}%
  \BibitemOpen
  \bibfield  {author} {\bibinfo {author} {\bibfnamefont {M.~B.}\ \bibnamefont
  {Hastings}}\ and\ \bibinfo {author} {\bibfnamefont {T.~A.}\ \bibnamefont
  {Loring}},\ }\href {\doibase https://doi.org/10.1016/j.aop.2010.12.013}
  {\bibfield  {journal} {\bibinfo  {journal} {Annals of Physics}\ }\textbf
  {\bibinfo {volume} {326}},\ \bibinfo {pages} {1699} (\bibinfo {year}
  {2011})},\ \bibinfo {note} {july 2011 Special Issue}\BibitemShut {NoStop}%
\bibitem [{\citenamefont {Alexandradinata}\ \emph {et~al.}(2021)\citenamefont
  {Alexandradinata}, \citenamefont {Nelson},\ and\ \citenamefont
  {Soluyanov}}]{alexandradinata2021}%
  \BibitemOpen
  \bibfield  {author} {\bibinfo {author} {\bibfnamefont {A.}~\bibnamefont
  {Alexandradinata}}, \bibinfo {author} {\bibfnamefont {A.}~\bibnamefont
  {Nelson}}, \ and\ \bibinfo {author} {\bibfnamefont {A.~A.}\ \bibnamefont
  {Soluyanov}},\ }\href {\doibase 10.1103/PhysRevB.103.045107} {\bibfield
  {journal} {\bibinfo  {journal} {Phys. Rev. B}\ }\textbf {\bibinfo {volume}
  {103}},\ \bibinfo {pages} {045107} (\bibinfo {year} {2021})}\BibitemShut
  {NoStop}%
\bibitem [{Note4()}]{Note4}%
  \BibitemOpen
  \bibinfo {note} {The model in Eq.~(\ref {HLudwig}) can be viewed as the
  mean-field description of a bulk topological superconductor, see also
  Sec.~\ref {sec:exp}. In particular, the Hamiltonian~(\ref {HLudwig}) defines
  a lattice representation of the DIII topological superfluid $^3$He-$B$,
  treated at the level of static mean-field theory~\cite {volovikbook}. In this
  interpretation, the Pauli matrices ${\tau }_a$ and ${\sigma }_b$ act on
  particle-hole and spin-1/2 space, respectively, and the $p$-wave pairing is
  encoded by the sine terms. The pairing amplitude has been set equal to unity,
  and the topology is controlled by the ``chemical potential'' $M$. The winding
  numbers are consistent with those of the continuum representation, where $\nu
  = 0$ corresponds to a topologically trivial BEC phase. In the $^3$He-$B$
  interpretation, the $\protect \mathcal {C}$-symmetry follows from the reality
  of the Balian-Werthammer spinor used to define the Bogoliubov-de Gennes
  Hamiltonian~\cite {ColemanBook,Foster2014}.}\BibitemShut {Stop}%
\bibitem [{Note5()}]{Note5}%
  \BibitemOpen
  \bibinfo {note} {In the numerical simulations disorder is introduced both in
  the bulk and surface layers. On the surface layers, however, it is stronger
  by a factor of $\times 5$. This is to enhance surface multifractality of the
  $E=0$ state.}\BibitemShut {Stop}%
\bibitem [{Note6()}]{Note6}%
  \BibitemOpen
  \bibinfo {note} {To see this, consider two gapped class-AIII Hamiltonians $H$
  and $H'$ with the the same winding numbers $\nu $ and with detached surface
  bands, having Chern numbers $\protect \mbox {Ch}$ and $\protect \mbox {Ch}'$,
  respectively. The difference $\protect \mbox {Ch} - \protect \mbox {Ch}'$ is
  the Chern number of the surface band of the direct sum $\delta H = H \oplus
  \protect \overline {H'}$, where $\protect \overline {H'}$ is the topological
  ``inverse'' of $H'$, {\protect \em i.e.}, a gapped Hamiltonian with detached
  surface band possessing the bulk winding number $-\nu $ and surface Chern
  number $-\protect \mbox {Ch}'$. Since, by construction, $\delta H$ has
  vanishing bulk winding number, its surface band can be continuously deformed
  to separate bands at positive and negative energy without closing the gap
  between surface band and bulk states during the deformation process. As the
  surface spectrum of $\delta H$ is symmetric around $E=0$, the Chern number
  $\delta \protect \mbox {Ch}$ of its (full) surface band is twice that of its
  positive-energy band and, hence, even. [Note that the same conclusion was
  reached in Eq.\ (\ref {eq:deltaCh}) using an explicit construction of the
  surface states.]}\BibitemShut {NoStop}%
\bibitem [{Note7()}]{Note7}%
  \BibitemOpen
  \bibinfo {note} {Close inspection of Fig.~\ref {fig:sigma} (solid curve)
  shows that $\theta (E) = \pi $ for a finite value $E \approx 0.8$ for
  $u_{{\protect \rm f},x} = -0.3 $, corresponding to delocalization of surface
  states at that energy according to the criterion~(\ref {eq:BerryCriterion}).
  This delocalization does not contradict our general conclusion that states
  are localizable in principle at $E \protect \neq 0$, as one finds, {\protect
  \em e.g.}, that all states with $E \protect \neq 0$ can be localized upon
  increasing the value of $ u_{{\protect \rm f},x}$ [dashed curve in Fig.~\ref
  {fig:sigma}(a)].}\BibitemShut {Stop}%
\bibitem [{\citenamefont {Trifunovic}\ and\ \citenamefont
  {Brouwer}(2019)}]{trifunovic2019b}%
  \BibitemOpen
  \bibfield  {author} {\bibinfo {author} {\bibfnamefont {L.}~\bibnamefont
  {Trifunovic}}\ and\ \bibinfo {author} {\bibfnamefont {P.~W.}\ \bibnamefont
  {Brouwer}},\ }\href {\doibase 10.1103/PhysRevB.99.205431} {\bibfield
  {journal} {\bibinfo  {journal} {Phys. Rev. B}\ }\textbf {\bibinfo {volume}
  {99}},\ \bibinfo {pages} {205431} (\bibinfo {year} {2019})}\BibitemShut
  {NoStop}%
\bibitem [{\citenamefont {Chalker}\ and\ \citenamefont
  {Coddington}(1988)}]{chalker1988}%
  \BibitemOpen
  \bibfield  {author} {\bibinfo {author} {\bibfnamefont {J.~T.}\ \bibnamefont
  {Chalker}}\ and\ \bibinfo {author} {\bibfnamefont {P.~D.}\ \bibnamefont
  {Coddington}},\ }\href {\doibase 10.1088/0022-3719/21/14/008} {\bibfield
  {journal} {\bibinfo  {journal} {Journal of Physics C: Solid State Physics}\
  }\textbf {\bibinfo {volume} {21}},\ \bibinfo {pages} {2665} (\bibinfo {year}
  {1988})}\BibitemShut {NoStop}%
\bibitem [{\citenamefont {Fulga}\ \emph {et~al.}(2014)\citenamefont {Fulga},
  \citenamefont {van Heck}, \citenamefont {Edge},\ and\ \citenamefont
  {Akhmerov}}]{fulga2014}%
  \BibitemOpen
  \bibfield  {author} {\bibinfo {author} {\bibfnamefont {I.~C.}\ \bibnamefont
  {Fulga}}, \bibinfo {author} {\bibfnamefont {B.}~\bibnamefont {van Heck}},
  \bibinfo {author} {\bibfnamefont {J.~M.}\ \bibnamefont {Edge}}, \ and\
  \bibinfo {author} {\bibfnamefont {A.~R.}\ \bibnamefont {Akhmerov}},\ }\href
  {\doibase 10.1103/PhysRevB.89.155424} {\bibfield  {journal} {\bibinfo
  {journal} {Phys. Rev. B}\ }\textbf {\bibinfo {volume} {89}},\ \bibinfo
  {pages} {155424} (\bibinfo {year} {2014})}\BibitemShut {NoStop}%
\bibitem [{\citenamefont {Nomura}\ \emph {et~al.}(2008)\citenamefont {Nomura},
  \citenamefont {Ryu}, \citenamefont {Koshino}, \citenamefont {Mudry},\ and\
  \citenamefont {Furusaki}}]{nomura2008}%
  \BibitemOpen
  \bibfield  {author} {\bibinfo {author} {\bibfnamefont {K.}~\bibnamefont
  {Nomura}}, \bibinfo {author} {\bibfnamefont {S.}~\bibnamefont {Ryu}},
  \bibinfo {author} {\bibfnamefont {M.}~\bibnamefont {Koshino}}, \bibinfo
  {author} {\bibfnamefont {C.}~\bibnamefont {Mudry}}, \ and\ \bibinfo {author}
  {\bibfnamefont {A.}~\bibnamefont {Furusaki}},\ }\href {\doibase
  10.1103/PhysRevLett.100.246806} {\bibfield  {journal} {\bibinfo  {journal}
  {Phys. Rev. Lett.}\ }\textbf {\bibinfo {volume} {100}},\ \bibinfo {pages}
  {246806} (\bibinfo {year} {2008})}\BibitemShut {NoStop}%
\bibitem [{\citenamefont {Ringel}\ \emph {et~al.}(2012)\citenamefont {Ringel},
  \citenamefont {Kraus},\ and\ \citenamefont {Stern}}]{ringel2012}%
  \BibitemOpen
  \bibfield  {author} {\bibinfo {author} {\bibfnamefont {Z.}~\bibnamefont
  {Ringel}}, \bibinfo {author} {\bibfnamefont {Y.~E.}\ \bibnamefont {Kraus}}, \
  and\ \bibinfo {author} {\bibfnamefont {A.}~\bibnamefont {Stern}},\ }\href
  {\doibase 10.1103/PhysRevB.86.045102} {\bibfield  {journal} {\bibinfo
  {journal} {Phys. Rev. B}\ }\textbf {\bibinfo {volume} {86}},\ \bibinfo
  {pages} {045102} (\bibinfo {year} {2012})}\BibitemShut {NoStop}%
\bibitem [{\citenamefont {Mong}\ \emph {et~al.}(2012)\citenamefont {Mong},
  \citenamefont {Bardarson},\ and\ \citenamefont {Moore}}]{mong2012}%
  \BibitemOpen
  \bibfield  {author} {\bibinfo {author} {\bibfnamefont {R.~S.~K.}\
  \bibnamefont {Mong}}, \bibinfo {author} {\bibfnamefont {J.~H.}\ \bibnamefont
  {Bardarson}}, \ and\ \bibinfo {author} {\bibfnamefont {J.~E.}\ \bibnamefont
  {Moore}},\ }\href {\doibase 10.1103/PhysRevLett.108.076804} {\bibfield
  {journal} {\bibinfo  {journal} {Phys. Rev. Lett.}\ }\textbf {\bibinfo
  {volume} {108}},\ \bibinfo {pages} {076804} (\bibinfo {year}
  {2012})}\BibitemShut {NoStop}%
\bibitem [{\citenamefont {Fu}\ and\ \citenamefont {Kane}(2012)}]{Fu2012}%
  \BibitemOpen
  \bibfield  {author} {\bibinfo {author} {\bibfnamefont {L.}~\bibnamefont
  {Fu}}\ and\ \bibinfo {author} {\bibfnamefont {C.~L.}\ \bibnamefont {Kane}},\
  }\href {\doibase 10.1103/PhysRevLett.109.246605} {\bibfield  {journal}
  {\bibinfo  {journal} {Physical Review Letters}\ }\textbf {\bibinfo {volume}
  {109}},\ \bibinfo {pages} {246605} (\bibinfo {year} {2012})},\ \bibinfo
  {note} {publisher: American Physical Society}\BibitemShut {NoStop}%
\bibitem [{\citenamefont {Rodriguez}\ \emph {et~al.}(2011)\citenamefont
  {Rodriguez}, \citenamefont {Vasquez}, \citenamefont {Slevin},\ and\
  \citenamefont {R\"omer}}]{rodriguez2011}%
  \BibitemOpen
  \bibfield  {author} {\bibinfo {author} {\bibfnamefont {A.}~\bibnamefont
  {Rodriguez}}, \bibinfo {author} {\bibfnamefont {L.~J.}\ \bibnamefont
  {Vasquez}}, \bibinfo {author} {\bibfnamefont {K.}~\bibnamefont {Slevin}}, \
  and\ \bibinfo {author} {\bibfnamefont {R.~A.}\ \bibnamefont {R\"omer}},\
  }\href {\doibase 10.1103/PhysRevB.84.134209} {\bibfield  {journal} {\bibinfo
  {journal} {Phys. Rev. B}\ }\textbf {\bibinfo {volume} {84}},\ \bibinfo
  {pages} {134209} (\bibinfo {year} {2011})}\BibitemShut {NoStop}%
\bibitem [{Note8()}]{Note8}%
  \BibitemOpen
  \bibinfo {note} {The computational cost of the sparse matrix diagonalization,
  carried out using the ARPACK library scales with the Hilbert space dimension
  of the 3d system, {\protect \em i.e.}, with $4\times N_x\times L^2\leq
  5\times 10^5$.}\BibitemShut {Stop}%
\bibitem [{\citenamefont {Ludwig}\ \emph {et~al.}(1994)\citenamefont {Ludwig},
  \citenamefont {Fisher}, \citenamefont {Shankar},\ and\ \citenamefont
  {Grinstein}}]{ludwig1994}%
  \BibitemOpen
  \bibfield  {author} {\bibinfo {author} {\bibfnamefont {A.~W.~W.}\
  \bibnamefont {Ludwig}}, \bibinfo {author} {\bibfnamefont {M.~P.~A.}\
  \bibnamefont {Fisher}}, \bibinfo {author} {\bibfnamefont {R.}~\bibnamefont
  {Shankar}}, \ and\ \bibinfo {author} {\bibfnamefont {G.}~\bibnamefont
  {Grinstein}},\ }\href {\doibase 10.1103/PhysRevB.50.7526} {\bibfield
  {journal} {\bibinfo  {journal} {Phys. Rev. B}\ }\textbf {\bibinfo {volume}
  {50}},\ \bibinfo {pages} {7526} (\bibinfo {year} {1994})}\BibitemShut
  {NoStop}%
\bibitem [{\citenamefont {Obuse}\ \emph {et~al.}(2008)\citenamefont {Obuse},
  \citenamefont {Subramaniam}, \citenamefont {Furusaki}, \citenamefont
  {Gruzberg},\ and\ \citenamefont {Ludwig}}]{Obuse2008}%
  \BibitemOpen
  \bibfield  {author} {\bibinfo {author} {\bibfnamefont {H.}~\bibnamefont
  {Obuse}}, \bibinfo {author} {\bibfnamefont {A.~R.}\ \bibnamefont
  {Subramaniam}}, \bibinfo {author} {\bibfnamefont {A.}~\bibnamefont
  {Furusaki}}, \bibinfo {author} {\bibfnamefont {I.~A.}\ \bibnamefont
  {Gruzberg}}, \ and\ \bibinfo {author} {\bibfnamefont {A.~W.~W.}\ \bibnamefont
  {Ludwig}},\ }\href {\doibase 10.1103/PhysRevLett.101.116802} {\bibfield
  {journal} {\bibinfo  {journal} {Phys. Rev. Lett.}\ }\textbf {\bibinfo
  {volume} {101}},\ \bibinfo {pages} {116802} (\bibinfo {year}
  {2008})}\BibitemShut {NoStop}%
\bibitem [{\citenamefont {Evers}\ \emph {et~al.}(2008)\citenamefont {Evers},
  \citenamefont {Mildenberger},\ and\ \citenamefont {Mirlin}}]{evers2008a}%
  \BibitemOpen
  \bibfield  {author} {\bibinfo {author} {\bibfnamefont {F.}~\bibnamefont
  {Evers}}, \bibinfo {author} {\bibfnamefont {A.}~\bibnamefont {Mildenberger}},
  \ and\ \bibinfo {author} {\bibfnamefont {A.~D.}\ \bibnamefont {Mirlin}},\
  }\href {\doibase 10.1103/PhysRevLett.101.116803} {\bibfield  {journal}
  {\bibinfo  {journal} {Phys. Rev. Lett.}\ }\textbf {\bibinfo {volume} {101}},\
  \bibinfo {pages} {116803} (\bibinfo {year} {2008})}\BibitemShut {NoStop}%
\bibitem [{\citenamefont {Zirnbauer}(2019)}]{zirnbauer2019}%
  \BibitemOpen
  \bibfield  {author} {\bibinfo {author} {\bibfnamefont {M.~R.}\ \bibnamefont
  {Zirnbauer}},\ }\href {\doibase
  https://doi.org/10.1016/j.nuclphysb.2019.02.017} {\bibfield  {journal}
  {\bibinfo  {journal} {Nuclear Physics B}\ }\textbf {\bibinfo {volume}
  {941}},\ \bibinfo {pages} {458 } (\bibinfo {year} {2019})}\BibitemShut
  {NoStop}%
\bibitem [{\citenamefont {Foster}\ \emph {et~al.}(2014)\citenamefont {Foster},
  \citenamefont {Xie},\ and\ \citenamefont {Chou}}]{Foster2014}%
  \BibitemOpen
  \bibfield  {author} {\bibinfo {author} {\bibfnamefont {M.~S.}\ \bibnamefont
  {Foster}}, \bibinfo {author} {\bibfnamefont {H.-Y.}\ \bibnamefont {Xie}}, \
  and\ \bibinfo {author} {\bibfnamefont {Y.-Z.}\ \bibnamefont {Chou}},\ }\href
  {\doibase 10.1103/PhysRevB.89.155140} {\bibfield  {journal} {\bibinfo
  {journal} {Phys. Rev. B}\ }\textbf {\bibinfo {volume} {89}},\ \bibinfo
  {pages} {155140} (\bibinfo {year} {2014})}\BibitemShut {NoStop}%
\bibitem [{\citenamefont {Chiu}\ \emph {et~al.}(2016)\citenamefont {Chiu},
  \citenamefont {Teo}, \citenamefont {Schnyder},\ and\ \citenamefont
  {Ryu}}]{chiu2016}%
  \BibitemOpen
  \bibfield  {author} {\bibinfo {author} {\bibfnamefont {C.-K.}\ \bibnamefont
  {Chiu}}, \bibinfo {author} {\bibfnamefont {J.~C.~Y.}\ \bibnamefont {Teo}},
  \bibinfo {author} {\bibfnamefont {A.~P.}\ \bibnamefont {Schnyder}}, \ and\
  \bibinfo {author} {\bibfnamefont {S.}~\bibnamefont {Ryu}},\ }\href {\doibase
  10.1103/RevModPhys.88.035005} {\bibfield  {journal} {\bibinfo  {journal}
  {Rev. Mod. Phys.}\ }\textbf {\bibinfo {volume} {88}},\ \bibinfo {pages}
  {035005} (\bibinfo {year} {2016})}\BibitemShut {NoStop}%
\bibitem [{\citenamefont {Wu}\ \emph {et~al.}(2020)\citenamefont {Wu},
  \citenamefont {Pal}, \citenamefont {Hosur},\ and\ \citenamefont
  {Foster}}]{Wu2020}%
  \BibitemOpen
  \bibfield  {author} {\bibinfo {author} {\bibfnamefont {T.~C.}\ \bibnamefont
  {Wu}}, \bibinfo {author} {\bibfnamefont {H.~K.}\ \bibnamefont {Pal}},
  \bibinfo {author} {\bibfnamefont {P.}~\bibnamefont {Hosur}}, \ and\ \bibinfo
  {author} {\bibfnamefont {M.~S.}\ \bibnamefont {Foster}},\ }\href {\doibase
  10.1103/PhysRevLett.124.067001} {\bibfield  {journal} {\bibinfo  {journal}
  {Phys. Rev. Lett.}\ }\textbf {\bibinfo {volume} {124}},\ \bibinfo {pages}
  {067001} (\bibinfo {year} {2020})}\BibitemShut {NoStop}%
\bibitem [{\citenamefont {Wang}\ \emph {et~al.}(2000)\citenamefont {Wang},
  \citenamefont {Fisher}, \citenamefont {Girvin},\ and\ \citenamefont
  {Chalker}}]{Wang2000}%
  \BibitemOpen
  \bibfield  {author} {\bibinfo {author} {\bibfnamefont {Z.}~\bibnamefont
  {Wang}}, \bibinfo {author} {\bibfnamefont {M.~P.~A.}\ \bibnamefont {Fisher}},
  \bibinfo {author} {\bibfnamefont {S.~M.}\ \bibnamefont {Girvin}}, \ and\
  \bibinfo {author} {\bibfnamefont {J.~T.}\ \bibnamefont {Chalker}},\ }\href
  {\doibase 10.1103/PhysRevB.61.8326} {\bibfield  {journal} {\bibinfo
  {journal} {Phys. Rev. B}\ }\textbf {\bibinfo {volume} {61}},\ \bibinfo
  {pages} {8326} (\bibinfo {year} {2000})}\BibitemShut {NoStop}%
\bibitem [{Note9()}]{Note9}%
  \BibitemOpen
  \bibinfo {note} {The `(d+0)' means that we are considering non-dynamical and
  non-interacting frameworks and thus can avoid the introduction of a time-like
  coordinate; we will always be considering a fixed reference
  energy.}\BibitemShut {Stop}%
\bibitem [{\citenamefont {Efetov}(1997)}]{Efetbook}%
  \BibitemOpen
  \bibfield  {author} {\bibinfo {author} {\bibfnamefont {K.~B.}\ \bibnamefont
  {Efetov}},\ }\href@noop {} {\emph {\bibinfo {title} {Supersymmetry in
  {Disorder} and {Chaos}}}}\ (\bibinfo  {publisher} {Cambridge University
  Press},\ \bibinfo {address} {Cambridge},\ \bibinfo {year} {1997})\BibitemShut
  {NoStop}%
\bibitem [{\citenamefont {Gruzberg}\ \emph {et~al.}(1999)\citenamefont
  {Gruzberg}, \citenamefont {Ludwig},\ and\ \citenamefont
  {Read}}]{gruzbergExactExponentsSpin1999}%
  \BibitemOpen
  \bibfield  {author} {\bibinfo {author} {\bibfnamefont {I.~A.}\ \bibnamefont
  {Gruzberg}}, \bibinfo {author} {\bibfnamefont {A.~W.~W.}\ \bibnamefont
  {Ludwig}}, \ and\ \bibinfo {author} {\bibfnamefont {N.}~\bibnamefont
  {Read}},\ }\href {\doibase 10.1103/PhysRevLett.82.4524} {\bibfield  {journal}
  {\bibinfo  {journal} {Physical Review Letters}\ }\textbf {\bibinfo {volume}
  {82}},\ \bibinfo {pages} {4524} (\bibinfo {year} {1999})},\ \bibinfo {note}
  {publisher: American Physical Society}\BibitemShut {NoStop}%
\bibitem [{\citenamefont {Gruzberg}\ \emph {et~al.}(2001)\citenamefont
  {Gruzberg}, \citenamefont {Read},\ and\ \citenamefont
  {Ludwig}}]{gruzbergRandombondIsingModel2001}%
  \BibitemOpen
  \bibfield  {author} {\bibinfo {author} {\bibfnamefont {I.~A.}\ \bibnamefont
  {Gruzberg}}, \bibinfo {author} {\bibfnamefont {N.}~\bibnamefont {Read}}, \
  and\ \bibinfo {author} {\bibfnamefont {A.~W.~W.}\ \bibnamefont {Ludwig}},\
  }\href {\doibase 10.1103/PhysRevB.63.104422} {\bibfield  {journal} {\bibinfo
  {journal} {Physical Review B}\ }\textbf {\bibinfo {volume} {63}},\ \bibinfo
  {pages} {104422} (\bibinfo {year} {2001})},\ \bibinfo {note} {publisher:
  American Physical Society}\BibitemShut {NoStop}%
\bibitem [{\citenamefont {Altland}\ and\ \citenamefont
  {Merkt}(2001)}]{Altland2001511}%
  \BibitemOpen
  \bibfield  {author} {\bibinfo {author} {\bibfnamefont {A.}~\bibnamefont
  {Altland}}\ and\ \bibinfo {author} {\bibfnamefont {R.}~\bibnamefont
  {Merkt}},\ }\href {\doibase http://dx.doi.org/10.1016/S0550-3213(01)00209-7}
  {\bibfield  {journal} {\bibinfo  {journal} {Nuclear Physics B}\ }\textbf
  {\bibinfo {volume} {607}},\ \bibinfo {pages} {511} (\bibinfo {year}
  {2001})}\BibitemShut {NoStop}%
\bibitem [{\citenamefont
  {Schulz-Baldes}(2016)}]{schulz-baldesTopologicalInsulatorsPerspective2016}%
  \BibitemOpen
  \bibfield  {author} {\bibinfo {author} {\bibfnamefont {H.}~\bibnamefont
  {Schulz-Baldes}},\ }\href {\doibase 10.1365/s13291-016-0142-5} {\bibfield
  {journal} {\bibinfo  {journal} {Jahresbericht der Deutschen
  Mathematiker-Vereinigung}\ }\textbf {\bibinfo {volume} {118}},\ \bibinfo
  {pages} {247} (\bibinfo {year} {2016})}\BibitemShut {NoStop}%
\bibitem [{\citenamefont {Pruisken}(1984)}]{Pruisken1984a}%
  \BibitemOpen
  \bibfield  {author} {\bibinfo {author} {\bibfnamefont {A.}~\bibnamefont
  {Pruisken}},\ }\href
  {http://www.sciencedirect.com/science/article/pii/0550321384901019}
  {\bibfield  {journal} {\bibinfo  {journal} {Nuclear Physics B}\ }\textbf
  {\bibinfo {volume} {235}},\ \bibinfo {pages} {277} (\bibinfo {year}
  {1984})}\BibitemShut {NoStop}%
\bibitem [{\citenamefont {Altland}\ \emph {et~al.}(2015)\citenamefont
  {Altland}, \citenamefont {Bagrets},\ and\ \citenamefont
  {Kamenev}}]{altlandTopologyAndersonLocalization2015}%
  \BibitemOpen
  \bibfield  {author} {\bibinfo {author} {\bibfnamefont {A.}~\bibnamefont
  {Altland}}, \bibinfo {author} {\bibfnamefont {D.}~\bibnamefont {Bagrets}}, \
  and\ \bibinfo {author} {\bibfnamefont {A.}~\bibnamefont {Kamenev}},\ }\href
  {\doibase 10.1103/PhysRevB.91.085429} {\bibfield  {journal} {\bibinfo
  {journal} {Physical Review B}\ }\textbf {\bibinfo {volume} {91}},\ \bibinfo
  {pages} {085429} (\bibinfo {year} {2015})},\ \bibinfo {note} {publisher:
  American Physical Society}\BibitemShut {NoStop}%
\bibitem [{\citenamefont {Gade}\ and\ \citenamefont
  {Wegner}(1991)}]{gadeReplicaLimitModels1991}%
  \BibitemOpen
  \bibfield  {author} {\bibinfo {author} {\bibfnamefont {R.}~\bibnamefont
  {Gade}}\ and\ \bibinfo {author} {\bibfnamefont {F.}~\bibnamefont {Wegner}},\
  }\href {\doibase 10.1016/0550-3213(91)90401-I} {\bibfield  {journal}
  {\bibinfo  {journal} {Nuclear Physics B}\ }\textbf {\bibinfo {volume}
  {360}},\ \bibinfo {pages} {213} (\bibinfo {year} {1991})}\BibitemShut
  {NoStop}%
\bibitem [{\citenamefont {Altland}\ and\ \citenamefont
  {Simons}(1999)}]{altlandFieldTheoryRandom1999a}%
  \BibitemOpen
  \bibfield  {author} {\bibinfo {author} {\bibfnamefont {A.}~\bibnamefont
  {Altland}}\ and\ \bibinfo {author} {\bibfnamefont {B.}~\bibnamefont
  {Simons}},\ }\href {\doibase 10.1088/0305-4470/32/31/101} {\bibfield
  {journal} {\bibinfo  {journal} {Journal of Physics A: Mathematical and
  General}\ }\textbf {\bibinfo {volume} {32}} (\bibinfo {year} {1999}),\
  10.1088/0305-4470/32/31/101}\BibitemShut {NoStop}%
\bibitem [{\citenamefont {Altland}\ \emph {et~al.}(2002)\citenamefont
  {Altland}, \citenamefont {Simons},\ and\ \citenamefont
  {Zirnbauer}}]{Altland2002}%
  \BibitemOpen
  \bibfield  {author} {\bibinfo {author} {\bibfnamefont {A.}~\bibnamefont
  {Altland}}, \bibinfo {author} {\bibfnamefont {B.}~\bibnamefont {Simons}}, \
  and\ \bibinfo {author} {\bibfnamefont {M.}~\bibnamefont {Zirnbauer}},\ }\href
  {\doibase 10.1016/S0370-1573(01)00065-5} {\bibfield  {journal} {\bibinfo
  {journal} {Physics Reports}\ }\textbf {\bibinfo {volume} {359}},\ \bibinfo
  {pages} {283} (\bibinfo {year} {2002})}\BibitemShut {NoStop}%
\bibitem [{\citenamefont {Nersesyan}\ \emph {et~al.}(1994)\citenamefont
  {Nersesyan}, \citenamefont {Tsvelik},\ and\ \citenamefont
  {Wenger}}]{nersesyanDisorderEffectsTwodimensional1994}%
  \BibitemOpen
  \bibfield  {author} {\bibinfo {author} {\bibfnamefont {A.~A.}\ \bibnamefont
  {Nersesyan}}, \bibinfo {author} {\bibfnamefont {A.~M.}\ \bibnamefont
  {Tsvelik}}, \ and\ \bibinfo {author} {\bibfnamefont {F.}~\bibnamefont
  {Wenger}},\ }\href {\doibase 10.1103/PhysRevLett.72.2628} {\bibfield
  {journal} {\bibinfo  {journal} {Physical Review Letters}\ }\textbf {\bibinfo
  {volume} {72}},\ \bibinfo {pages} {2628} (\bibinfo {year} {1994})},\ \bibinfo
  {note} {publisher: American Physical Society}\BibitemShut {NoStop}%
\bibitem [{\citenamefont {Bernard}\ and\ \citenamefont
  {LeClair}(2002)}]{BernardLeClair2002}%
  \BibitemOpen
  \bibfield  {author} {\bibinfo {author} {\bibfnamefont {D.}~\bibnamefont
  {Bernard}}\ and\ \bibinfo {author} {\bibfnamefont {A.}~\bibnamefont
  {LeClair}},\ }\href {\doibase 10.1088/0305-4470/35/11/303} {\bibfield
  {journal} {\bibinfo  {journal} {J. Phys. A}\ }\textbf {\bibinfo {volume}
  {35}},\ \bibinfo {pages} {2555} (\bibinfo {year} {2002})}\BibitemShut
  {NoStop}%
\bibitem [{\citenamefont {Song}\ \emph {et~al.}(2014)\citenamefont {Song},
  \citenamefont {Fine},\ and\ \citenamefont {Prodan}}]{song2014}%
  \BibitemOpen
  \bibfield  {author} {\bibinfo {author} {\bibfnamefont {J.}~\bibnamefont
  {Song}}, \bibinfo {author} {\bibfnamefont {C.}~\bibnamefont {Fine}}, \ and\
  \bibinfo {author} {\bibfnamefont {E.}~\bibnamefont {Prodan}},\ }\href
  {\doibase 10.1103/PhysRevB.90.184201} {\bibfield  {journal} {\bibinfo
  {journal} {Phys. Rev. B}\ }\textbf {\bibinfo {volume} {90}},\ \bibinfo
  {pages} {184201} (\bibinfo {year} {2014})}\BibitemShut {NoStop}%
\bibitem [{\citenamefont {Volovik}(2003)}]{volovikbook}%
  \BibitemOpen
  \bibfield  {author} {\bibinfo {author} {\bibfnamefont {G.~E.}\ \bibnamefont
  {Volovik}},\ }\href@noop {} {\emph {\bibinfo {title} {The Universe in a
  Helium Droplet}}}\ (\bibinfo  {publisher} {Oxford University Press, Oxford,
  UK},\ \bibinfo {year} {2003})\BibitemShut {NoStop}%
\bibitem [{\citenamefont {Coleman}(2015)}]{ColemanBook}%
  \BibitemOpen
  \bibfield  {author} {\bibinfo {author} {\bibfnamefont {P.}~\bibnamefont
  {Coleman}},\ }\href@noop {} {\emph {\bibinfo {title} {Introduction to
  Many-Body Physics}}}\ (\bibinfo  {publisher} {Cambridge University Press,
  Cambridge, England},\ \bibinfo {year} {2015})\BibitemShut {NoStop}%
\bibitem [{\citenamefont {Wegner}(1979)}]{Wegner1979}%
  \BibitemOpen
  \bibfield  {author} {\bibinfo {author} {\bibfnamefont {F.}~\bibnamefont
  {Wegner}},\ }\href {\doibase 10.1007/BF01319839} {\bibfield  {journal}
  {\bibinfo  {journal} {Zeitschrift fuer Physik B Condensed Matter and Quanta}\
  }\textbf {\bibinfo {volume} {35}},\ \bibinfo {pages} {207} (\bibinfo {year}
  {1979})},\ \bibinfo {note} {publisher: Springer-Verlag}\BibitemShut {NoStop}%
\bibitem [{\citenamefont {Ostrovsky}\ \emph {et~al.}(2007)\citenamefont
  {Ostrovsky}, \citenamefont {Gornyi},\ and\ \citenamefont
  {Mirlin}}]{ostrovsky2007}%
  \BibitemOpen
  \bibfield  {author} {\bibinfo {author} {\bibfnamefont {P.~M.}\ \bibnamefont
  {Ostrovsky}}, \bibinfo {author} {\bibfnamefont {I.~V.}\ \bibnamefont
  {Gornyi}}, \ and\ \bibinfo {author} {\bibfnamefont {A.~D.}\ \bibnamefont
  {Mirlin}},\ }\href {\doibase 10.1103/PhysRevLett.98.256801} {\bibfield
  {journal} {\bibinfo  {journal} {Phys. Rev. Lett.}\ }\textbf {\bibinfo
  {volume} {98}},\ \bibinfo {pages} {256801} (\bibinfo {year}
  {2007})}\BibitemShut {NoStop}%
\bibitem [{\citenamefont {Puschmann}\ \emph {et~al.}(2021)\citenamefont
  {Puschmann}, \citenamefont {Hernang\'omez-P\'erez}, \citenamefont {Lang},
  \citenamefont {Bera},\ and\ \citenamefont {Evers}}]{puschmann2021}%
  \BibitemOpen
  \bibfield  {author} {\bibinfo {author} {\bibfnamefont {M.}~\bibnamefont
  {Puschmann}}, \bibinfo {author} {\bibfnamefont {D.}~\bibnamefont
  {Hernang\'omez-P\'erez}}, \bibinfo {author} {\bibfnamefont {B.}~\bibnamefont
  {Lang}}, \bibinfo {author} {\bibfnamefont {S.}~\bibnamefont {Bera}}, \ and\
  \bibinfo {author} {\bibfnamefont {F.}~\bibnamefont {Evers}},\ }\href
  {\doibase 10.1103/PhysRevB.103.235167} {\bibfield  {journal} {\bibinfo
  {journal} {Phys. Rev. B}\ }\textbf {\bibinfo {volume} {103}},\ \bibinfo
  {pages} {235167} (\bibinfo {year} {2021})}\BibitemShut {NoStop}%
\bibitem [{\citenamefont {Foster}\ and\ \citenamefont
  {Ludwig}(2008)}]{foster2008}%
  \BibitemOpen
  \bibfield  {author} {\bibinfo {author} {\bibfnamefont {M.~S.}\ \bibnamefont
  {Foster}}\ and\ \bibinfo {author} {\bibfnamefont {A.~W.~W.}\ \bibnamefont
  {Ludwig}},\ }\href {\doibase 10.1103/PhysRevB.77.165108} {\bibfield
  {journal} {\bibinfo  {journal} {Phys. Rev. B}\ }\textbf {\bibinfo {volume}
  {77}},\ \bibinfo {pages} {165108} (\bibinfo {year} {2008})}\BibitemShut
  {NoStop}%
\end{thebibliography}%

\appendix

\section{Field theory analysis}
\label{sec:FieldTheoryAppendix}
In this section we discuss the field theoretical analysis of the disordered
surface. To make the paper self-contained, we start with a quick review of the
derivation of the intermediate representation
Eq.~\eqref{eq:FieldTheoryIntermediate}, here formulated for class A for
concreteness. In a  second step we then show that the expansion of that
action establishes a connection between the $\theta$-term describing the real space surface topology in the presence of disorder and the integrated momentum space Berry curvature.  
\subsection{Replica field theory}
We begin by adding disorder a potential $\hat{V}(\textbf{x})$ with variance $ \langle \hat{V}(\textbf{x}) \hat{V}(\textbf{x}')
\rangle = \frac{\gamma_0}{2}\delta( \textbf{x} -\textbf{x}')$ to the clean  Hamiltonian $\hat{H}$
. Transport observables such as the longitudinal or transverse conductance at characteristic energy $E$  may
be computed from the $R$-fold replicated partition sum~\cite{Wegner1979}
 \begin{align*}
 Z^R &= \int D\psi \exp(-S[\psi]),  \\    
 S[\psi] &= -i\int dV  \,\bar{\psi} (E + i\delta \hat{\tau}_3 - \hat{H}-\hat{V}) \psi, 
\end{align*}
where   $\psi = \{ \psi_{s,\mu}^r (\mathbf{x}) \}$ is a Grassmann field,  the index $\mu=1,\dots,4$ labels the components of
the lattice spinor, $r=1,\dots,R$ is a replica index,   $s=1,2$ distinguishes between  advanced and
retarded components, and $\hat{\tau}_3$ is a Pauli matrix in advanced/retarded space. We average the partition sum over disorder to obtain a quartic interaction potential
between replicas,
\begin{align}
        S[\psi] &= S_0[\psi]+\frac{\gamma_0}{2} \int dV (\bar{\psi}\psi)^2, 
\end{align}
where $S_0$ is the clean action. To decouple the quartic term, we introduce a Hubbard-Stratonovich matrix field $B(\textbf{x})=\{ B_{ss',ii'}^{r r'}(\textbf{x}) \}$. Integrating
out the $\psi$ fields yields, 
\begin{align*}
    \left< Z^R \right> &= \int DB \exp\left(-\frac{1}{2 \gamma_0} \int dV  \tr B^2 + \tr \ln \hat{G}[B]\right),  
\end{align*}
with $\hat{G}[B] = (E + i \delta \hat{\tau}_3 - \hat{H} - B )^{-1}$. A
variation of the action in $B$ leads to  $\bar{B}(\textbf{x}) = \gamma_0 \tr
\hat{G}[\bar{B}](\textbf{x},\textbf{x})$, which has the structure of a
self-consistent Born equation. According to it, the mean field $\bar B$ pays the
role of an impurity scattering ``self energy'' whose strength is determined by the
impurity-broadened local spectral density,
$\hat{G}[\bar{B}](\textbf{x},\textbf{x})$. The equation is solved by the
diagonal ansatz $\bar{B}= -i\kappa \hat{\tau}_3$, where $\kappa$ an effective
scattering rate determined by the bare strength $\gamma_0$. For our purposes, we
need not discuss the self-consistent dependence $\kappa(\gamma_0)$ in detail.
However, what does matter is that the stationarity equation affords a whole
manifold of solutions besides the matrix-diagonal one,   
 $B= -i \kappa T \hat{\tau}_3 T^{-1}=-i\kappa Q$, where $T \in U(2R)/(U(R) \times
U(R))$, and $Q=T \hat{\tau}_3 T^{-1}$. Physically, these are the Goldstone modes
associated to the ``spontaneous symmetry breaking'' $i \delta \to i \kappa$
reflected in the upgrade of the infinitesimal causal parameter $i \delta$ to the
finite damping $i \kappa$. 

Substituting  these modes into the action, and noting that $\tr(Q^2)=\textrm{const}.$ is a constant (vanishing in the replica limit), we arrive at the soft mode action \eqref{eq:FieldTheoryIntermediate} 
which will be our starting point for all further considerations. 

\subsection{Gradient expansion}

We now discuss the steps required to advance from
Eq.~\eqref{eq:FieldTheoryIntermediate} and its equivalent representation
\eqref{eq:TrLnMoyal} to a local action containing of lowest non-vanishing order
in gradient operators. There are only two of these consistent with the
symmetries of the model, namely $\tr(\partial_i Q \partial_i Q)$ and
$\epsilon_{ij} \,\tr(Q \partial_i Q \partial_j Q)$. The derivation of an action
containing the first via expansion of the tr ln is textbook material~\cite{Efetbook} (see also Ref.~\cite{moreno2023topological} for the specific case of the two-dimensional topological class A insulator) and is not of primary relevance to our present discussion. However, the 
construction of a topological action containing the second terms is
concerned, we need to start afresh; previous derivations of this action
 where specific to the quantum Hall effect~\cite{Pruisken1984a}, or other
genuinely two-dimensional materials~\cite{ostrovsky2007}. By contrast, we here want to allow for
situations where the effective $H(\mathbf{k})$ is given by a general spectral
decomposition as in Eq.~\eqref{eq:SurfaceDecomposition}. In this way, we will address all class A situations relevant to our
discussion (bulk two-dimensional insulators for the sake of comparison, and the
AIII surface at finite energies) in one go.       

The second order expansion of the tr ln in the combinations $F_i\Phi_i$ defined
after Eq.~\eqref{eq:TrLnMoyal} leads to two terms, $S\to
S_\textrm{top}=S^{(1)}+S^{(2)}$, where $S^{(1)}=-\,\tr(\hat{G}F_i \Phi_i)$, and
$S^{(2)}= \frac{1}{2}\,\tr(\hat{G}F_i \Phi_i)^2$, and the subscript ``top'' indicates
that we wish to isolate the topological contribution to the action. Naively, one
might think that the first order term drops out by symmetry. However, this is
not so because the trace over a single Green function leads to ultraviolet
divergent expressions, i.e. we are facing a $0 \times \infty $ situation. The
way out, originally suggested by Pruisken, is to process the first order term as 
\begin{widetext}
    \begin{align*}
        S^{(1)}&=- \int_{E}^{\infty} d \epsilon \tr(\hat{G} F_i \Phi_i \hat{G}) \approx 
        - \frac{i}{2} \int dx dk \int_{E}^{\infty} d \epsilon \,\tr((\partial_j\hat{G}) F_i (\partial_j \Phi_i) \hat{G}- \hat{G} F_i (\partial_j \Phi_i) \partial_j \hat{G} )=\cr 
        &=- \frac{i}{4}   \int_{E}^{\infty} d \epsilon \int dk\sum_s s\,\tr([\hat{G}^s,\partial_j\hat{G}^s] F_i) \int dx \,\tr (\hat{\tau}_3\partial_j \Phi_i),  
    \end{align*}
where in the first equality we used $\hat{G}(E)=-\int_E^\infty
d\epsilon\,\hat{G}^2(\epsilon)$ to increase the number of Green functions, thereby
mitigating the UV issues. To keep the notation slim, we omit the energy
arguments throughout. In the second equality we applied another Moyal
expansion (with $\partial_i \hat{G} = \partial_{k_i}\hat{G}(\mathbf{k})$), and used that
only the imaginary part $F(\hat{G}) \to \frac{1}{2}(F(\hat{G}^+)-F(\hat{G}^-))
\hat{\tau}_3=\frac{1}{2}\sum_s s F(\hat{G}^s)\hat{\tau}_3$ will contribute to a non-vanishing trace. With
the second of the two auxiliary identities
\begin{align*}
    -4 \epsilon_{lm}\sum_{s}\,\mathrm{tr}( s P^s \Phi_l P^{-s} \Phi_m)=
    4 \epsilon_{lm}\mathrm{tr}(\hat{\tau}_3\partial_l \Phi_m)=  \epsilon_{ij}\tr(Q \partial_i Q \partial_j Q
    )\equiv \mathcal{L}_\mathrm{top}(Q),
  \end{align*}
we reduce this expression to $S^{(1)}=I_1 \int dx\,
\mathcal{L}_\mathrm{top}(Q)$, with the energy-momentum integral 
\begin{align*}
    I_1= \frac{i}{32} \int_{E}^{\infty} d  \epsilon \int dk\sum_s s \epsilon_{ij}\,\tr([\hat{G}^s,\partial_j\hat{G}^s] F_i).
\end{align*}
Turning to second order term and using the first of the above auxiliary relations,
it is straightforward to derive an analogous expression, $S^{(2)}=I_2 \int dx\,
\mathcal{L}_\mathrm{top}(Q)$, where
\begin{align*}
    I_2 = -\frac{1}{32} \epsilon_{ij}\int dk\sum_s s\,\tr(\hat{G}^s F_i \hat{G}^{-s} F_j),
\end{align*}
and we again retain only contributions which combine to a non-vanishing trace.
It remains to make sense of the momentum integrals, $I_1$ and $I_2$. To this end, we engage the  eigenfunction representation, 
\begin{align*}
    \hat{G}(E)=\sum_\alpha  |\alpha\rangle \frac{1}{E+is \kappa -\epsilon_\alpha}\langle \alpha |,\qquad F_i =-i \partial_i \hat{H} = -i \sum_\alpha \epsilon_\alpha \partial_i |\alpha\rangle \langle \alpha|, 
\end{align*}
where we anticipate that momentum derivatives of energies will not contribute to
an expression of  topological significance (this can be checked by
explicit computation). Substituting the first of these identities into $I_1$, the
energy dependent denominators can all be pulled out and integrated over. As a
result, we obtain
\begin{align*}
    I_1=\frac{i \pi}{16}\int dk  \epsilon_{ij} 
    \sum_{\alpha \beta} \frac{1}{\epsilon_\alpha-\epsilon_\beta}
    \left( \delta(E-\epsilon_\alpha)+ \delta(E-E_\beta)-
    \frac{2}{\epsilon_\alpha-\epsilon_\beta}
    \left( \Theta(\epsilon_\alpha-E)- \Theta(\epsilon_\beta-E) \right)\right) \langle \alpha |\partial_j \hat{H} | \beta \rangle \langle \beta | \partial_i \hat{H} |\alpha \rangle,
\end{align*}  
where we assumed the disorder to be weak enough to justify the approximation $
\delta(E-\epsilon_\alpha)=-\frac{1}{\pi} \mathrm{Im}(E + i \kappa -
\epsilon_\alpha)^{-1}$. In $I_2$ no energy integral needs to be done, and the
substitution of the spectral decomposition leads to 
\begin{align*}
    I_2 = -\frac{i\pi }{16} \epsilon_{ij} \sum_{\alpha \beta} \int (dk) 
    \frac{1}{\epsilon_\alpha-\epsilon_\beta}
     \left(\delta(E-\epsilon_\alpha)+\delta(E-\epsilon_\beta)\right)\langle \alpha |\partial_i \hat{H} | \beta \rangle \langle \beta | \partial_j \hat{H} |\alpha \rangle.
\end{align*}
We observe that in the combination $S_\textrm{top}=(I_1+I_2)\int dx\,
\mathcal{L}_\textrm{top}(Q)$ the on--Fermi--shell term $I_2$ cancels against the
on--shell contributions of $I_1$, a phenomenon  which in the context of the
quantum Hall effect is known as the cancellation of the Streda I Fermi surface
conductance against a contribution to the Streda II conductance~\cite{Pruisken1984a}. In a
final step, we substitute the second of the above spectral decompositions to
compute the matrix elements as $\langle \alpha|\partial_i \hat{H}
|\beta\rangle=\epsilon_\beta \langle \alpha|\partial_i \beta \rangle +
\epsilon_\alpha \langle \partial_i \alpha|\beta \rangle =
(\epsilon_\alpha-\epsilon_\beta)\langle \partial_i \alpha|\beta \rangle$.
Substitution into $I_1+I_2$ leads to the final result
\begin{align}
    \label{eq:TopologicalAction}
    S_\textrm{top}[Q]= \frac{1}{16 \pi}\int d^2k\, \Omega_k \Theta(\epsilon_\alpha-E) \,\int dx\, \epsilon_{ij}\tr(Q \partial_i Q \partial_j Q), 
\end{align}
where $\Omega_k = i\langle d \alpha |\wedge d \alpha\rangle= i \epsilon_{ij}  \langle \partial_i \alpha| \partial_j \alpha \rangle$.
\end{widetext}

\newpage
$\phantom{0}$
\newpage
$\phantom{0}$
\newpage

\section{Details of numerical calculation of the multifractal spectra}
\label{app:Numerics}

\subsection{Effective multifractal exponent}

In practice to analyze the convergence of the numerical calculation in the
linear surface dimension $L$ we define an effective $L$-dependent multifractal
dimension
\begin{equation}
    \tau_q^E(L)=-\frac{\ln P_q^E(L) -\ln P_q^E(L/2)}{\ln L -\ln L/2}.
\end{equation}
For a critical point with  multifractal scaling, this quantity will converge to
the true multifractal exponent $\lim_{L\to\infty}\tau_q^E(L)\to \tau_q^E$ according to Eq.~\eqref{eq:powerlawIPR}. \cite{evers2008}
However, away from a critical point, where   wave functions localize for
sufficiently large system sizes, $L$,  its value will not converge until  $L$ is
larger than the localization length and $\tau_q^E \to 0$. In this way the
effective exponent  allows us to distinguish between a localizing and critical
behavior. In the latter case, it also   quantifies the convergence of the
multifractal spectrum.

\subsection{Distribution functions of inverse participation ratios}
\label{app:distros}

\begin{figure}
    \centering
    \includegraphics[width=\linewidth]{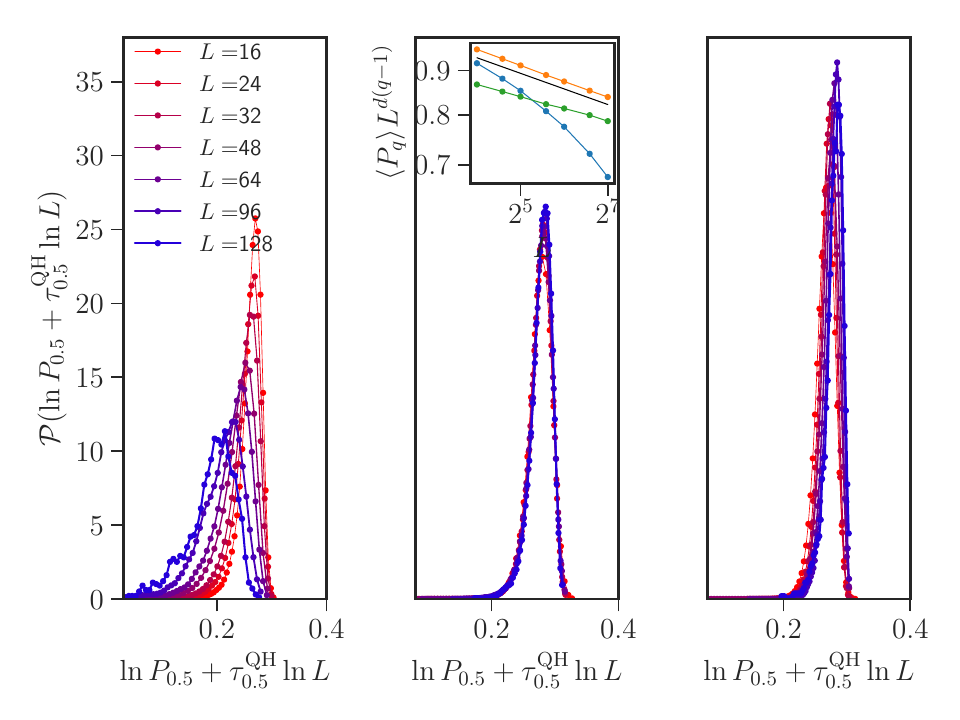}
    \caption{Distribution functions of $P_q$ at $q=0.5$ for   $u_\mathrm{f}=0.3$ (left), $u_\mathrm{f}=0$ (middle), and $u_\mathrm{f}$ random (right) with $\mathrm{rms}\, u_\mathrm{f}=0.5$. (Here, the topological control parameter 
    is $M=2.0$ and the disorder strength is $W=0.2$.) The horizontal axis has been rescaled as discusseed in the text.  
    Inset: the convergence of the mean of the IPRs, scaled by the trivial
    scaling exponent of fully extended states for
    constant (blue), zero (yellow) and random (green) $u_\mathrm{f}$.  }
    \label{fig:distros}
\end{figure}

In a numerical experiment, the moments $P_q$ are randomly distributed
 quantities, whose mean values are shown in  Fig.~\ref{fig:mfa}. The probability
 distribution of the moments $P_{0.5}$ is shown in Fig.~\ref{fig:distros}, again
 for the three cases  constant, zero and random $u_{\rm f}$. For ideal quantum
 Hall criticality, we expect $P_{0.5}= c \, L^{-\tau_{0.5}}$, with the exponent
 [cf.~Eqs.~\eqref{eq:DeltaqDef} and ~\eqref{eq:qh}]  $\tau_{0.5}\approx
 -1+\frac{1}{16}$. This implies that the  variable $\ln[P_{0.5}+\tau_{0.5}\ln
 (L)]$ should be distributed around the non-universal constant $\ln
 c$~\cite{evers2008,puschmann2021}. The figure shows for $u_{\rm f}=0$ (center) this
 variable is indeed narrowly distributed around a maximum, with data collapse
 for all values of $L$. Qualitatively similar behavior is found for random
 $u_\mathrm{f}$ (right). While, for the numerically accessible system sizes the
 scaling limit is harder to reach in this case, the  collapse becomes more
 pronounced for our largest values of $L$, shown in blue. However, for non-vanishing
 constant $u_\mathrm{f}$ (left) the data cannot be scaled to collapse, including
 for different values of $\tau_{0.5}$. In this way, the absence of  criticality
 reveals itself.  
 
 In the inset we show the scaling of the mean of the moments with system size,
rescaled by the trivial extended scaling behavior. Blue (yellow, green) data
represents the finite (vanishing, random) $u_\mathrm{f}$, where the latter two
asymptote towards the QH scaling indicated as a black dotted line.

\subsection{Convergence in transverse direction}

\label{app:conv}

\begin{figure}
    \centering
    \includegraphics[width=\linewidth]{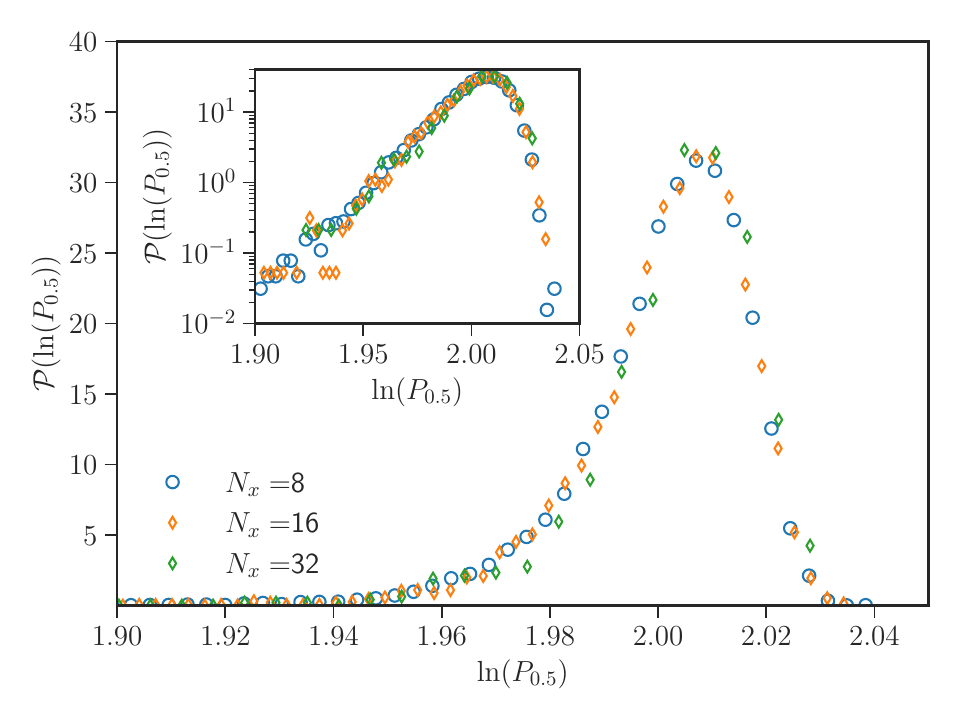}
    \caption{Distribution functions of $P_q$ at $E=0.1$, $u_\mathrm{f}=0$, $W=0.15$ and linear surface extension $L=64$ for different slab widths $N_x$ on a linear  (main) and semi-logarithmic scale (inset).}
    \label{fig:slab}
\end{figure}

We here demonstrate that already  small slab widths $L_x$  are sufficient to
make quantitative statements about the localization properties of the surface
states. As an example, Fig. \ref{fig:slab} shows the distrubtion function of the
moments $P_{0.5}$ for  $u_\mathrm{f}=0$ and a surface extension $L=64$ for
different values of $L_x$. The full distribution function, including mean and
tails, coincide, indicating that for all shown slab widths the distribution
function are already converged. 

The reason for the observed width--independence is  the exponential transverse localization of surface states, with a decay length of order one layer or less. Here,
and in the  numerical calculations shown above, only surface states
with a surface weight of more than $75\%$ are taken into account to avoid
artefacts due to low lying bulk states. (These are rare but they exist due to disorder
inside the clean bulk gap.)

\section{Class AIII superconductors}
\label{sec:AIIISC}

In a superconductor with a spin rotational invariance around a fixed axis, the Bogoliubov-de Gennes (BdG) Hamiltonian $H = \mbox{diag}\, (h_{\uparrow},h_{\downarrow})$ splits into two blocks, corresponding to ``spin up'' and ``spin down'' sectors. (The BdG Hamiltonian acts on four-component spinors with spin and particle-hole degrees of freedom.) Particle-hole conjugation $\mathcal{C}$ and time-reversal symmetry $\mathcal{T}$ map these two blocks onto each other,
\begin{equation}
  h_{\uparrow} = -\mathcal{C}^{-1} h_{\downarrow} \mathcal{C} = \mathcal{T}^{-1} h_{\downarrow} \mathcal{T},
\end{equation}
so that it is sufficient to consider the ``spin-up'' block $h \equiv h_{\uparrow}$ only.
The product $\mathcal{S} = \mathcal{C T}$ acts as an antisymmetry constraint on the Hamiltonian, 
\begin{equation}
  h = - \mathcal{S}^{-1} h \mathcal{S}.
\end{equation}
It follows that a superconductor with time-reversal symmetry and a remnant U(1) of spin SU(2) rotational invariance resides in class AIII \cite{schnyder2008,foster2008}. We now make these arguments more explicit, using a formulation in terms of fermion creation and annihilation operators, so that particle-hole symmetry is automatically encoded in the fermion anticommutation relations and need not be implemented explicitly.

For a system of spin-1/2 electrons, we can form the spin-triplet Cooper-pair annihilation operator in position space
\begin{align}\label{eq:baDef}
    b_a(\vex{r},\vex{r'}) 
    \equiv&\, 
    c_{\sigma}(\vex{r})
    \,
    c_{\sigma'}(\vex{r'})
    \,
    \left(\sigma_2 \, \sigma_a\right)_{\sigma,\sigma'}
\nonumber\\
    =&\,
    c^\mathsf{T}(\vex{r})
    \,
    \sigma_2 \, \sigma_a
    \,
    c(\vex{r'}).
\end{align}
Here the electron annihilation operator $c_{\sigma}(\vex{r})$ carries spin indices $\sigma \in \{\uparrow,\downarrow\}$,
and the Pauli matrices $\sigma_a$ act on this space; repeated indices are summed.
On the second line of Eq.~(\ref{eq:baDef}), we suppress indices and $\mathsf{T}$ denotes the transpose, viewing $c$ ($c^{\mathsf{T}}$) as a column (row) spinor.
The pair operator is antisymmetric (``$p$-wave'') under the exchange of $\vex{r} \leftrightarrow \vex{r'}$, and transforms like a vector under spin SU(2) rotations of the fermions.
Under the physical $\mathcal{T}^2 = -1$ antiunitary time-reversal transformation, 
\begin{align}\label{eq:T}
    c(\vex{r}) \rightarrow i \sigma_2 \, c(\vex{r}),
    \quad 
    i \rightarrow -i,
\end{align}
the pair operator $b_a(\vex{r},\vex{r}')$ is \emph{invariant}. By contrast the ordinary magnetization density inverts under $\mathcal{T}$.

In Bogoliubov-de Gennes (BdG) static mean field theory, the Hamiltonian for a spin-triplet superconductor can be expressed as 
\begin{align}\label{eq:HBdG}
    H
    =&\,
    \int 
    \frac{d^d \vex{k}}{(2 \pi)^d}
    \,
    \tilde{\varepsilon}(\vex{k})
    \,
    c^\dagger(\vex{k}) 
    \,    
    c(\vex{k})
\nonumber\\
&\,
    +
    \frac{1}{2}
    \sum_{\vex{r},\vex{r'}}
    \,
    \left[
        \Delta_a(\vex{r} - \vex{r'})
        \,
        b_a^\dagger(\vex{r},\vex{r'})
        +
        \mathrm{H.c.}
    \right],
\end{align}
where $\mathrm{H.c.}$ denotes the Hermitian conjugate.
Here $c(\vex{k})$ is the Fourier transform of the position-space annihilation spinor, and
$\tilde{\varepsilon}(\vex{k})$ denotes the normal-state band structure, incorporating the chemical potential.
The vector-valued function $\Delta_a(\vex{r}) = - \Delta_a(-\vex{r})$ is the mean-field
BCS order parameter. With a particular gauge choice, Eq.~(\ref{eq:HBdG}) is invariant under the time-reversal transformation in Eq.~(\ref{eq:T}) if the band structure is invariant and 
$\Delta_a(\vex{r}) = \Delta_a^*(\vex{r})$.

If we restrict to a real-valued $\Delta_a(\vex{r}) = \delta_{a,3} \, \Delta(\vex{r})$, then 
Eq.~(\ref{eq:HBdG}) describes a time-reversal invariant superconductor with a remnant U(1) of spin SU(2) invariance, corresponding to rotations about the $z$-axis in spin space. To see that this is class AIII, we reformulate in terms of the Nambu spinor,
\begin{align}
    \eta(\vex{r})
    \equiv  
    \begin{bmatrix}
    c_{\uparrow}(\vex{r})\\
    c_{\downarrow}^\dagger(\vex{r})
    \end{bmatrix},
    \quad
    \eta^\dagger(\vex{r})
    =
    \begin{bmatrix}
    c_\uparrow^\dagger(\vex{r})
    &
    c_{\downarrow}(\vex{r})
    \end{bmatrix}.
\end{align}
In the Nambu language, a $z$-axis spin rotation becomes the U(1) transformation 
$\eta \rightarrow e^{i \phi/2} \, \eta$,
$\eta^\dagger \rightarrow \eta^\dagger \, e^{-i \phi/2}$.
Time-reversal [Eq.~(\ref{eq:T})] becomes 
\begin{align}
\label{eq:NambuT}
    \eta(\vex{r})
    \rightarrow
    \begin{bmatrix}
    c_{\downarrow}(\vex{r}) 
    \\
    -
    c_{\uparrow}^\dagger(\vex{r})
    \end{bmatrix}
    =
    i \tau_2
    \left[
        \eta^\dagger(\vex{r})
    \right]^\mathsf{T}.
\end{align}
Here the Pauli matrix $\tau_2$ acts on the components of the Nambu spinor,
and  $\left[\eta^\dagger\right]^\mathsf{T}$ is the column spinor corresponding to the row $\eta^\dagger$. 
Eq.~(\ref{eq:NambuT}) implies that time-reversal acts like an \emph{antiunitary
particle-hole transformation} in the Nambu language, because spin (unlike electric charge) inverts under $\mathcal{T}$. This is in fact chiral symmetry in second quantization. 
To see this, we recast Eq.~(\ref{eq:HBdG}) compactly as
\begin{align}
    H = \frac{1}{2} \eta^\dagger h \eta,
\end{align}
where $h$ is the Hermitian BdG Hamiltonian that acts on position and Nambu components. This form is manifestly invariant under spin U(1) rotations. Imposing invariance under $\mathcal{T}$ in Eq.~(\ref{eq:NambuT}) leads to the chiral condition on $h$,
\begin{align}
    - \tau_2 \, h \, \tau_2 = h.
\end{align}
Physical time-reversal symmetry is thus transmuted into a chiral condition on the BdG Hamiltonian. Since there are no other constraints on $h$, the superconductor resides in class AIII.

The other topological superconductor classes in three dimensions are CI and DIII; both require physical time-reversal symmetry \cite{schnyder2008,ryu2010}. Class CI, in addition, possesses a $\mathcal{C}^2 = -1$ particle-hole symmetry. In the superconductor interpretation, this encodes invariance under $\pi$-rotations along $x$ and $y$ axes in spin space, which is tantamount to full SU(2) symmetry. Class DIII by contrast has no spin symmetry, and is usually cast in terms of a real (Balian-Werthammer \cite{ColemanBook}) spinor that encodes both spin and particle-hole degrees of freedom. Physical time-reversal also appears as a chiral condition, while $\mathcal{C}^2 = +1$ particle-hole symmetry is imposed on the BdG Hamiltonian by the reality condition on the spinor.

\end{document}